\documentclass[11 pt]{article}

\expandafter\let\csname equation*\endcsname\relax
\expandafter\let\csname endequation*\endcsname\relax
\usepackage[dvips]{graphicx}
\usepackage{paralist}
\usepackage{bbding}
\usepackage{url}
\usepackage{fancyhdr,array}
\usepackage{footmisc}
\usepackage{listings}
\usepackage{fancyvrb}
\usepackage{amsmath,epsfig,amssymb,bbm,stmaryrd,mathrsfs}
\usepackage{pstricks}
\usepackage{pstricks}
\usepackage{subfigure}
\usepackage{wrapfig}
\usepackage{lipsum}
\usepackage{mdframed,algorithmic}
\usepackage[boxruled]{algorithm2e}

\usepackage{cite}
\usepackage{amsmath,amssymb,subfigure,epsfig,bbm,stmaryrd,MnSymbol,mathrsfs,url}
\usepackage{pstricks}
\usepackage{subfigure}
\usepackage{booktabs}
\usepackage{multirow}
\usepackage{amsmath,amssymb}
\usepackage{pstricks}
\usepackage{amsmath,amssymb,subfigure,epsfig,bbm,stmaryrd,MnSymbol,mathrsfs}
\usepackage{subfigure}
\DeclareMathAlphabet{\mathpzc}{OT1}{pzc}{m}{it}

\newcommand{\h}{\mathpzc{H}}
\newcommand{\bdelta}{\boldsymbol{\delta}}
\newcommand{\bmu}{\boldsymbol{\mu}}
\newcommand{\argmin}{\operatornamewithlimits{arg\;\!min\;}}
\newcommand{\maxi}{\operatornamewithlimits{max\;}}

\newcommand{\x}{\mathbf{x}}
\newcommand{\p}{\mathbf{p}}
\newcommand{\y}{\boldsymbol{y}}

\begin{document}

\newcommand{\BigFig}[1]{\parbox{12pt}{\Huge #1}}
\newcommand{\BigZero}{\BigFig{0}}

\title{\bf A Geometric Approach to Joint Inversion with Applications to Contaminant Source Zone Characterization}   

\author{Alireza Aghasi${}^{1,2}$, Itza Mendoza-Sanchez${}^{3}$,
 Eric L. Miller${}^{2}$,\\ C. Andrew Ramsburg${}^{3}$
 and Linda M. Abriola ${}^{3}$ \thanks{${}^{1}$School of Electrical and Computer Engineering, Georgia Institute of Technology, Atlanta, GA, USA\newline
${}^{2}$Department of Electrical and Computer Engineering, Tufts University, Medford, MA, USA\newline
${}^{3}$Department of Civil and Environmental Engineering, Tufts University,
Medford, MA, USA\newline Email contacts: \tt aghasi@gatech.edu, \tt \{itza.mendoza-sanchez, eric.miller, andrew.ramsburg, linda.abriola\}@tufts.edu.}}

\date{}
\maketitle

\begin{abstract}

This paper presents a new joint inversion approach to shape-based inverse problems. Given two sets of data from distinct physical models, the main objective is to obtain a unified characterization of inclusions within the spatial domain of the physical properties to be reconstructed. Although our proposed method generally applies to many types of inverse problems, the main motivation here is to characterize subsurface contaminant source-zones by  processing down  gradient hydrological data and cross-gradient electrical  resistance tomography  (ERT) observations. Inspired by Newton's  method for multi-objective optimization, we present an iterative  inversion scheme in which descent steps are chosen to simultaneously reduce both data-model misfit terms. Such an approach, however, requires solving a non-smooth convex problem at  every iteration, which is computationally expensive for a pixel-based inversion over the whole domain. Instead, we employ a parametric level set (PaLS) technique that substantially reduces the number of underlying parameters, making the inversion  computationally tractable. The  performance of the technique is examined and discussed through the reconstruction of source zone architectures that are representative of dense non-aqueous phase liquid (DNAPL) contaminant release in a statistically homogenous sandy aquifer. In these examples, the geometric configuration of the DNAPL mass is considered along with additional information about its spatial variability within the contaminated zone, such as the identification of low and high saturation regions.    Comparison of the reconstructions with the true DNAPL  architectures highlights the superior performance of the model-based technique and joint inversion scheme.

\end{abstract}

\section{Introduction} \label{sec1}
In recent years, there has been increasing interest, especially with respect to subsurface sensing applications, in the development of inversion methods that process data from highly heterogeneous sets of sensors to obtain a unified characterization of a region of space \cite{vozoff1975joint, haber1999joint, linde2006improved, hinnell2010improved, hoversten2006direct, hyndman1994coupled}. These \emph{joint inversion} techniques are motivated by the idea that a variety of modalities, each sensitive to a different set of constitutive parameters, combined with appropriate regularization to mathematically relate one set of parameters to the others, can significantly improve characterization relative to what can be achieved through the processing of individual modalities. The potential of such a joint inversion approach has been clearly demonstrated for various modality combinations, e.g., hydrological and seismic \cite{cardiff2009bayesian}, electromagnetic and elastic \cite{abubakar2012joint}, electromagnetic and seismic \cite{hu2009joint}, gravity and seismic \cite{haber1999joint}, gravity and magnetic \cite{gallardo2011structure}, as well as magneto-telluric, gravity and seismic data \cite{moorkamp2010joint}.

In the present work, we consider a joint inversion approach based on a geometric parameterization of the problem and employing a new multi-objective optimization scheme to combine the data from disparate sensor types. While the method is general, we are specifically concerned with exploring its performance in the context of an environmental remediation problem, namely the characterization of a subsurface contaminant source zone based on both geophysical and hydrological data.  This work is motivated by the problem of remediating sites contaminated by dense non-aqueous  phase liquids (DNAPLs), such as chlorinated solvents including trichloroethylene (TCE) or tetrachloroethylene (PCE), which are used in dry cleaning, degreasing operations, and gas production.  Successful treatment and management of source zones contaminated by such compounds typically is predicated on knowledge of the mass distribution of the DNAPLs in the subsurface.

Partitioning tracer tests (PTTs) are among the most developed methods for DNAPL source zone characterization \cite{jin1995partitioning,enfield2005design}. In this technique, boreholes are used to inject and extract tracer fluids. Based on the physics of the problem and a knowledge of observed tracer concentration data in the extraction well (and sometimes in boreholes along the flow path), an inverse problem is solved to characterize the source zone.  There are a number of drawbacks with PTTs.  For example, this technique only provides spatially averaged estimates of DNAPL saturation over flow paths defined by the typically sparse distribution of injection and pumping or observation wells \cite{moreno2006influence}.  Its reliance on the use of boreholes not only may fail to provide full coverage of the affected zone but also increases the risk of mobilizing the DNAPL mass, if observation points within the contaminated zone are employed.  Partitioning  tracer tomography \cite{Yeh2007} is a more recent  technique which attempts to address the shortcoming of PTT  by considering the full transport model and characterizing the architecture through a rigorous stochastic inversion. However, this method still suffers from the coverage and mobilization risks of PTT.

In addition  to  these hydrological methods, geophysical modalities including seismic methods  \cite{temples2001noninvasive}, ground  penetrating  radar  (GPR) \cite{brewster1994ground}, and electrical impedance/resistance tomography (EIT/ERT) \cite{daily1998electrical, chambers2004noninvasive, goes2004effective} have also received recent attention as potential  tools for DNAPL  characterization.  Less invasive than hydrologic modalities, the utility of these methods for the DNAPL problem arises from the contrast in electrical or seismic properties of the contaminant relative to the nominal subsurface \cite{ajo2006survey}. Even with these methods, data tend to be collected sparsely and one must still solve a challenging inverse problem to develop an image of the subsurface. As is typically the case, the need for regularization tends to result in images with rather coarse spatial resolution \cite{engl1996regularization}.

Joint inversion techniques may hold promise in overcoming the challenges discussed in the preceding paragraphs.  Indeed, a substantial literature has developed exploring joint inversion techniques for a variety of related subsurface characterization problems. For example, in \cite{koch2009joint} two-dimensional  ERT is performed over various ground transects to provide estimates of the structure and water content of the subsurface.  These images were then interpreted with respect to hydrological investigations of the  same basin using both tracer  methods and groundwater level observations.  Researchers have also investigated coupled inversions based on electrical resistance  tomography data and hydrological models (e.g., see  \cite{hinnell2010improved, pollock2012fully}).   In  these techniques, inversion is performed simultaneously by combining the data sets and forming a joint  model, instead of constraining the hydrological interpretations via the electrical tomography results or vice versa.  A joint  inversion over these types of  data  sets significantly  improves  the  estimation  and  reduces the  uncertainty encountered with uncoupled techniques.  Similarly, GPR data have been used along with hydrological data to perform a coupled inversion to identify soil structure \cite{Finsterle2008Joint}. In the latter study, the uniqueness and stability of the inversion  process was also analyzed numerically.

Although significant strides have been made, the majority  of the joint inversion techniques presented in the literature for subsurface characterization have been for limited scenarios (e.g. 2D problems) or simplified physical models \cite{Finsterle2008Joint, cardiff2009bayesian, koch2009joint, pollock2008temporal}. Thus, joint inversion techniques that can address large scale, complex, fully three-dimensional physical domains are highly desirable.  Inspired by the idea in \cite{fliege2010newton}, in this work, we propose a multi-objective optimization formulation for a fully three dimensional joint inversion problem addressing DNAPL source zone characterization.

The DNAPL problem is one example of a much broader class of inverse problems whose primary objective is the identification of a region of interest (source zone, tumor, crack in a material sample, etc.) embedded in an inhomogeneous background. For such problems shape-based methods, such as active contours and level sets, have received considerable attention \cite{dorn2006level}. Level sets are employed in various applications including  geophysical and hydrologic inverse problems \cite{cardiff2009bayesian, dorn2008history, lu2006parameter}. Despite their broad applicability, our prior work in this area has demonstrated that in the case of ill-posed problems (such as the one of concern in this paper), traditional level set methods require delicate regularization to perform well \cite{aghasi2011parametric}. The method presented herein employs a new parameterization for joint inverse methods, which is an extension of the alternative parametric level set (PaLS) approach, originally developed in the context of single-modality inverse methods \cite{aghasi2011parametric}.  In this work we also extend this approach to employ a pair of PaLS functions to allow for the identification of both high and low saturation regions in the source zone, known as \emph{pools} and \emph{ganglia}.

The DNAPL source zone characterization problem addressed here encompasses the hydro-geophysical inversion problem illustrated in Figure \ref{fig1}. The hydrologic measurements for this problem correspond to the observations of contaminant concentration collected on a plane (transect) orthogonal to the nominal direction of groundwater flow and located down gradient of the source zone. Electrical resistance tomography (ERT) data are also collected orthogonal to the flow direction but across planes that intersect the source zone.  The specific form of this problem is motivated by a number of factors. First, the down-gradient transect concentration data are readily available in practice and unlike PTT data, their acquisition does not risk any additional mobilization of the contaminant.  In terms of the geophysical modality, our choice of ERT was driven by the fact that there is some contrast between water and DNAPL resistance as well as the ease of modeling this modality relative to GPR or seismic data.  While GPR may provide a more robust signature than ERT for DNAPL in the field \cite{knight2001ground}, a rigorous, physics based inversion of radar data using the full Maxwell’s equations is a daunting task.  ERT on the other hand requires only the solution of Poisson's equation, which is computationally far more tractable.

A preliminary result associated with the problem posed in this paper is reported in our recent review \cite{miller2012environmental}; however the material here represents a substantial advance over that work. In \cite{miller2012environmental} a PaLS- based approach was used to characterize a simple source zone structure whose saturation profile was quite close to uniform.  Moreover the  average saturation  value was considered  known a priori  so that  only the geometry of the source zone was to be  reconstructed.  We also assumed perfect knowledge of the hydraulic permeability field.  As a result, a straightforward scalarized optimization formulation performed adequately.  In lifting the simplifying assumptions cited above to handle more realistic scenarios, we found that  a scalarization approach was no longer applicable, thereby leading to the development of the multi-objective ideas presented here. Additionally, the use of  multiple  level set functions to identify  both pool and ganglia is entirely new to this paper. In short, the work in this paper presents a fundamentally new variational approach to joint inversion. It extends the PaLS ideas, and demonstrates the overall performance of these concepts on a far more challenging form of the DNAPL  source zone characterization problem than was considered in \cite{miller2012environmental}.

The remainder of this paper is organized as follows. Section 2 provides a general formulation of the problem, discussing the electrical and hydrologic models and their relationship to source zone physical properties. Section 3 provides a brief overview of inversion techniques and presents the PaLS representation and the proposed multi- objective  minimization method.  Section 4 discusses implementation of the method and presents  and discusses  some illustrative  simulation results.  Finally,  Section 5 provides some general conclusions and implications.

\section{General Problem Formulation}
\begin{figure}\centering
\includegraphics[width=32pc]{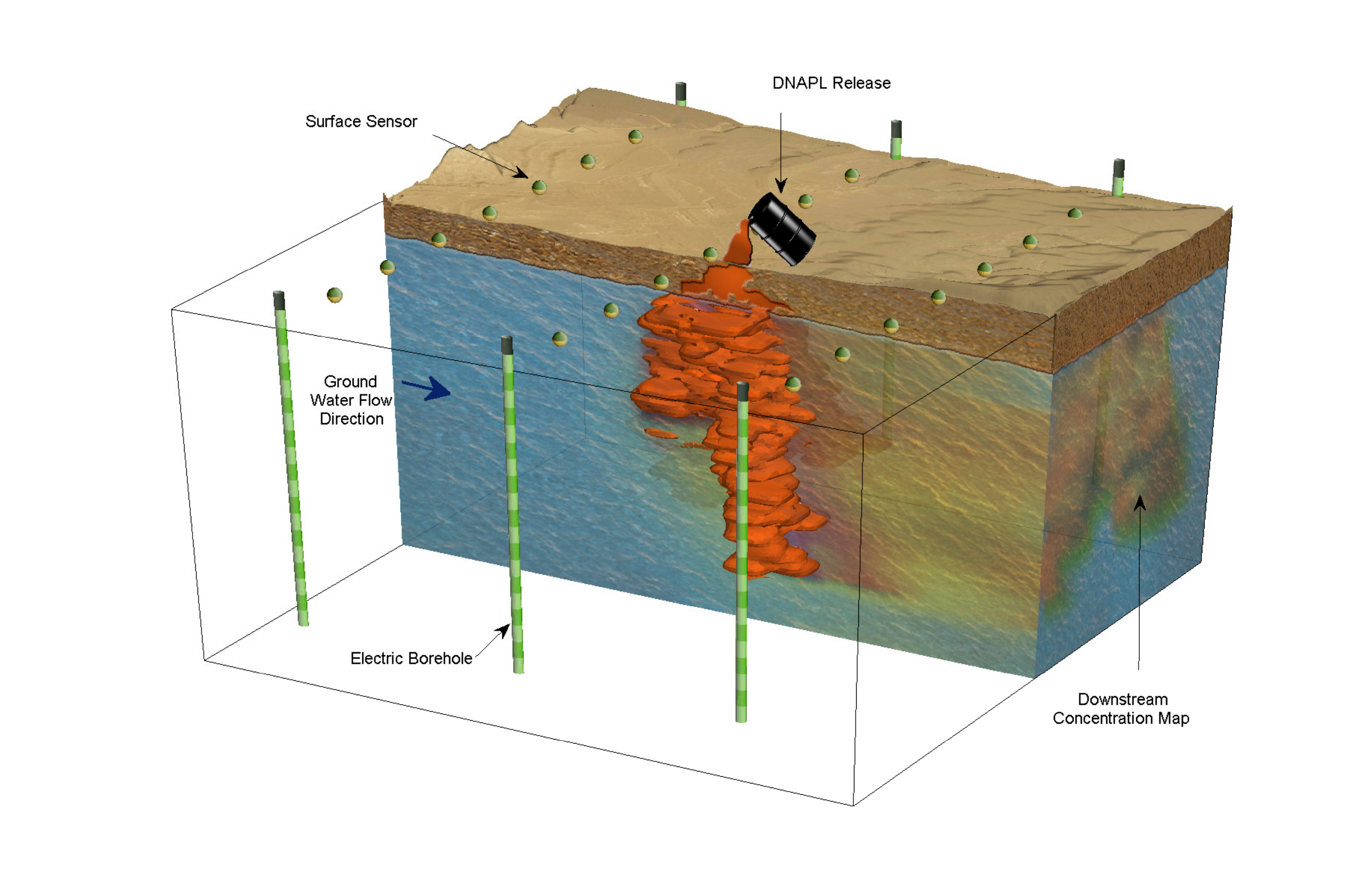}
\caption{DNAPL source zone and associated plume instrumented for hydro-geophysical assessment using down gradient concentration observations and cross-gradient electrical resistance tomography}\label{fig1}
\end{figure}

As we discussed in Section \ref{sec1}, to address source zone characterization, we consider a combination of hydrologic and geophysical measurements as illustrated in Figure \ref{fig1}. The ultimate goal of this characterization effort is to extract useful information about the source zone geometry and the underlying saturation map. In the sequel we provide a brief description of each modality and ultimately link them to the proposed joint inversion and shape-based characterization techniques.

\subsection{Multi-Phase Transport and Dissolution Model}
Simulation of DNAPL infiltration and subsequent mass dissolution in the saturated zone requires the solution of both phase and component mass balance equations. The phase mass balance equations are of the form \cite{abriola1989modeling}:
\begin{align}\label{eq1}
\frac{\partial}{\partial t}(\rho_\alpha \varphi s_\alpha)-\nabla \cdot (\rho_\alpha \boldsymbol{q}^\alpha )=\sum_{\alpha'}\sum_i E_{\alpha \alpha_i'} ,
\end{align}
where
\begin{equation}\label{eq1b}
\boldsymbol{q}^\alpha=  \varphi s_\alpha \mathbf{v}^\alpha =\frac{\mathbf{k}k_{r\alpha}}{\mu_\alpha}(\nabla P_\alpha -\rho_\alpha \mathbf{g}).
\end{equation}
Here $\rho_\alpha$ is the intrinsic mass density of the $\alpha$-phase, $\varphi$ is the matrix porosity, $s_\alpha$ is the saturation, $\boldsymbol{q}^\alpha$  is the $\alpha$-phase seepage velocity vector, $\mathbf{v}^\alpha$ is the $\alpha$-phase pore velocity vector, $\mathbf{k}$ is the intrinsic permeability tensor of the medium, $k_{r\alpha}$ is the relative permeability, $\mu_\alpha$ is the fluid phase dynamic viscosity, $P_\alpha$ is the phase pressure, $\mathbf{g}$ is the gravity vector, and $E_{\alpha \alpha_i'}$ is a mass exchange term, representing the increase in $\alpha$-phase mass due to the interphase transfer of component $i$ from the $\alpha'$ phase to the $\alpha$ phase. The first term in (\ref{eq1}) accounts for the accumulation of mass, the second term represents the change in mass due to the advective flux, and the right hand side represents the net change in $\alpha$-phase mass due to interphase mass transfer of all constituents $i$ to and from the phase. Equation (\ref{eq1}) neglects intra-phase compositional transformations, e.g., chemical reactions that take place within each phase, and for slightly soluble compounds (such as DNAPLs), the right hand side typically has negligible influence on the bulk phase flow.

Equation (\ref{eq1b}), which quantifies $\boldsymbol{q}^\alpha$, is a constitutive equation commonly known as the modified Darcy law. Other constitutive equations that relate saturation and relative permeability to the pressure differential between the fluid phases (capillary pressure, $P_c$) are required to close the phase mass balance equation system. The $P_c$-$s_\alpha$-$k_{r\alpha}$ relationships implemented in this work combine the Brooks-Corey \cite{brooks1964hydraulic} $P_c$-$s_\alpha$ model with the Burdine $k_{r\alpha}$-model \cite{burdine1953relative}, integrated into a model developed by Parker and Lenhard \cite{parker1987model} to account for entrapment hysteresis (see \cite{bradford1998flow} for a complete presentation of the hysteretic model equations).

Within each bulk fluid phase, the spatial-temporal distribution of a component $i$ is described by a component mass balance equation written in terms of the mass concentration of component $i$ in the $\alpha$-phase ($C_i^\alpha$):
\begin{align}\label{eq2}
\varphi\frac{\partial}{\partial t}(s_\alpha C_i^\alpha)+\varphi\nabla \cdot s_\alpha (C_i^\alpha \mathbf{v}^\alpha -\boldsymbol{D}_i^\alpha \cdot \nabla C_i^\alpha )=\sum_{\alpha'} E_{\alpha \alpha_i'}\ .
\end{align}
Here $\boldsymbol{D}_i^\alpha$ is the three-dimensional hydrodynamic dispersion tensor for component $i$ in phase $\alpha$ \cite{bear1988dynamics}. In this work, the interphase mass exchange of component $i$ from the $\alpha'$-phase to the $\alpha$-phase (i.e., $E_{\alpha \alpha_i'}$) is represented using a linear driving force expression \cite{WeberJr1996}, where interphase transfer (dissolution) is controlled by diffusion across a thin (stagnant) aqueous boundary layer that surrounds the entrapped DNAPL phase. The mass exchange is assumed proportional to the difference between component concentrations across this layer, with the proportionality coefficient (an effective mass transfer coefficient) determined by a mass transfer correlation expression.  Simulations presented in this work employed the laboratory-validated correlation for DNAPL dissolution presented in \cite{powers1992experimental}.

For the applications presented here, only two fluid phases are modeled, the aqueous phase ($\alpha=w$) and the DNAPL ($\alpha=n$), which requires that $s_n=1-s_w$. Because the focus of this study is on the DNAPL mass, sorption to the solid phase is neglected and no independent equation is written for the solid phase. Representative DNAPL source zone saturation distributions were developed by solving a coupled system of equations of the form (\ref{eq1}) for fluid phase pressures, to describe DNAPL infiltration and redistribution. For the plume transport simulations used in the inversion, the redistributed DNAPL is assumed immobile and composed of a single component (i.e., $i$ indexing is not needed). Thus, solution of only a single (aqueous) flow equation is required and this is coupled to a single transport equation of the form (\ref{eq2}). Given that sorption has been neglected, dissolution is the only interphase mass transfer process considered.

\subsection{Electrical Resistance Tomography}
The ERT model is based on introduction of electrical current into a medium and measuring the electrical potential at the periphery of the medium to analyze how the electrical conductivity is distributed throughout the medium. The underlying partial differential equation which relates the potential, $u(\x)$, to the conductivity $\sigma(\x)$ and the electric current distribution $j(\x)$ is
\begin{equation}\label{eq3}
\nabla \cdot (\sigma \nabla u)=j\ ,
\end{equation}
with the boundary conditions
\begin{eqnarray}
\hspace{1.5cm}\mathbf{n}\cdot \nabla u &=& 0, \qquad \qquad\x\in \Gamma_n\ ,\nonumber \\
\hspace{1.5cm}\mathbf{n}\cdot \nabla u +\zeta u  & = & 0, \qquad \qquad\x\in \Gamma_{mix}\ .\label{eq4}
\end{eqnarray}
In (\ref{eq4}), $\mathbf{n}$ is the surface normal and $\Gamma_n$ is a no-current boundary corresponding to the air-soil interface modeled using a Neumann boundary condition. Over $\Gamma_{mix}$ an infinite half space is approximated using a mixed boundary condition by appropriately choosing the function $\zeta$ (see \cite{dey1979resistivity, pollock2008temporal}). The introduction of current is usually represented as point source dipoles of the form
\begin{equation}\label{eq5}
j(\x)=J_0\big(\delta(\x-\x^+)-\delta(\x-\x^-) \big)\ ,
\end{equation}
where $\delta(.)$ is the Dirac delta function, $J_0$  is a DC current and $\x^\pm$ are the current electrode coordinates. To obtain a full data set, multiple experiments are performed with different electrodes acting as current sources.

\subsection{Petro-physical Relationship}
Petro-physical relationships link $\sigma(\x)$, the conductivity distribution measure of the domain, and in this case, the saturation of the DNAPL, $s_n(\x)$ (i.e., $\sigma=P(s_n)$). The most widely used petro-physical model is the Archie’s law \cite{archie1942electrical}, which for this two phase aqueous- DNAPL system takes the form
\begin{eqnarray}\label{eq6}
\hspace{2cm}\sigma(s_n)& = & a\sigma_w \varphi^m s_w^q \nonumber \\
        & = & a\sigma_w \varphi^m (1-s_n)^q\ .
\end{eqnarray}
Here $\sigma_w$ is electrical conductivity of the aqueous phase, $\varphi$ is the porosity of the medium, $a$ is a fitting parameter, $m$ is a fitting parameter that commonly referred to as the cementation index, and $q$ is the saturation index. At large saturations of an electrically conductive aqueous phase, the value of the saturation index can be approximated as 2.0 \cite{ewing2006dependence}. Archie’s Law assumes that the solid and DNAPL do not contribute to the electrical conductivity. Hunt in \cite{hunt2004continuum} provided theoretical justification for the form of this model using continuum percolation theory. In this paper we use Archie's law, although the inversion approach developed is not specific to the selected petro-physical model.

\section{Inversion Strategy}
\subsection{Pixel Based and Shape Based Methods}
The goal of most inverse problems is to extract information about a physical property in space, $p=p(\x)$, using data that are linked to $p$ via a physical model. Consider $\boldsymbol{\mathcal{M}}$ as the computational model that maps $p$ to the data vector $\boldsymbol{d}$. For simplicity we start with a single modality and later extend the notion to more than one model. A straightforward strategy to obtain an estimate $p^*$ of $p$ is to minimize model-data mismatch in a variational sense:
\begin{equation}\label{eq7}
p^*=\argmin_p \frac{1}{2}\| \boldsymbol{d}-\boldsymbol{\mathcal{M}}(p)\|_{\boldsymbol{R}}^2\ ,
\end{equation}
where for a vector $\boldsymbol{u}$ and a symmetric positive definite matrix $\boldsymbol{R}$
\begin{equation}\label{eq8}
\|\boldsymbol{u}\|_{\boldsymbol{R}}^2=\boldsymbol{u}^T \boldsymbol{R} \boldsymbol{u}\ .
\end{equation}
The matrix $\boldsymbol{R}$ usually contains the noise statistics and a pattern for weighting the data. To perform the inversion using conventional approaches, $p$ is discretized over a dense grid of pixels in the region of interest and the minimization is carried out to find the corresponding pixel values. Given the practical limitations in acquiring dense, rich sets of data, many problems of this kind are ill-posed and require regularization. Well known regularizations typically take the form of added penalties to the inversion cost function to control the amplitude and smoothness of the of reconstruction \cite{Tikhonov1977, acar1994analysis}.

Shape-based methods are another class of techniques that are capable of better posing the problem. A shape-based approach proceeds by partitioning the domain of interest into a number of zones defined by similar property values. The inverse problem then amounts to determining the boundaries of each of the zones, along with a (generally low-order) representation for the spatial distribution of the property in each zone. This technique specifically suits inverse problems where the main objective is the characterization of an inclusion within a background domain.

The most well known shape-based technique is the level set method \cite{osher1988fronts}, in which the shape boundaries are represented via the zero level set of a higher dimensional surface. Consider the basic binary case in which the domain of interest, $D$, is composed of two regions $D_1$ and $D_2$, where $p(\x)=p_1$ in $D_1$ and $p(\x)=p_2$ in $D_2$. For a shape-based representation, one can characterize both zones using a level set function $\phi(.)$, such that
\begin{equation}\label{eq9}
\left\{
     \begin{array}{lr}
       \phi(\x)>0 & \quad \x \in D_1\\
       \phi(\x)<0 &  \quad \x \in D_2
     \end{array}
   \right.
\end{equation}
and accordingly rewrite the property of interest in terms of $\phi$, $p_1$ and $p_2$ as
\begin{equation}\label{eq10}
p(\x)=p_1 H\big ( \phi (\x) \big )+ p_2 \Big ( 1- H\big ( \phi (\x) \big )\Big)\ ,
\end{equation}
where $H(.)$ represents the Heaviside step function. The scalar anomaly coefficients $p_1$ and $p_2$ may in general be functions of $\x$, representing some low order representation of the anomaly texture in each zone (e.g., see \cite{kilmer2003three}). The binary case discussed above can be generalized to multiple regions by using more than one level set function (e.g., see \cite{cardiff2009bayesian}).

\begin{figure}[t]\centering
\begin{center}$
\begin{array}{cc}
\includegraphics[width=16pc]{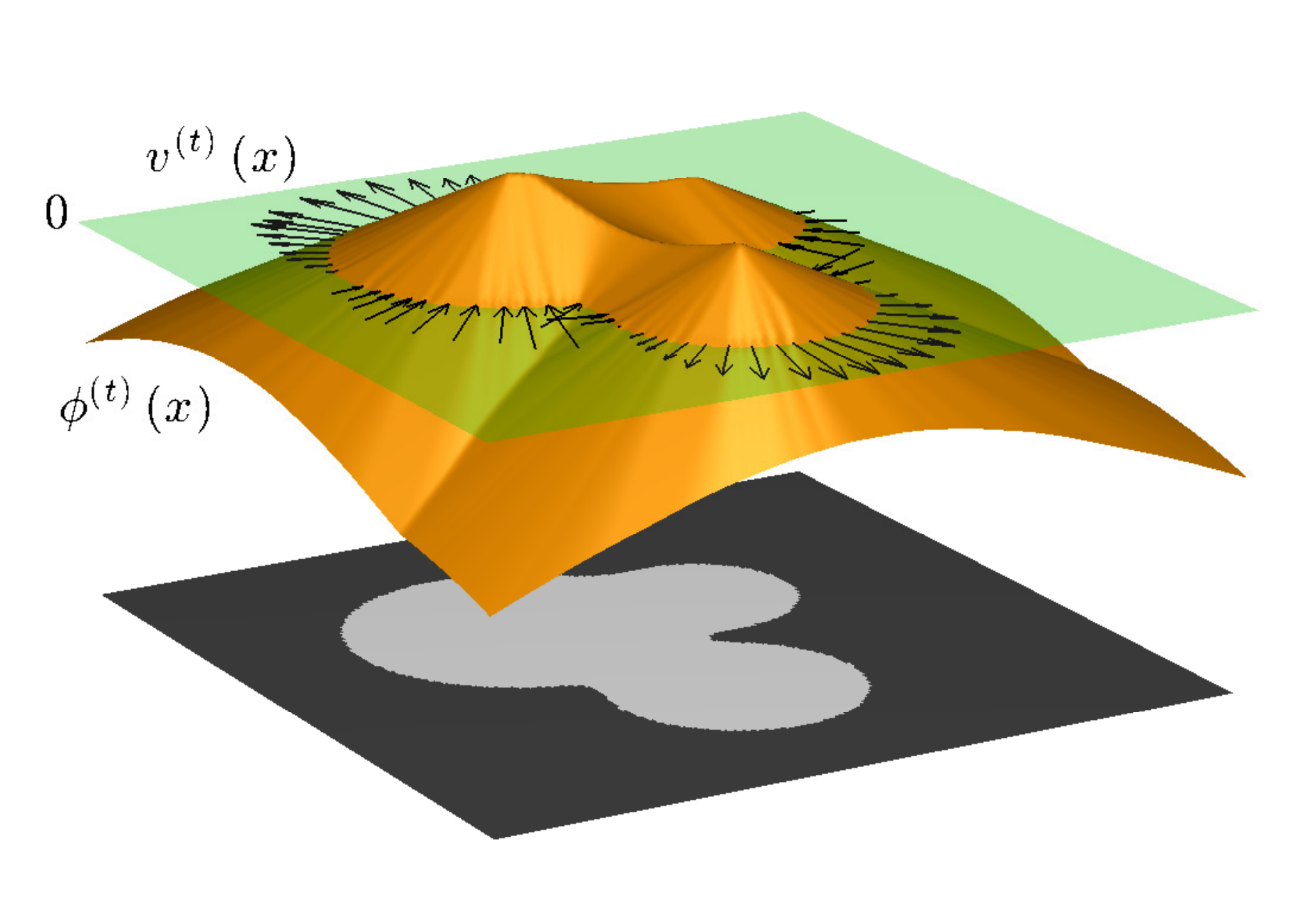}
\includegraphics[width=16pc]{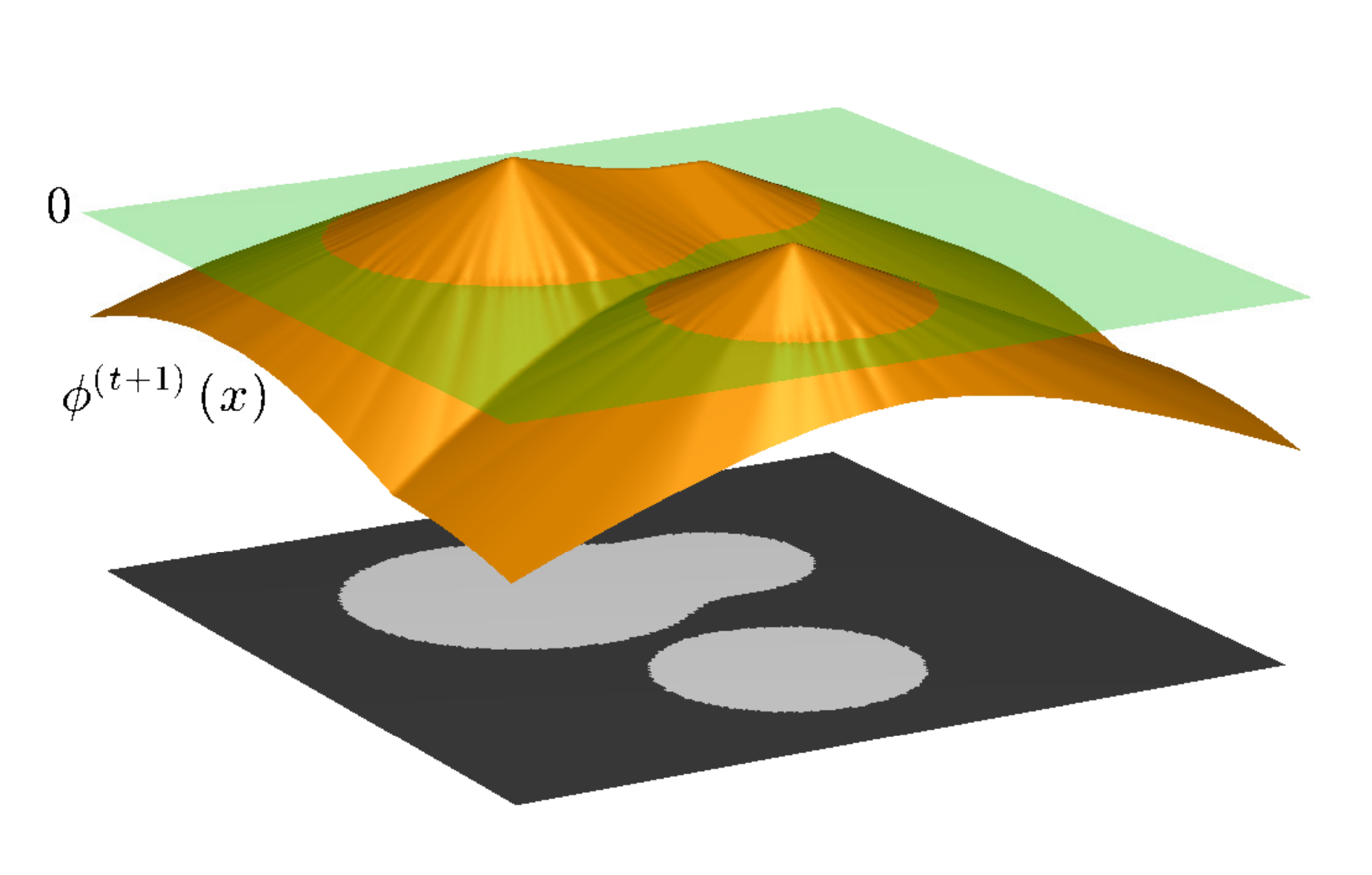}
\end{array}$
\end{center}
\caption{Level set function and topological flexibility. The orange surface represents the level set function $\phi$ and the green plane indicates the zero-level plane. The dark gray area of the lower plane is the set of points for which $\phi(\x)<0$. while the lighter gray represent the points for which $\phi(\x)>0$. In three dimensions, this latter set will represent the region of the source zone occupied by DNAPL. The bottom frame is obtained from the top by ``evolving'' the level set function according to the velocity field indicated by the vectors in the top. Under this motion, the connectivity of the underlying zero level set is able to naturally change with an unknown number of components.}\label{fig2}
\end{figure}

Minimization of (\ref{eq7}) for a level-set based property model of the form (\ref{eq10}) is performed in an evolutionary fashion. Starting with some initial level set function $\phi_0$, the function evolves to attain a state such that its zero level set best describes the true shape. The resulting time-discretized Hamilton-Jacobi type of evolution equation takes the form
\begin{equation}\label{eq11}
\phi^{(t+\Delta t)}(\x)=\phi^{(t)}(\x)-\Delta t \|\nabla \phi^{(t)}(\x) \|v^{(t)}(\x)\ ,
\end{equation}
and is initialized as $\phi^{(0)}(\x)=\phi_0 (\x)$. In this equation $t$ represents the artificial time in the evolutionary process, $\Delta t$ is the corresponding time increment and $v^{(t)}(\x)$ is a normal speed function (shown in Figure \ref{fig2}). At every iteration, the speed function is calculated based upon the sensitivity of the cost function to the current shape state \cite{dorn2006level}. As illustrated in Figure \ref{fig2}, one attractive feature of the level set technique is its topological flexibility, which allows for the identification of disjoint objects without the need to know a priori the number of components \cite{dorn2006level, burger2005survey}.

Despite the positive features, there are usually additional complexities associated with implementation of level sets, especially for inverse problems \cite{dorn2006level}. To guarantee proper convergence, the level set function should retain a certain form (usually a signed distance function). Also, since the acquired speed function only applies to the zero level set of $\phi$, speed extension methods to globally deform the level set function need to be applied \cite{osher2003level}. Moreover, from a numerical perspective, a level set function is still represented in terms of discrete grid values (pixels) which potentially increases the dimensionality of the problem. Such dimensionality can again cause problems in tackling ill-posed inverse problems and pose the challenge of applying traditional or geometric regularizations \cite{dorn2006level, ben2007projection}.

\begin{figure}[!htb]
\begin{center}$
\begin{array}{cc}
\includegraphics[width=16pc]{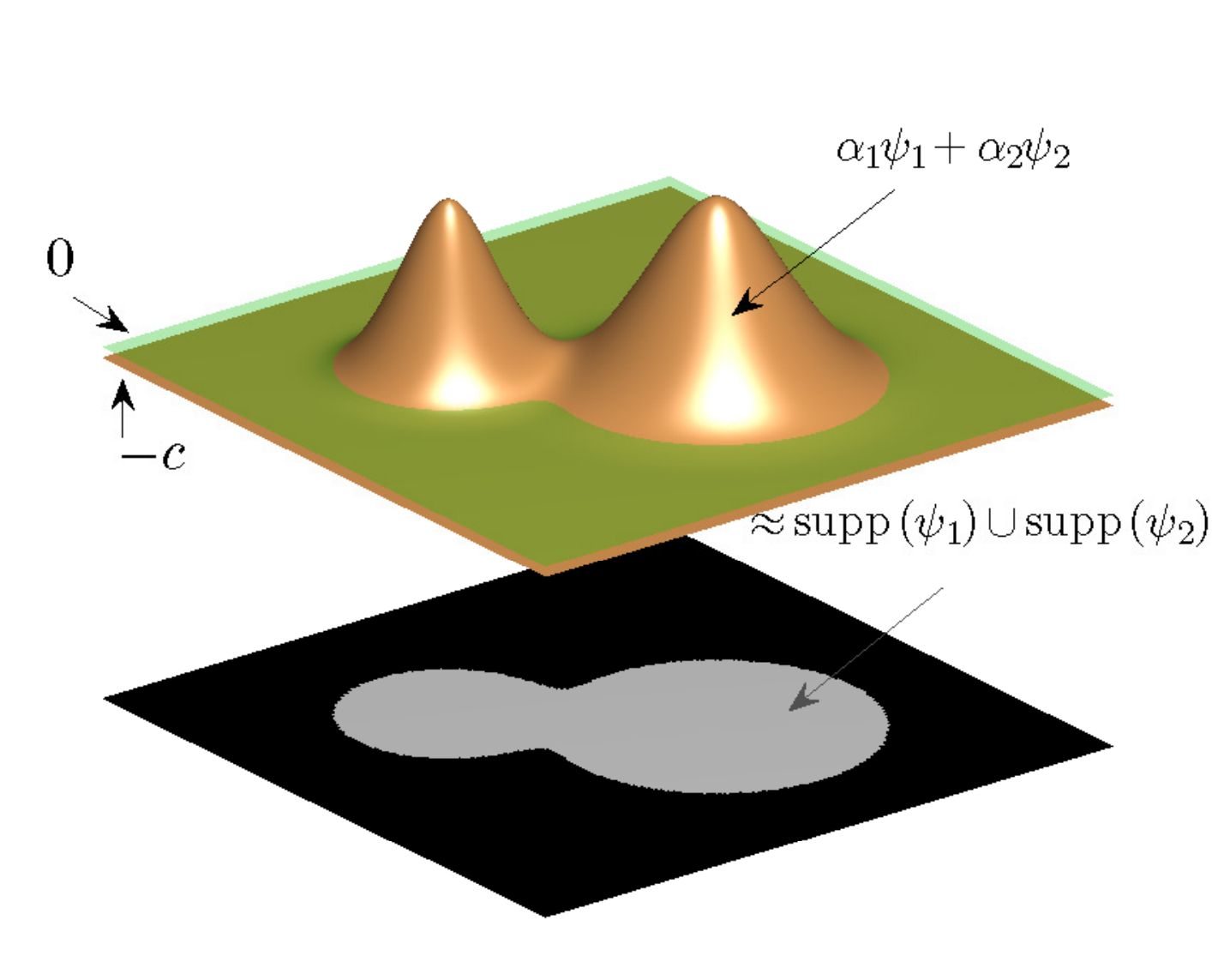}
\includegraphics[width=16pc]{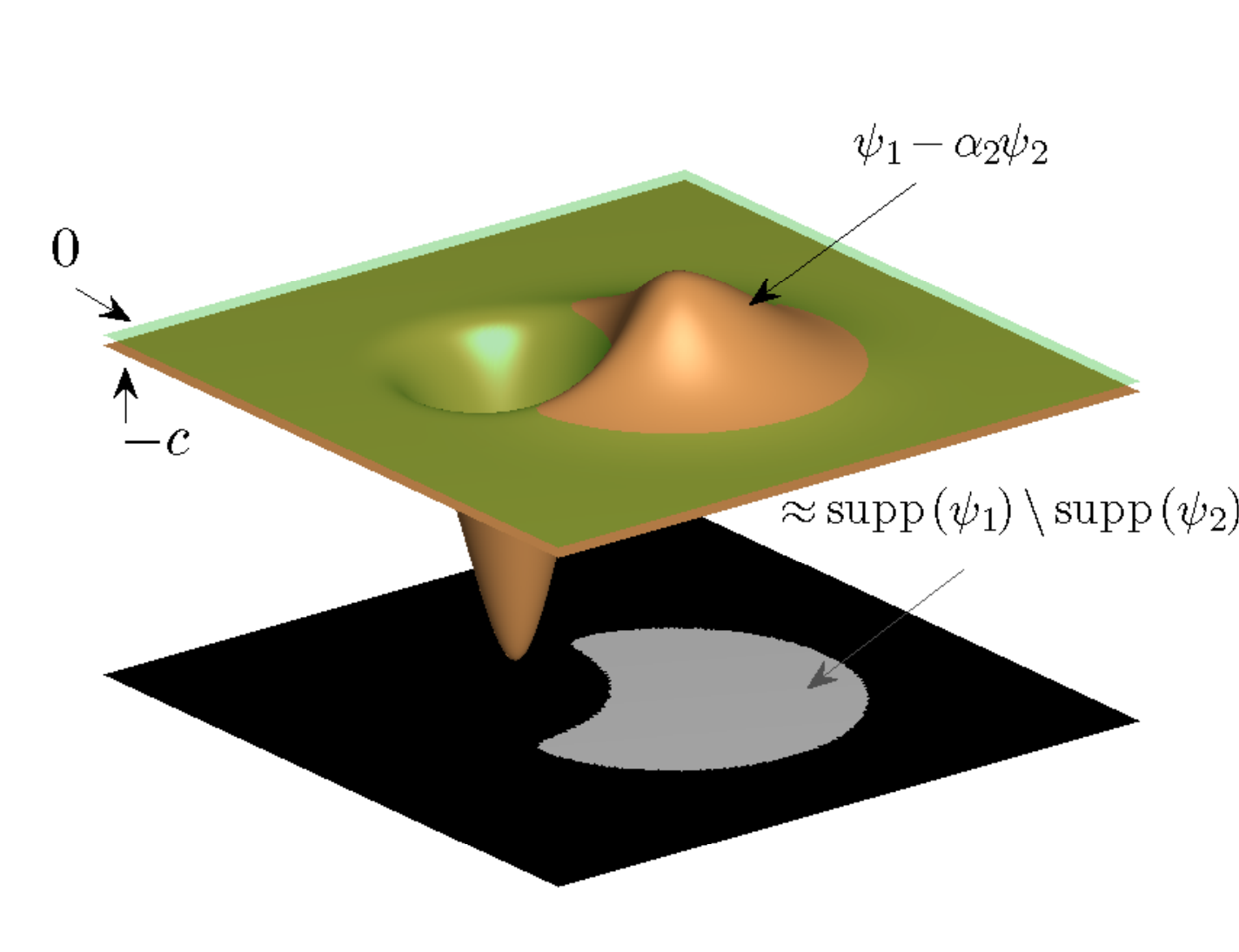}
\end{array}$
\end{center}
\caption{Illustration of the pseudo-logical behavior of the parametric level set functions. Top: set union; Bottom: set exclusion}\label{fig3}
\end{figure}

To remedy the problems associated with applying level set techniques to ill-posed inverse problems and yet take advantage of their topological flexibility, recent work by Aghasi \emph{et al.} \cite{aghasi2011parametric} proposes using a parametric level set (PaLS) function. In the PaLS technique, the level set function is parameterized in terms of a parameter vector $\bmu_\p=[\mu_1,\mu_2,\cdots,\mu_M]$, with $M$ much less than the number of pixels involved in the discrete representation of the problem. Its deformations are controlled by changing the elements of ${\mu}_\p$. This type of parameterization thereby induces a form of regularization by parameterization (e.g., see \cite{chavent2009nonlinear, feng2003curve, kilmer2003three}). In \cite{aghasi2011parametric, aghasi2013sparse} the authors also present a \emph{pseudo-logical} approach, which approximately models applying set operations on simple geometries to form more complex structures. More specifically, considering $\psi(.)$ to be a compactly supported radial basis function (simply called a bump) they propose a PaLS form as
\begin{equation}\label{eq12}
\phi(\x,\bmu_\p)=-c+\sum_{i=1}^M \alpha_i \psi_{\beta_i,\chi_i}(\x)\ ,
\end{equation}
where the constant $c$ is a positive scalar close to zero, $\psi_{\beta_i,\chi_i} (\x)=\psi(\|\beta_i (\x-\chi_i)\|)$ and $\bmu_\p=\{α\alpha_i,\beta_i,\chi_i \}_{i=1}^M$ is the parameter vector controlling weights, radii, and centers of the bumps. Note that $c$ is required to obtain nontrivial results as the radial basis functions are themselves exactly zero after a certain radius. As illustrated in Figure \ref{fig3}, this model exhibits a pseudo-logical behavior. For example, the sum of two positive bumps having comparable size (shown in the top panel of Figure \ref{fig3} as the orange surface) approximates the union operation on their zero level sets (grey shapes in the bottom black plane of the figure). Similarly, summation of a positive and a relatively large negative bump can approximate the set exclusion operator (Figure \ref{fig3}, bottom). Using this concept, the basic algebraic summation in (\ref{eq12}) can imply set operations on the support of the bumps and make the level set function capable of expressing a large class of geometries .

By using $\phi(\x,\bmu_\p)$ in (\ref{eq10}) and employing a smooth version of the Heaviside function (as discussed more fully in \cite{aghasi2011parametric}), the large-scale minimization problem in (\ref{eq7}) reduces to a minimization over the PaLS parameters and the texture parameters, $p_1$ and $p_2$  in (\ref{eq10}) as
\begin{align}\label{eq13}
\{\bmu_\p^*,p_1^*,p_2^*\}=\!\argmin_{\{\bmu_\p,p_1,p_2\}}\! \frac{1}{2} \big \| \boldsymbol{d}- \boldsymbol{\mathcal{M}}\big(p(\x, \bmu_\p,p_1,p_2) \big)\big\|_{\boldsymbol{R}}^2\ .
\end{align}

The number of unknown parameters in (\ref{eq13}) tends to be much smaller than those in pixel based and conventional level set methods. The low dimensionality of the PaLS approach makes the inverse problem less ill-posed without sacrificing much in terms of flexibility in shape representation. Moreover, it establishes a foundation that supports using quadratic minimization techniques such as the Newton methods. These methods are faster than gradient descent techniques and robust against the scaling of different variables appearing in the minimization \cite{gill1981practical, aghasi2011parametric, Polydorides2012}.

\subsection{Joint Inversion and Multi-Objective Minimization}
In Section 2 we presented two different modalities; the hydrological and electrical models. In the hydrological inversion we are interested in reconstructing the DNAPL saturation values based on the measurements of the contaminant concentration in a down gradient transect. On the other hand, in the ERT inversion we seek to extract the electric conductivity of the domain given a limited number of potential measurements. Using a petro-physical model, the problem can be expressed entirely in terms of saturation and inversion can be cast as the solution to the multi-objective optimization problem
\begin{equation}\label{eq14}
s_n^*=\argmin_{s_n} \left\{
     \begin{array}{c}
       \frac{1}{2} \| \boldsymbol{d}_\h- \boldsymbol{\mathcal{M}}_\h(s_n)\|_{\boldsymbol{R}_\h}^2 \\[.2cm]
       \frac{1}{2} \big \| \boldsymbol{d}_{\mathpzc{E}}- \boldsymbol{\mathcal{M}}_{\mathpzc{E}}\big(P(s_n)\big)\big \|_{\boldsymbol{R}_{\mathpzc{E}}}^2
     \end{array}\ ,
   \right.
\end{equation}
where indices $\h$ and ${\mathpzc{E}}$ correspond to the hydrological and electrical modalities, and $P(.)$ represents the petro-physical relationship that links saturation and conductivity (here, the Archie’s law).

A generalization for the notion of optimality in multi-objective minimization is the \emph{Pareto optimality} \cite{censor1977pareto}. For multiple costs $G_j(\bmu):\mathbb{R}^M\to\mathbb{R}$, where $j=1,2,\cdots ,m_c$, a vector $\bmu^*\in\mathbb{R}^M$ is called Pareto optimal, if there exist no $\bmu\neq\bmu^*$ such that $G_j(\bmu)\leq G_j(\bmu^*)$ for $j=1,2,\cdots ,m_c$ with a strict inequality for at least one $j$. Accordingly, $\bmu^*$ is called \emph{locally Pareto optimal} if there exists a neighborhood of $\bmu^*$ in which $\bmu^*$ is Pareto optimal.

For the problem of interest here, using a PaLS shape-based approach as in (\ref{eq13}), $s_n (\x)$ can be parameterized as $s_n(\x,\bmu)$, where $\bmu$ is a vector containing the PaLS and texture parameters. In this case the inverse problem amounts to finding a locally Pareto optimal point for the multi-objective minimization problem
\begin{equation}\label{eq15}
\bmu^*=\argmin_{\bmu}\left\{
     \begin{array}{c}
       \mathcal{G}_\h(\bmu)\\
       \mathcal{G}_{\mathpzc{E}}(\bmu)
     \end{array}\ ,
   \right.
\end{equation}
where
\begin{eqnarray}\label{eq16}
\hspace{1cm}\mathcal{G}_\h(\bmu)&=& \frac{1}{2} \big \| \boldsymbol{d}_\h- \boldsymbol{\mathcal{M}}_\h\big(s_n(\x,\bmu)\big)\big \|_{\boldsymbol{R}_\h}^2\ ,\\ \hspace{1cm}\mathcal{G}_{\mathpzc{E}}(\bmu)&=& \frac{1}{2} \Big \| \boldsymbol{d}_{\mathpzc{E}}- \boldsymbol{\mathcal{M}}_{\mathpzc{E}}\Big(P\big(s_n(\x,\bmu)\big)\Big)\Big\|_{\boldsymbol{R}_{\mathpzc{E}}}^2\ .\label{eq17}
\end{eqnarray}

One of the main solution strategies in multi-objective optimization is the scalarization approach, for which a single objective function is obtained as a linear combination of the underlying costs. For our case, a scalarized version of the multi-objective problem would be
\begin{equation}\label{eq18}
\mathcal{G}_T(\bmu)=\mathcal{G}_\h(\bmu)+\mathcal{G}_{\mathpzc{E}}(\bmu)\ ,
\end{equation}
where $\boldsymbol{R}_\h$ and $\boldsymbol{R}_{\mathpzc{E}}$ are fixed at the beginning of the minimization to roughly make the two cost terms comparable. In addition to the general problem of balancing the terms, scalarization is not always an efficient approach for multi-objective problems (e.g., see Section 7 in \cite{fliege2010newton}). In employing gradient-based techniques to iteratively minimize (\ref{eq18}), there is no guarantee that a descent direction based on $\mathcal{G}_T$ simultaneously reduces $\mathcal{G}_\h$ and $\mathcal{G}_{\mathpzc{E}}$. Moreover, the balance between the two terms may change substantially, as the iterations progress, thereby, yielding a solution that in a sense does not maximally exploit all of the information in the various data sources.

\begin{figure}[!htb]
\centering
\subfigure{\includegraphics[width=90mm]{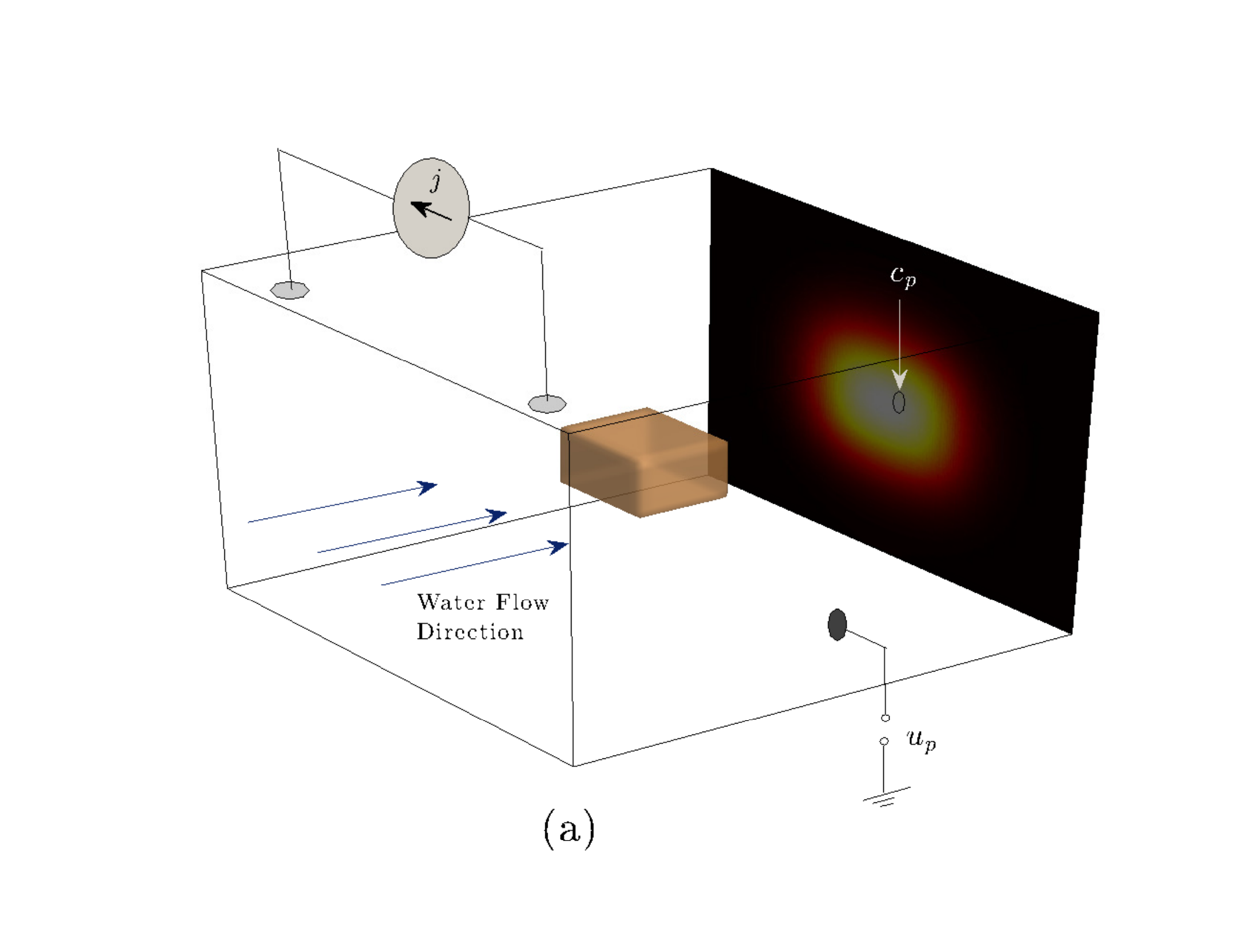}}\\[-.9cm]
\subfigure[][]{\includegraphics[width=60mm]{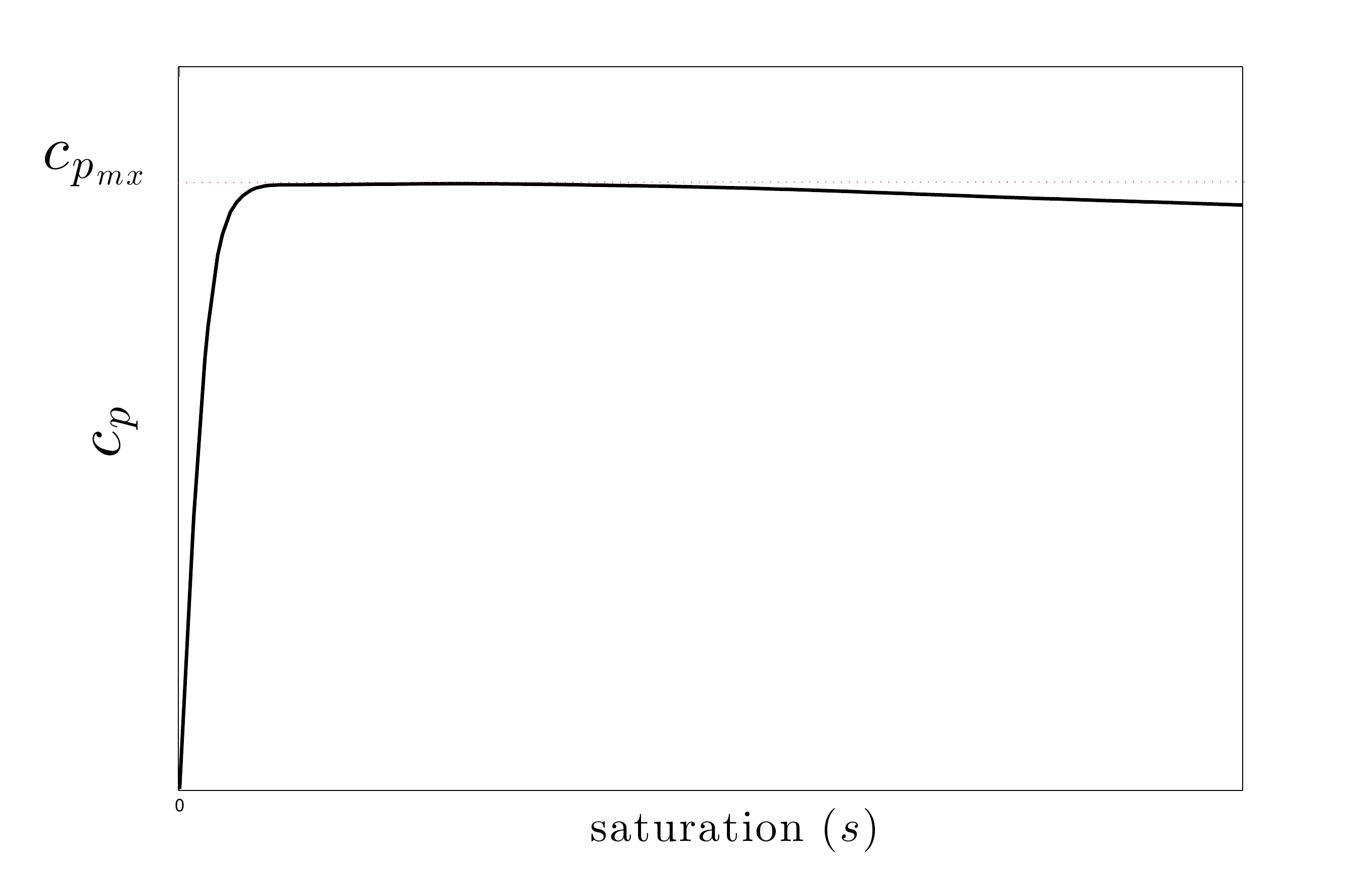}}
\subfigure[][]{\includegraphics[width=60mm]{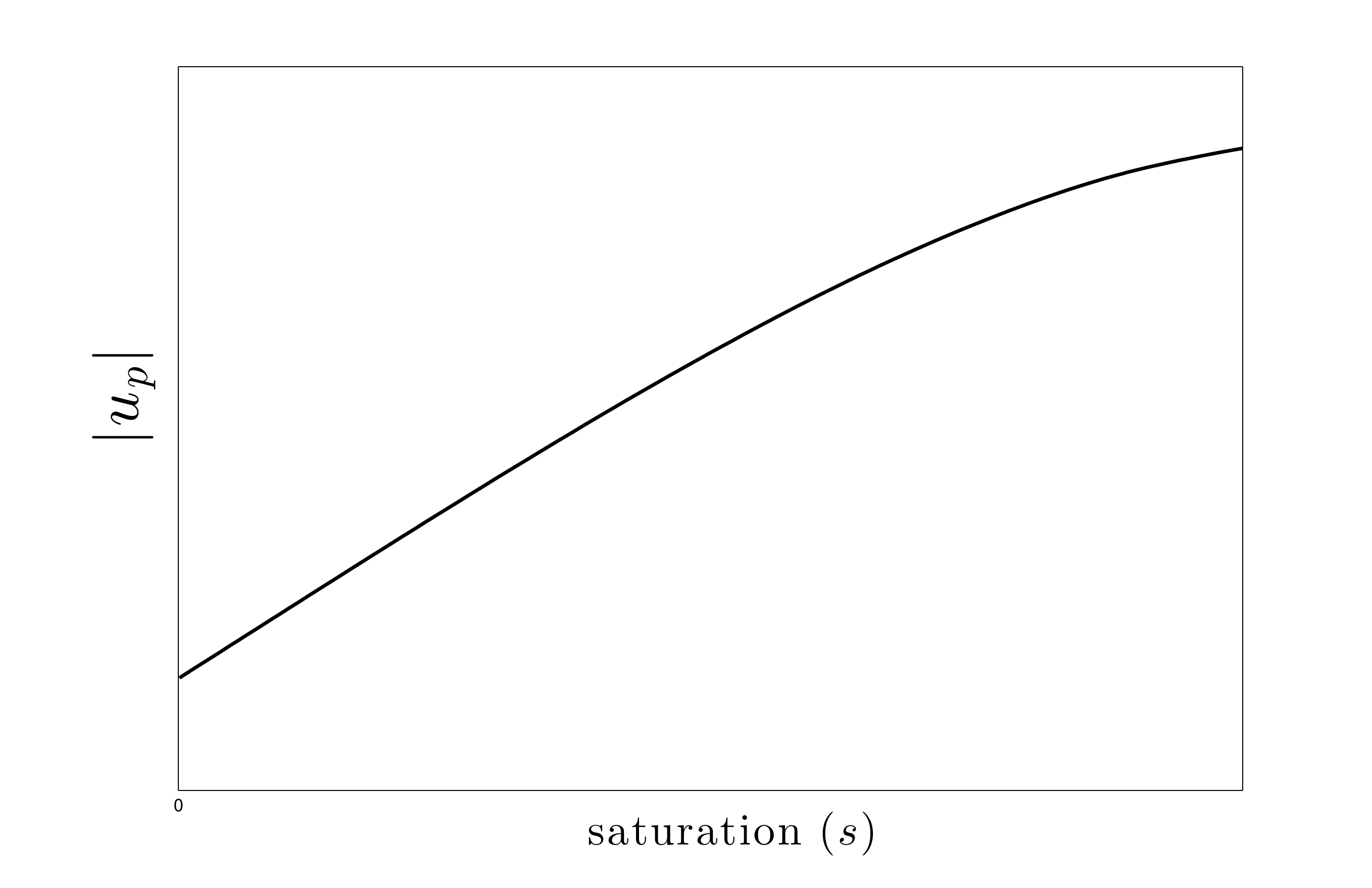}}
\caption{(a) An experiment setup to observe the contrast between the sensitivities of electrical and hydrological modalities to the source zone saturation (b) The typical downstream measurement for a single point as the bulk saturation increases (c) The typical change in the magnitude of potential measurements as the bulk saturation increases}%
\label{fig3p5}%
\end{figure}

For the particular problem of interest here such behavior is in fact observed. Specifically, DNAPL saturation values may increase without subsequent change to the down gradient contaminant concentration, as concentrations along a given flow path approach the aqueous solubility. This phenomenon is better demonstrated in Figure \ref{fig3p5}(a). Specifically in this demonstration, for a fixed volume of DNAPL (here a cube of uniform saturation $s$), we measure $c_p$, the quasi-steady state downstream concentration of a single point for different values of $s$. We also have an electric potential measurement, $u_p$, corresponding to a fixed source of current.

Typical plots of $c_p$ and $|u_p|$ in terms of $s$ are shown in Figures \ref{fig3p5}(b) and \ref{fig3p5}(c). Technically speaking, once the aqueous solubility of water is reached, for a wide range of saturation values, $c_p$ stays almost constant\footnote{After this range, further increase in $s$ would result $c_p$ to start stepping below $c_{p_{mx}}$. This reduction is due to the fact that as $s$ increases, the permeability to the water phase decreases. At high values of $s$, the water does not flow to any appreciable extent through the contaminated zone and will start to flow around it, which decreases the rate of dissolution. As this second phenomenon only happens at very large saturations, we only emphasized on the first phenomenon.} close to a threshold $c_{p_{mx}}$. For the electrical measurement, according to the Archie's law and the Poisson's equation, increasing $s$ causes an overall reduction in the electrical conductivity of the domain which causes a monotonic increase in $|u_p|$ as depicted in Figure \ref{fig3p5}(c).

This simple experiment reveals that the two physical models would not in general maintain a balanced sensitivity to the saturation values in a course of reconstruction. In these circumstances the scalarized cost function effectively ignores one data set due to larger decreases that can be obtained by ``listening to'' the other data. While a heuristic way of approaching this problem is iteratively rebalancing the cost terms, a proper convergence may not be guaranteed for such an approach. Furthermore, rebalancing the cost terms does not necessarily balance the corresponding sensitivities and still one of the costs may be neglected in determining a descent direction. Inspired by the idea presented in \cite{fliege2010newton}, in the sequel we present an iterative scheme that the step determined in the course of reconstruction would simultaneously reduce both misfit terms.

\subsubsection{A Joint Newton Type Minimization}\-

\noindent Classic Newton techniques provide an iterative procedure to converge to a stationary point. More specifically, for a given cost $\mathcal{G}(\bmu)$, initializing the process with $\bmu_0$, at the $k$-th iteration a suitable scalar multiple of a descent direction $\bdelta_k$ is added to the current estimate $\bmu_k$ to generate the subsequent estimate. To determine the descent direction, a quadratic Taylor approximation is considered as
\begin{equation}\label{eq19}
\mathcal{G}(\bmu_k+\bdelta)\simeq \mathcal{G}(\bmu_k)+ \bdelta^T\nabla \mathcal{G}(\bmu_k) + \frac{1}{2}\bdelta^T\nabla^2 \mathcal{G}(\bmu_k)\bdelta\ ,
\end{equation}
and to reduce the cost to the maximum extent, $\bdelta_k$ is obtained by minimizing $g(\bdelta)=\mathcal{G}(\bmu_k+\bdelta)-\mathcal{G}(\bmu_k)$, i.e.,
\begin{equation}\label{eq20}
\bdelta_k= \argmin_{\bdelta} \bdelta^T\nabla \mathcal{G}(\bmu_k)+\frac{1}{2}\bdelta^T \nabla^2 \mathcal{G}(\bmu_k)\bdelta\ .
\end{equation}
When the Hessian of the cost, $\nabla^2  \mathcal{G}(\bmu_k)$, is positive definite, (\ref{eq20}) can be solved in closed form as $\bdelta_k=\nabla^2 \mathcal{G}(\bmu_k)^{-1} \nabla \mathcal{G}(\bmu_k)$, which represents the well-known Newton direction.

\begin{figure}[!htb]\centering
\centering\includegraphics[width=19pc]{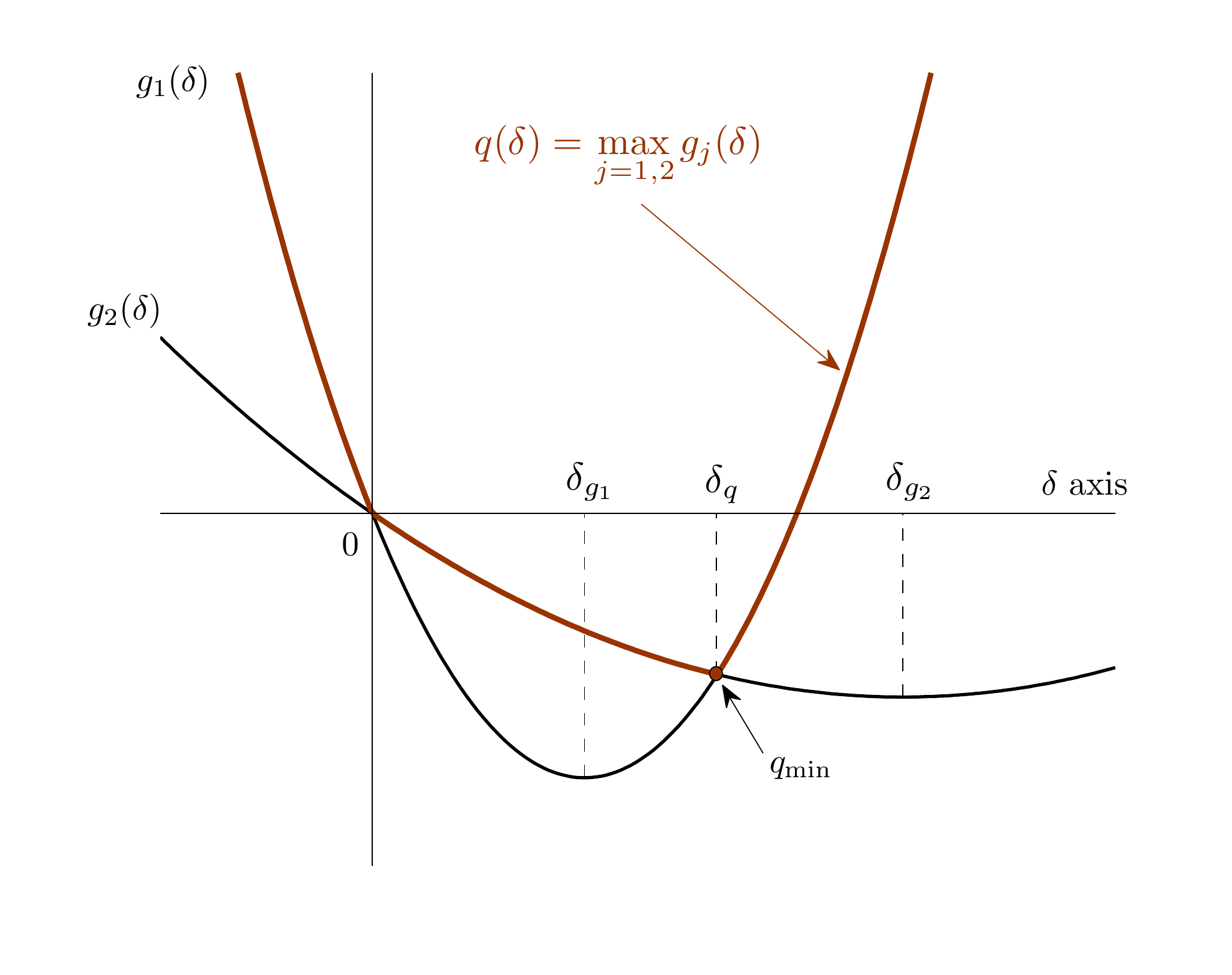}
\caption{Determining a Newton descent direction for multiple costs; the figure shows quadratic approximations $g_1$ and $g_2$ of the costs $\mathcal{G}_1$ and $\mathcal{G}_2$ according to (\ref{eq21}), for which the corresponding Newton steps are $\delta_{g_1}$ and $\delta_{g_2}$. Choosing $\delta_{q}$ as the ultimate direction guarantees that $g_1 (\delta_{q} )<0$ and $g_2 (\delta_{q} )<0$ (i.e., simultaneously reducing both costs $\mathcal{G}_1$ and $\mathcal{G}_2$). Moreover, as observable in the figure, $\delta_{q}$ is in some sense the optimal direction that performs this concurrent reduction.}\label{fig4}
\end{figure}

In the case of multiple costs $\mathcal{G}_j (\bmu)$ (where $j:{\mathpzc{E}},\h$ here), for every cost we have the quadratic approximation
\begin{equation}\label{eq21}
g_j(\bdelta)= \bdelta^T\nabla \mathcal{G}_j(\bmu_k)+\frac{1}{2}\bdelta^T\nabla^2 \mathcal{G}_j(\bmu_k)\bdelta\ .
\end{equation}
Fliege \emph{et al.} in \cite{fliege2010newton} discuss that an optimal direction in this case may be acquired by minimizing the function $q(\bdelta)=\max_j g_j(\bdelta)$. Figure \ref{fig4} illustrates how a direction acquired by minimizing $q(\bdelta)$ is the optimal direction to simultaneously reduce both costs $\mathcal{G}_j(\bmu)$. Finding this direction requires the solution of the following non-smooth min-max problem
\begin{equation}\label{eq22}
\bdelta_k= \argmin_{\bdelta} \maxi_j \bdelta^T\nabla \mathcal{G}_j(\bmu_k)+\frac{1}{2}\bdelta^T \nabla^2 \mathcal{G}_j(\bmu_k)\bdelta\ .
\end{equation}
At the expense of adding an auxiliary variable $z$, problem (\ref{eq22}) can be framed as the constrained convex optimization problem
\begin{equation}\label{eq23}
       \left\{
     \begin{array}{c}
       \underset{{(z,\bdelta)}} {\min} \; z\\[.3cm]
       \mbox{s.t.:}\quad\bdelta^T\nabla \mathcal{G}_j(\bmu_k)+\frac{1}{2}\bdelta^T \nabla^2 \mathcal{G}_j(\bmu_k)\bdelta-z\leq 0
     \end{array}\ .
   \right.
\end{equation}
Although solving this convex program per iteration is computationally expensive for a pixel-based inversion, it suits well into a PaLS framework that $\bmu$ is of moderate dimensionality. In other words, the dimensionality reduction that PaLS brings into the problem makes taking the $\bdelta_k$ steps in (\ref{eq20}) tractable and obtain a direction that simultaneously reduces both cost terms.

One of the main limiting assumptions of the multi-objective Newton technique in \cite{fliege2010newton} is the positivity of the Hessians, which does not necessarily hold for $\mathcal{G}_\h$ and $\mathcal{G}_{\mathpzc{E}}$. However, we can take advantage of the least squares nature of $\mathcal{G}_\h$ and $\mathcal{G}_{\mathpzc{E}}$ in (\ref{eq16}) and (\ref{eq17}) and obtain positive definite approximations to the Hessian.

Motivated by Gauss-Newton techniques \cite{madsen1999methods}, denoting $\boldsymbol{J}\!_j=\partial\boldsymbol{\mathcal{M}}_j/\partial \bmu$ as the model Jacobian for $j:{\mathpzc{E}},\h$, we have $\nabla \mathcal{G}_j={\boldsymbol{J}\!_j}\!^T \boldsymbol{R}_j (\boldsymbol{\mathcal{M}}_j-\boldsymbol{d}_j)$ and $\nabla^2  \mathcal{G}_j\simeq {\boldsymbol{J}\!_j}\!^T \boldsymbol{R}_j  \boldsymbol{J}\!_j$. The matrix ${\boldsymbol{J}\!_j}\!^T \boldsymbol{R}_j  \boldsymbol{J}\!_j$ is at least a positive semi-definite approximation to $\nabla^2  \mathcal{G}_j$ which by adding a small positive multiple of the identity matrix, (i.e., ${\boldsymbol{J}\!_j}\!^T \boldsymbol{R}_j  \boldsymbol{J}\!_j+\lambda_j \boldsymbol{I}$), becomes a strictly positive definite matrix. In the Levenberg-Marquardt (LM) algorithm, at every iteration an optimal value of $\lambda$ is determined based on trust region or damped techniques \cite{dennis1996numerical}. Accordingly, the LM algorithm of \cite{madsen1999methods} can be generalized to the multi-objective case, where at every iteration by choosing suitable values of $\lambda_j$, reasonable approximations to $\nabla^2  \mathcal{G}_j$ matrices are obtained and replaced in (\ref{eq23}) to find a descent direction. In \ref{app}, we have provided the details for generalizing the basic LM algorithm to a multi-objective scheme. Furthermore, a solution strategy to address the convex program (\ref{eq23}) is provided in \ref{app2}.

\section{Simulation and Discussion}
In this section the performance of the joint inversion scheme is examined for reconstructions of realistic source zone architectures. We first discuss how the models are treated in generating synthetic data and then technical details associated with the inversion scheme are provided.

\subsection{Data Generation and Underlying Modeling}
Each realization used in this work is representative of a suite of those generated by numerically simulating a PCE-DNAPL release in a statistically-homogenous sandy aquifer medium. Here the University of Texas Chemical Flooding Simulator UTCHEM 9.0 \cite{Delshad2000} was used to solve a coupled system of equations of the form (\ref{eq1}) for known release and flow conditions. Distributions of hydraulic permeability and related capillary parameters were produced using sequential Gaussian simulation (SGS) \cite{lemke2004dense}, based upon the general characteristics of an aquifer at the Bachman Road glacial outwash site in Michigan \cite{abriola2005pilot}). Simulation parameters and boundary conditions are presented in Table \ref{tab1}. Note that, for this work, the average permeability and porosity employed are representative of the Bachman site, but a larger permeability variance and shorter correlation length were applied in the SGS algorithm to produce more variability in the source zone DNAPL saturation distributions. The interested reader is referred to \cite{christ2005comparison} and the references therein, for more detail related to the infiltration simulations.

\begin{table}[!htb]
\caption{Source Zone Scenario Simulation Parameters}
\centering
\begin{tabular}{@{}lcc}
\hline
\textbf{Fluid Properties}&\\
\hline
               & Water & PCE  \\
    Density $\rho_\alpha$ (g/cm${}^3$)${}^{\dag}$ & 0.999 & 1.625 \\
Dynamic Viscosity  (cP)${}^{\dag}$ &  1.121 & 0.89 \\
Compressibility  (Pa${}^{-1}$)${}^{\dag}$ &  4.4$\times$10${}^{-10}$ &0.0 \\
Aqueous Diffusivity  (cm${}^2$/s)${}^{\ddag}$ &  -- & 8.6$\times$10${}^{-6}$ \\
Aqueous Solubility  (g/L)${}^{\natural}$ &  -- &  0.150 \\
Initial Saturation   &  1.0 &  0.0 \\

\hline
 \textbf{Spill Scenario}&\\
\hline

             Spill Volume (L)&&	128  \\
   Spill Duration (d)&&	400 \\
    Release Rate (L/m${}^2$-d)&&	0.32  \\

\hline
$P_c$-$S_\alpha$-$K_{r\alpha}$ \textbf{Model Parameters} &\\
\hline

Air Entry Pressure (kPa)	&& 2.809\\
Pore Size Index	&& 2.0773\\
Interfacial Tension && \\
\hspace{.3cm}Air/Water (dyn/cm) && 72.75\\
\hspace{.3cm}PCE/Water (dyn/cm)	&& 47.8\\
Irreducible Water Saturation	&&0.080\\
Max Residual Organic Saturation && 0.151\\
Reference Permeability ($\mu$m${}^2$)	&& 19.7\\
\hline
\textbf{Matrix Properties}&\\
\hline

       Variogram Parameters& Horizontal & Vertical  \\
    Nugget & 0.333 & 0.333 \\
Range (m) &  4.66 & 0.72 \\
Integral Scale (m) &  1.55 &0.24 \\
\hline
Mean Hydraulic Conductivity, $\bar K$  (m/d)${}^{\dag}$ && 16.8\\
Anisotropy Ratio, $k_v/k_h$	&& 0.5\\
Lognormal Transformed $\bar K$ variance ($\sigma^2 \log(K)$)${}^{\dag}$ && 1.5\\
Applied Hydraulic Gradient (m/m)&&	0.01\\
Longitudinal Dispersivity, $\omega_m$ (m)${}^{\dag}$	&& 0.30\\
Horizontal Transverse Dispersivity, $\omega_p$ (m)${}^{\sharp}$	&&0.10\\
Vertical Transverse Dispersivity, $\omega_p$ (m)${}^{\sharp}$	&&0.0075\\
Median Grain Size, d${}_{50}$ ($\mu$m)${}^{\dag}$ &&	295\\
Uniformity Index, $U_i$ ${}^{\dag}$&&	1.86\\
Uniform Porosity, $\varphi$ ${}^{\dag}$ &&	0.36\\
$\Delta x$ (m) ($N_x$ = 26)	&& 0.3048\\
$\Delta y$ (m) ($N_y$ = 26)	&& 0.3048\\
$\Delta z$ (m) ($N_z$ = 128) &&	0.0726\\
\hline
${}^\dag$ \cite{lemke2004dense}, ${}^\ddag$ \cite{dekker2000influence}, ${}^\natural$ \cite{horvath1982halogenated}, ${}^\sharp$ \cite{EPA1986Background}
\end{tabular}\label{tab1}
\end{table}

\begin{figure}[!htb]
\centering
\begin{minipage}[b]{0.75\linewidth}
\centering
\includegraphics[width=\textwidth]{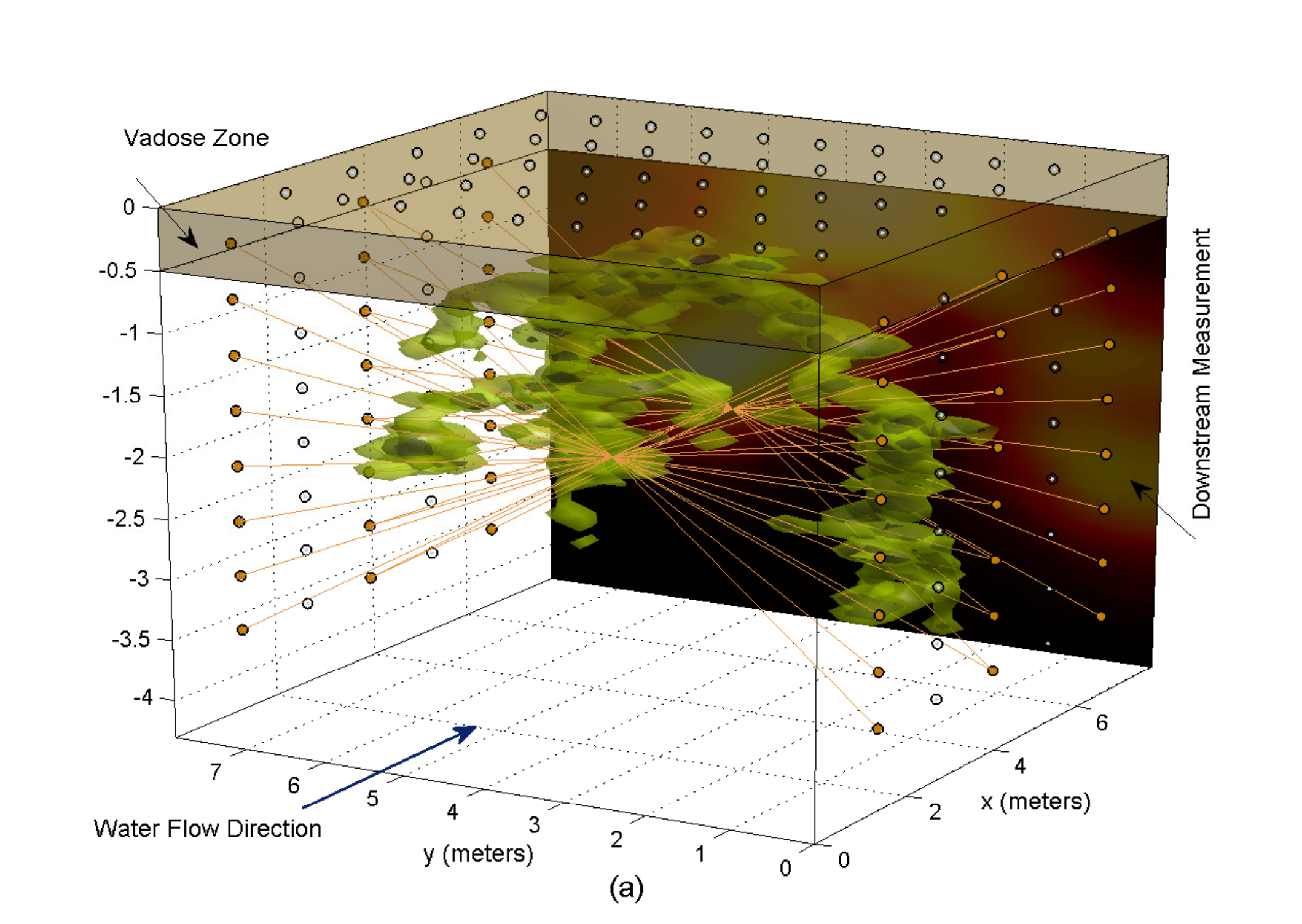}
\end{minipage}
\\
\begin{minipage}[b]{0.38\linewidth}
\begin{flushleft}
\includegraphics[width=\textwidth]{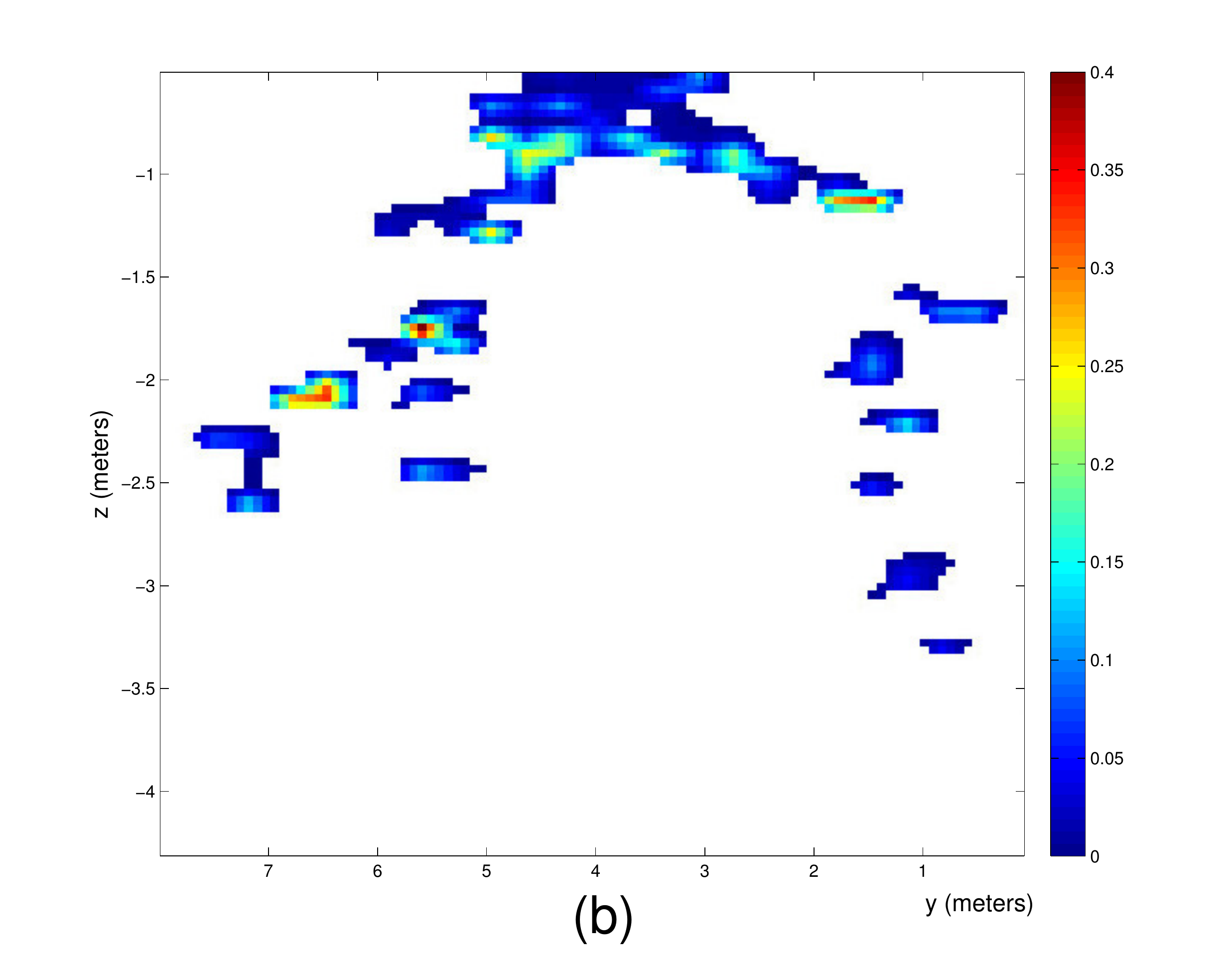}
\end{flushleft}
\end{minipage}
\begin{minipage}[b]{0.45\linewidth}
\includegraphics[width=\textwidth]{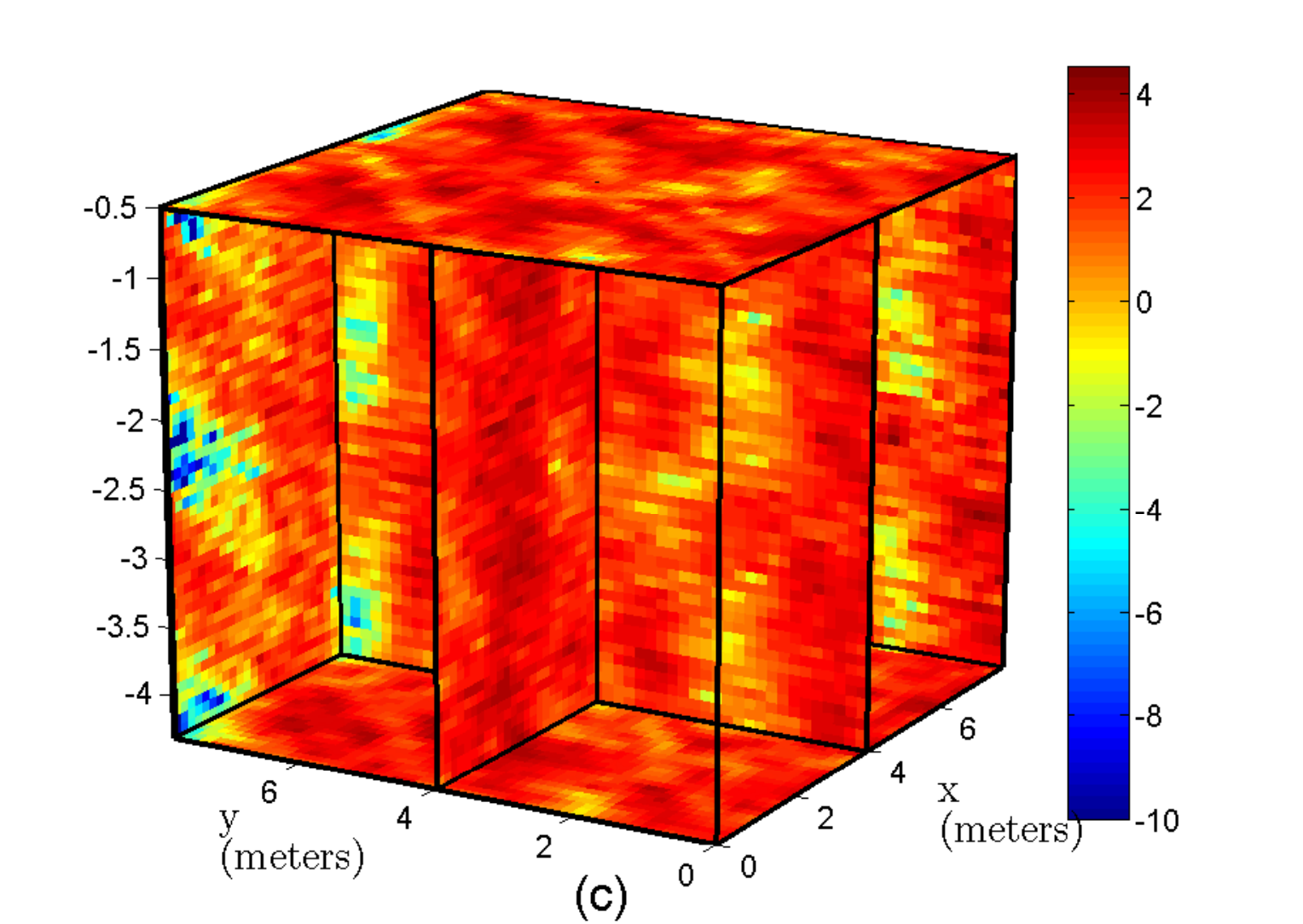}
\end{minipage}
\caption{General setting of the problem; a) the saturation profile where the green iso-surface corresponds to low saturation values and darker spots correspond to high saturations. The electrical setting, water concentration measurements and vadose zone are also shown in the figure. b) A slice of the saturation at x=3.97 m showing the DNAPL saturation texture in this plane. c) The log valued permeability field corresponding to the release. }\label{fig5}
\end{figure}

Figure \ref{fig5} depicts one of the simulated source zones. Here the green isosurface roughly delimits the configuration of the contaminated zone, corresponding to a saturation of 1$\%$. The brown (darker) spots or surfaces correspond to high saturation zones which exceed the residual saturation of these sands (15$\%$) and are typically classified as pools. Also shown in Figure \ref{fig5} (panel b) is a representative 2-D slice of the source zone taken at $x =$3.97 m, illustrating the high degree of heterogeneity in the saturation distribution. Note that the spatial domain depicted in the figure encompasses all of the infiltrated PCE mass, but is smaller than the actual computational domain used to simulate the distribution, which extended an additional 5.48 meters in depth. This smaller domain, with the same grid spacing, was employed for all subsequent hydrologic model computations in the inversion modeling process described below.

The ERT test configuration for the problem is also shown in Figure \ref{fig5}(a). The top plane ($z = 0$) represents the air interface (ground surface) where a Neumann boundary condition is applied. At the remaining planes, an absorbing boundary condition (\ref{eq4}) is applied to simulate an infinite half space. To account for the potential influence of the unsaturated zone in the electrical conductivity measurements, a thin vadose zone was added above the saturated domain and beneath the ground surface. The sensor placement and configuration are also shown in the figure. In total 130 sensors are located on the periphery of the imaging domain. The ten vertical boreholes each accommodate 8 equally spaced sensors and surface sensors are placed as linear arrays, each with 10 sensors filling the gap between a pair of boreholes. To generate a set of electrical measurements, 32 simulation experiments are carried out within the electrode array. In each experiment two sensors act as the current dipole (connected via the lines in Figure \ref{fig5}(a)) and the remaining sensors measure the corresponding electric potential. Current sources are placed in six boreholes. The remaining four boreholes, and the surface sensors (indicated with lighter color), serve as potential electrodes in all the experiments. The dipoles are chosen in a cross medium configuration to make the data more sensitive to the presence of DNAPL . Measurement data are generated by a finite difference solution of Poisson's equation, over the domain shown, which is discretized into 36$\times$36$\times$36 grid blocks, in the $x$, $y$ and $z$ directions respectively.

With respect to the hydrological modality, the groundwater flow is assumed to be in $+x$ direction. To create the hydrologic observations associated with a particular source zone realization, a modified version of MT3DMS \cite{Zheng1999}, which accounts for rate-limited dissolution and relative permeability effects \cite{Parker2004, christ2006estimating} was used to simulate the quasi-steady contaminant (PCE) concentrations in a down gradient plane ($x=x_{\max}$ transect) perpendicular to the groundwater flow (Figure \ref{fig5}(a)). These concentration data represent a single snapshot in time captured during the period of quasi-steady dissolution behavior that is characteristic of most DNAPL sites. Here we are less interested in the late-stage behavior of a source approaching exhaustion/clean-up but rather are interested in characterizing the source zone prior to remediation. The permeability field shown in Figure \ref{fig5}(c) was used to generate the ``true'' DNAPL source zone realization (Figure \ref{fig5}(a)) and its associated quasi-steady down gradient plume transect concentration ``measurements''. Relevant simulation parameters are presented in Table \ref{tab1}. Further detail on the simulation methods may be found in \cite{Christ2009}. Note that, in the hydrological model, aqueous phase concentrations measurements are assumed known at all grid points in the $x=x_{\max}$ transect.

We acknowledge that, for this method to become field-practicable, we must eventually address the issue that concentration measurements are likely to be sparse. However, in this paper we focus on assessing the general performance of the technique before considering the more difficult problem based upon sparse data.

\subsection{Inversion}
This section provides details on the inversion strategy. For the transport simulations employed in the inversion process, the saturated zone shown in Figure \ref{fig5}(a) was discretized into 26$\times$26$\times$50 grid blocks in the $x$, $y$ and $z$ directions. For all inversion simulations the permeability field was assumed unknown. Thus, the flow, transport, and electrical aspects of the inversion assume a uniform, average permeability and porosity. Within the inversion, for a given source zone configuration, the groundwater flow field and corresponding contaminant concentrations in the down gradient transect were generated using a version of MT3DMS \cite{Zheng1999}, modified in this work to support parallel computing.

Archie’s law was employed for the petrophysical model assuming a uniform porosity within the saturated zone. The Archie parameters and porosity shown in Table \ref{tab2} are based upon measurements conducted in our laboratory using Ottawa sands and PCE-DNAPL. These parameters fall within the range established within the literature \cite{archie1942electrical, ewing2006dependence}. Ottawa sands were used here because they provide a good physical surrogate for the Bachman aquifer material (e.g., \cite{Ramsburg2001, Ramsburg2004, abriola2005pilot}). The electrical simulations employ a vadose zone containing an uniform water saturation. Within this vadose zone the water saturation was assumed to be 8$\%$. This saturation is consistent with the field capacity of the porous media employed in \cite{Gorman1990}, and represents a second feature (i.e., in addition to DNAPL) within the domain that is electrically resistive. Because the vadose zone is assumed to have constant porosity and constant water saturation (8$\%$), we calculated the effective conductivity shown in Table \ref{tab2} using Archie’s Law with the saturation exponent reported by \cite{Gorman1990}. The bulk phase conductivity in all electrical simulations was assumed to be 0.05 Sm${}^{-1}$, which is analogous to an aqueous solution of approximately 250 mg/L $\mbox{CaCl}_2$. Larger values of the bulk phase conductivity could be employed, but would increase the contrast between the DNAPL and aqueous phase, and decrease the difficulty of the problem. Use of the relatively low value of electrical conductivity in the aqueous phase represents a more stringent demonstration of the methods described herein.

One of the advantages of the PaLS technique is that it is mesh-less, i.e., the underlying parameters are independent of the discretization. Thus, by using PaLS in a joint framework, every modality can have its own discretization method, and the grid points do not need to be co-located. This is certainly not the case with pixel-based methods where joint inversion on two different set of voxels generally requires some level of interpolation to provide a unified representation of unknowns, a process that can be complex.

An effort was made to account for some of the model and measurement uncertainty by applying random noise to the electrical conductivity and concentration data, prior to application of the PaLS method. Specifically, 0.1$\%$ additive Gaussian noise was added to the electrical conductivity data and 2$\%$ Gaussian noise to the concentration measurements. Although the current electrical noise level is relatively small compared to the total signal, this corresponds to more than 20$\%$ uncertainty in the scattered field. By definition, the scattered field is the measurement variation caused by including the anomaly (in this case DNAPL) in the system \cite{ben2007projection}. Thus, all the information about the inclusion is buried in the scattered field and the 0.1$\%$ noise actually adds considerable uncertainty to the data upon which the inversion is based. Using GPR as the geophysical modality may help here, as the contrast we get in the dielectric properties of water and contaminants using GPR is larger than the electrical conductivity contrast of DNAPL relative to the groundwater \cite{knight2001ground}. However, fully inverting the GPR data using the Maxwell's equations is computationally expensive. Given that the primary goal of this paper is to demonstrate the initial utility of our approach for joint inversion, we feel that the adaptation required for the processing of field data based on the GPR is a task best left to the future. As a result, the noise level in the ERT data is less than what would be expected from currently fielded geophysical instrumentation although it is in line with electrical systems employed in medical imaging \cite{smith1995real, hamilton2013direct}.

There are currently no explicit adjoint forms to aid extraction of the sensitivity values from the hydrologic model. However, use of PaLS significantly reduces the dimensionality of the problem, and permits application of a basic finite difference approximation for these sensitivity calculations. For ERT on the other hand, the Jacobian and model sensitivities for the inversion were acquired using the adjoint field technique as described in \cite{aghasi2011sensitivity, aghasi2011parametric, polydorides2012high}.

To examine the ability of the method to characterize the extent of the source we consider a single level set function for the DNAPL representation. In addition, we also consider the case where high saturation features are identified within the DNAPL representation. To identify these high saturation features we employ two level set functions; one corresponding to low saturation regions (ganglia) and one related to the high saturation regions (pools).

\subsubsection{Using a Single PaLS Function to Invert for DNAPL Structure}\-

\noindent Based upon the PaLS idea we model the DNAPL saturation as
\begin{equation}\label{eq24}
s_n(\x)=s_i H_\epsilon \big(\phi(\x,\bmu_{\p}) \big)+s_o \Big( 1-H_\epsilon \big(\phi(\x,\bmu_{\p}) \big)\Big)\ ,
\end{equation}
where $s_i$ is a scalar representing DNAPL texture value within the source zone and $s_o$ a similar quantity for the texture outside of the source zone and $\bmu_{\p}$ is the vector of PaLS parameters, $\boldsymbol{\alpha}$, $\boldsymbol{\beta}$ and $\boldsymbol{\chi}$. Since the DNAPL saturation outside of the source zone is zero, $s_o$ vanishes and simplifies the inversion such that only the PaLS parameters and $s_i$ need to be quantified. The function $H_\epsilon$ is a twice differentiable approximation of the Heaviside step function and $\epsilon$, a parameter which controls the transient width of $H_\epsilon$ and is a positive number smaller than $c$ (see \cite{aghasi2011parametric}).

\begin{figure*}[!htb]
\hspace{-.35cm}
\begin{tabular}{cc}
\includegraphics[width=16pc]{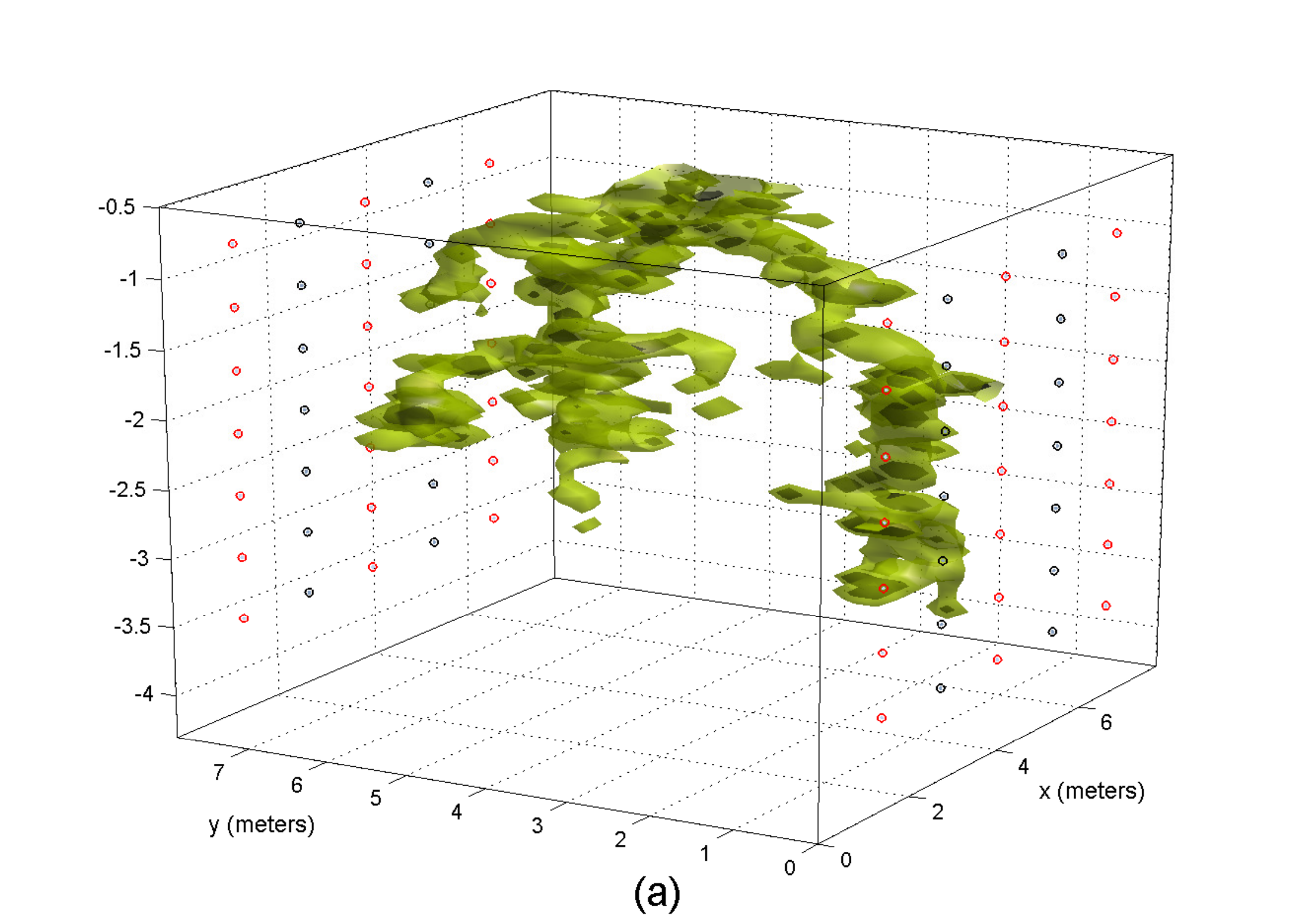}
&
\includegraphics[width=16pc]{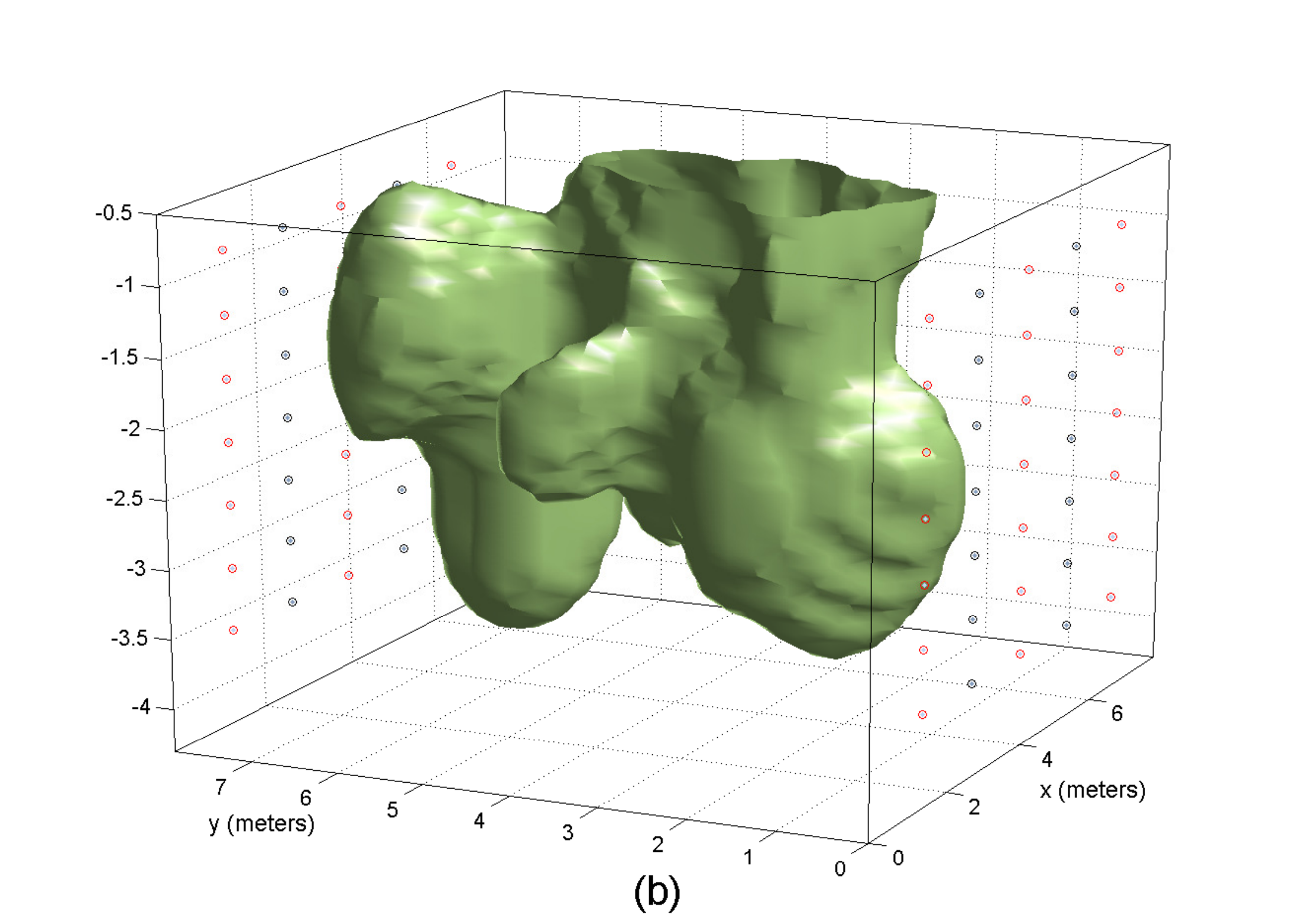}
\\

\includegraphics[width=16pc]{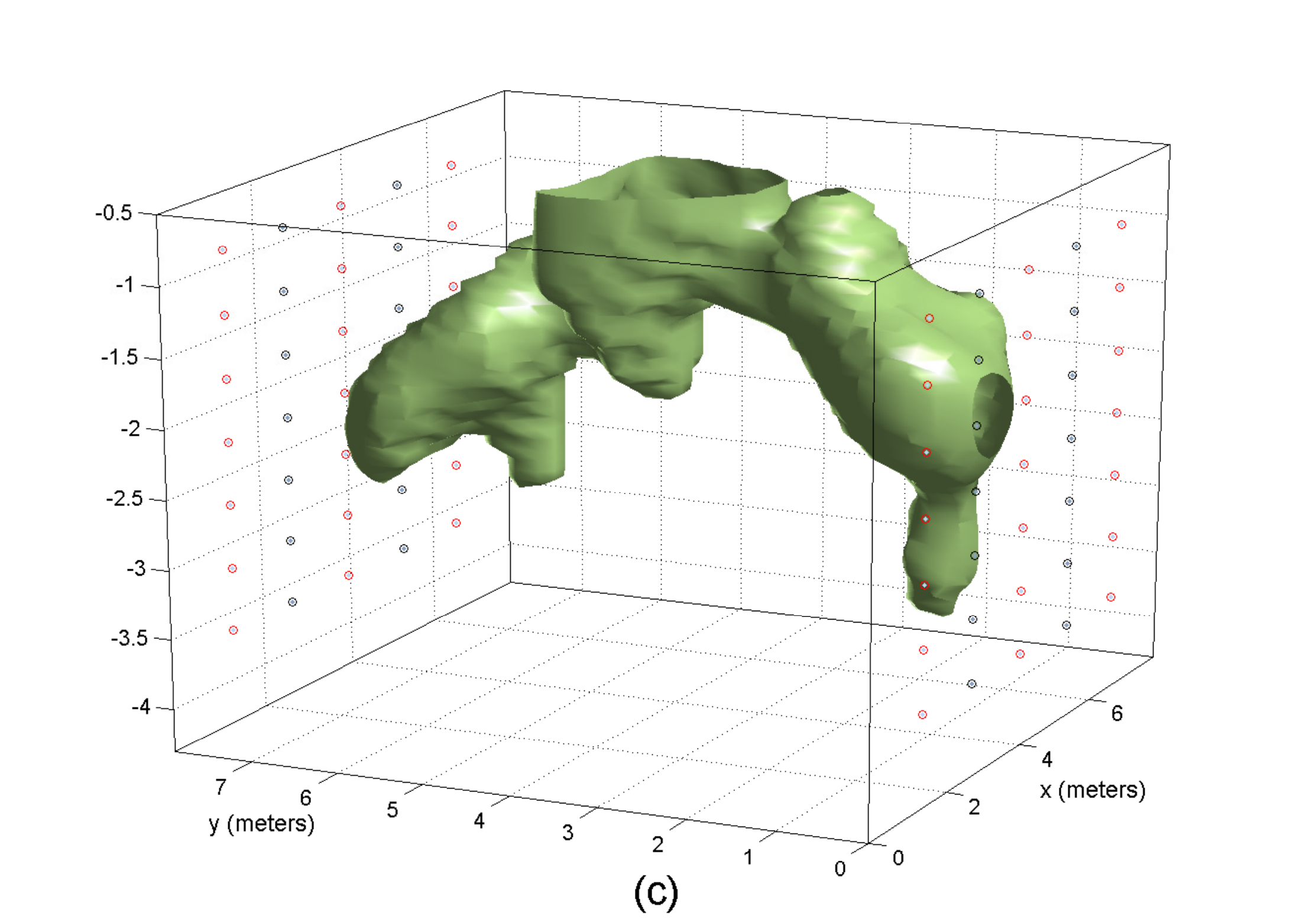}
&
\includegraphics[width=16pc]{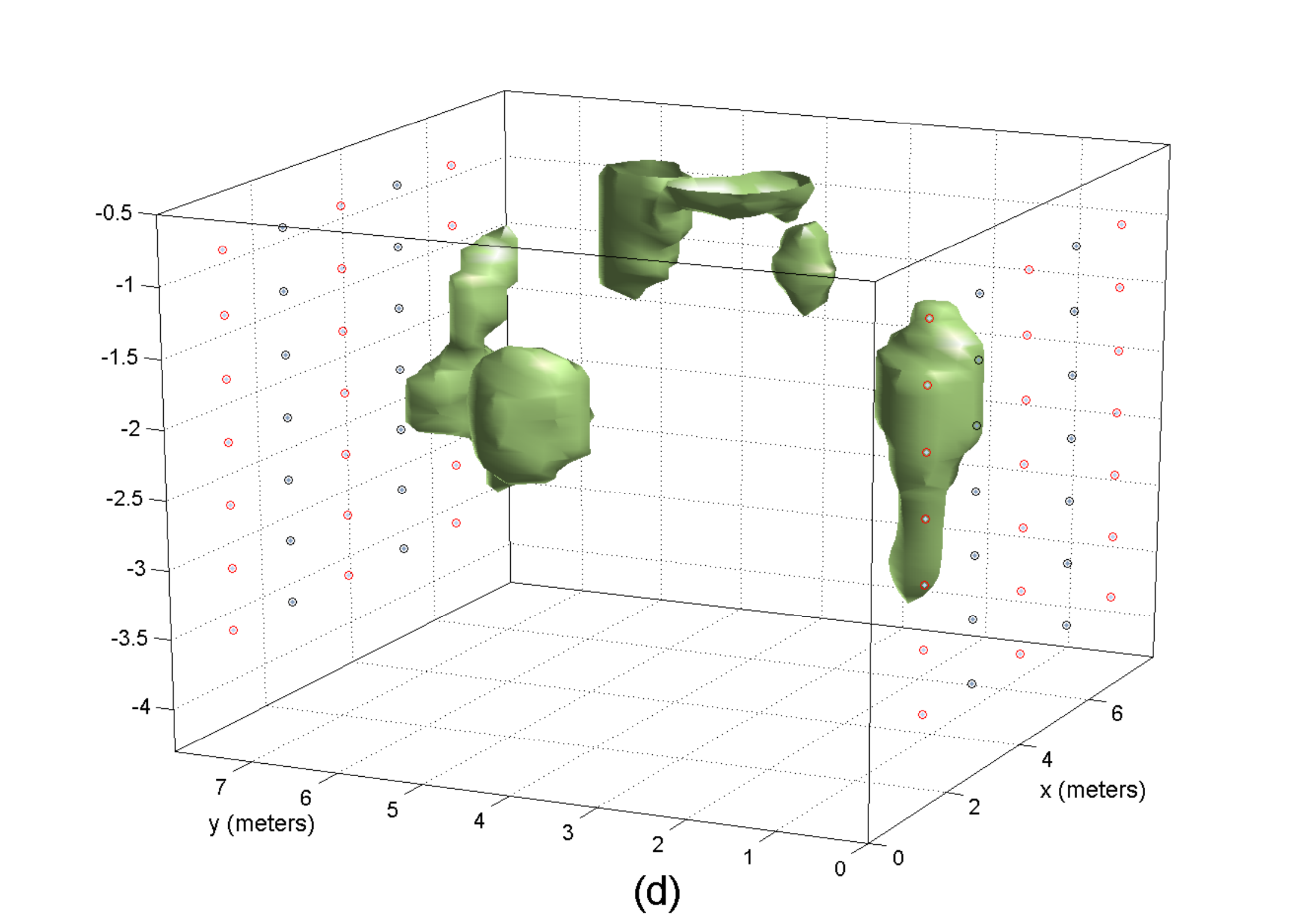}
\\
\includegraphics[width=16pc]{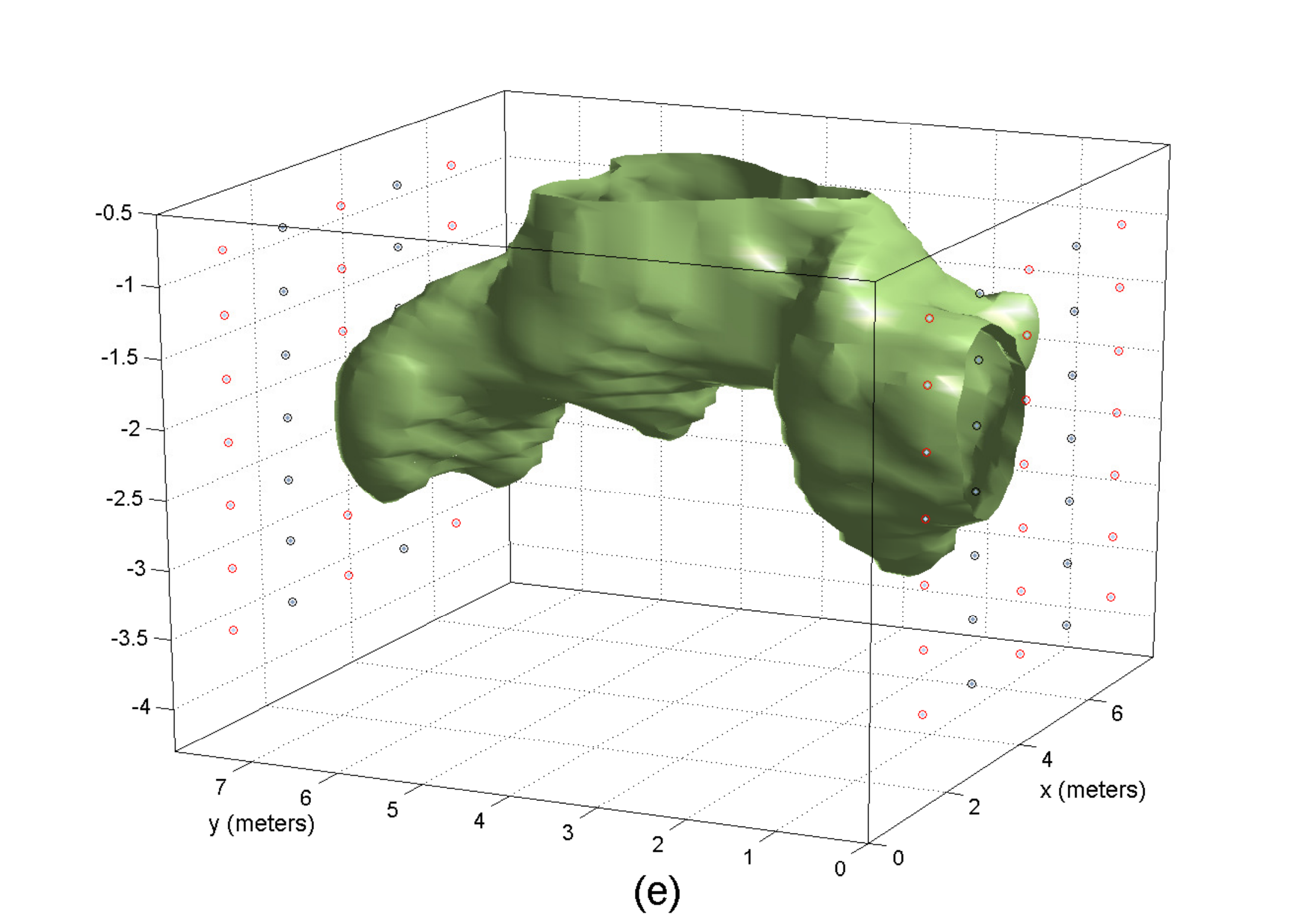}
&
\includegraphics[width=16pc]{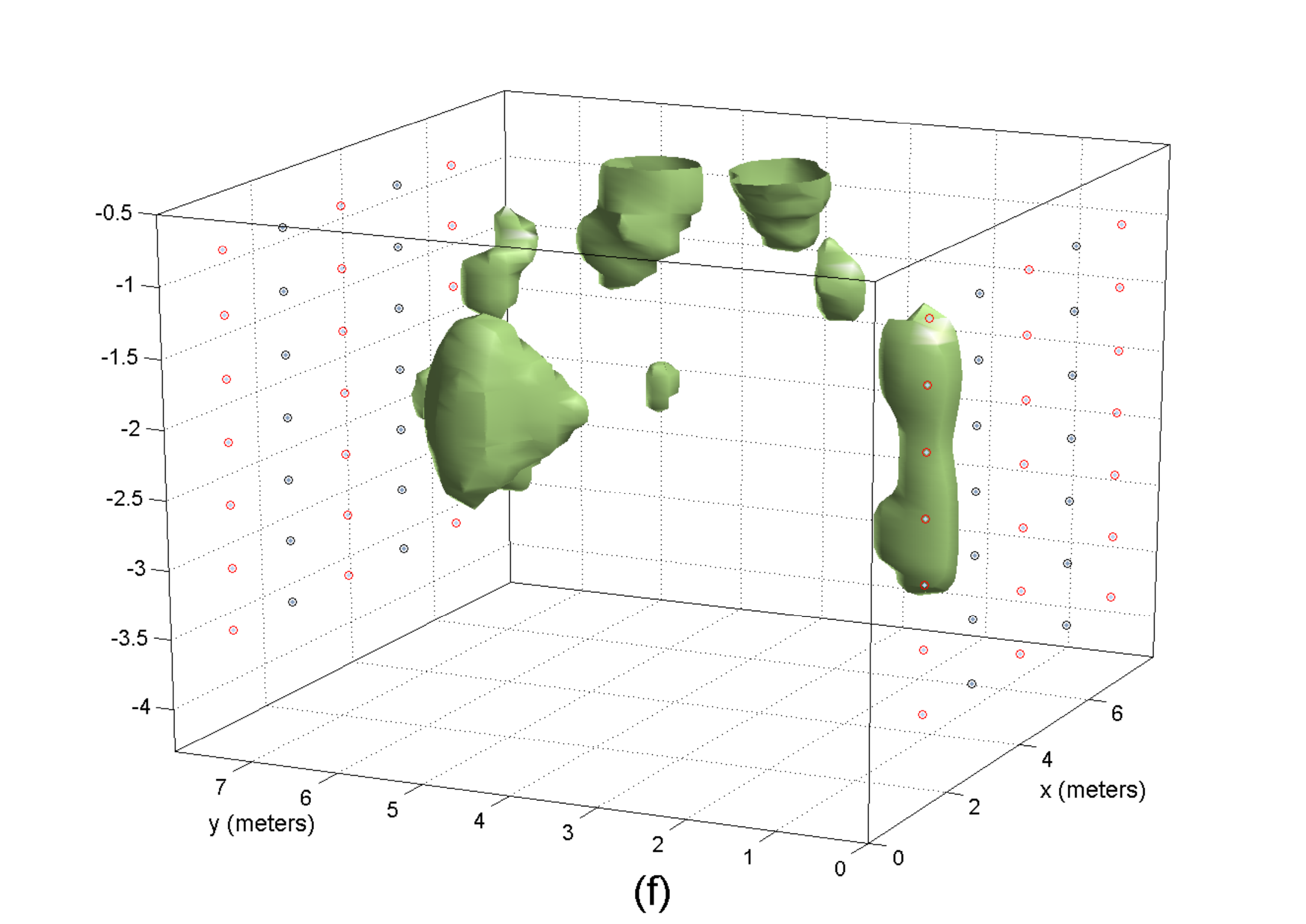}
\end{tabular}
\caption{Source zone reconstruction using a single level set function: a) original source zone structure; b) initialization; c) joint reconstruction of the source zone using proposed technique; d) joint reconstruction using a scalarization approach; e) reconstruction using only the ERT data; f) reconstruction using only the hydraulic data }\label{fig6}
\end{figure*}

For the PaLS function, we choose $M=$45 bumps (Table \ref{tab2}). The parameter initializations are intended to be general, and this selection of $M$ is guided by previous work that suggested a range of values low enough to ensure computational tractability and high enough to capture details \cite{aghasi2011parametric}. The initial centers $\chi_i$ are randomly chosen around the central parts of the imaging box. Values of $\beta_i$ are roughly set to make the initial shape be of comparable size to the domain being characterized. The weights $\alpha_i$ are randomly initialized to be ±1 such that the method starts with a balanced number of positive and negative bumps. The texture value $s_i$ is initially set to 1$\%$, and is updated through the iterative inversion process. Through some experimentation we have found that $s_i$ is most effectively updated in a sub-iteration using only the ERT data in a classic Newton step. The PaLS parameters $\alpha$, $\beta$ and $\chi$, however are updated based on the joint data. The generic initialization results in the shape shown in Figure \ref{fig6}(b).

\begin{table}[!htb]
\caption{A Summary of Underlying Inversion Parameters}
\centering
\begin{tabular}{@{}lcc}
\hline
{Saturated Zone}:	&Size (m)	& 7.93$\times$7.93$\times$3.81 \cr
	& Porosity, $\varphi$	& 0.36\\
	& $\sigma_w$ (S/m)	& 0.05 \\

\hline

{Vadose Zone}:	&Width (m)	& 0.5 \\
	& Electrical Conductivity (S/m)	& 2.5$\times$ 10${}^{-4}$ \\
\hline
{Archie's Law}:		& $a$	& 1\\
& $m$	& 1.4\\
& $q$	& 2.0\\
\hline
{PaLS Initialization/Setting}:	&Functions $\phi$ and $\phi_1$	& Function $\phi_2$\\
& $M=45$ &	$M=35$\\
& $\epsilon=0.1$, $c=0.11$  & $\epsilon=0.1$, $c=0.11$\\
& $\chi_i:$ random & $\chi_i:$ random\\
& $\beta_i=0.6$ &	$\beta_i=2$\\
& $\alpha_i=\pm1$ &$\alpha_i=\pm1$\\
\hline
\end{tabular}\label{tab2}
\end{table}

\begin{figure}[!htb]\centering
\subfigure[]{\includegraphics[width=12pc]{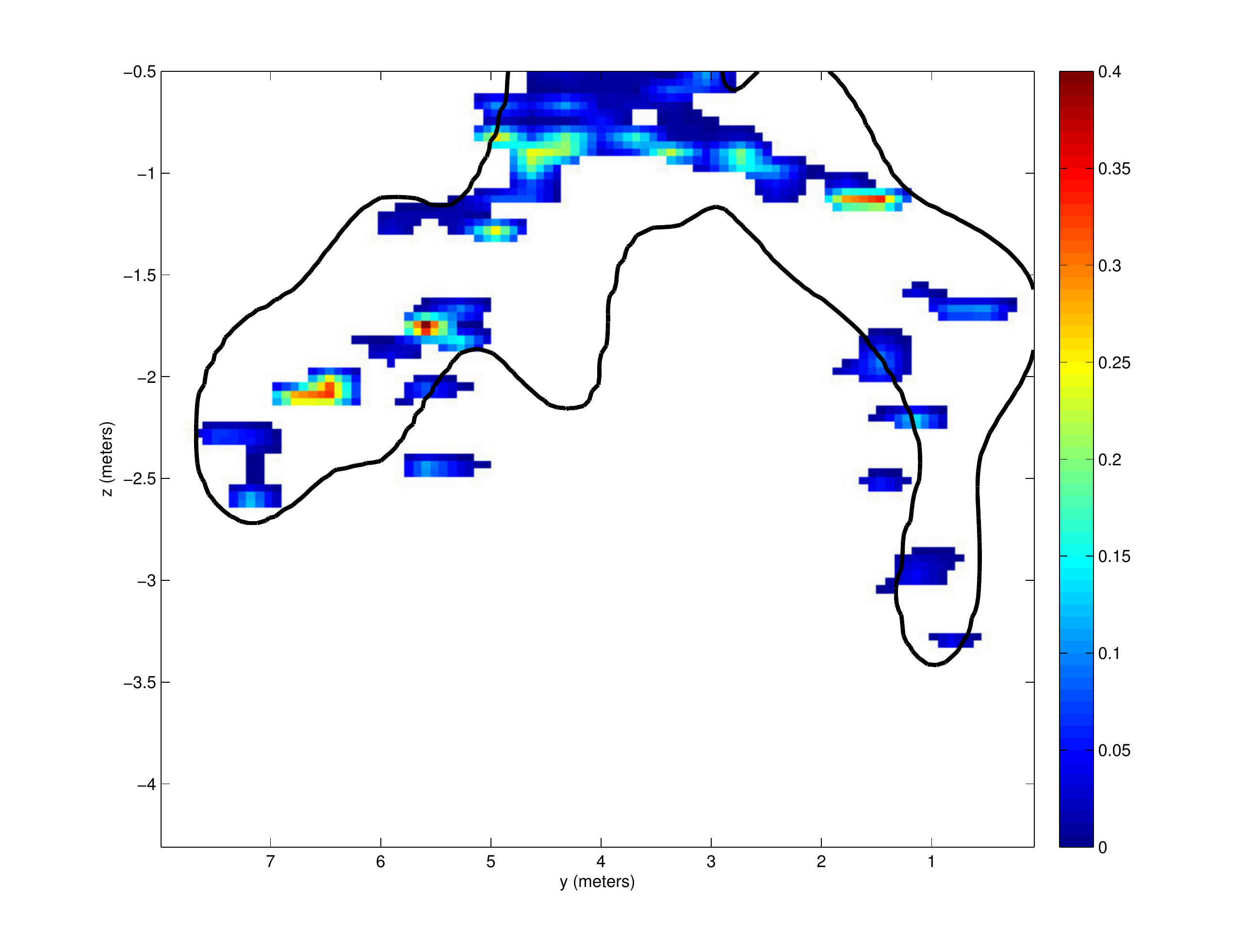}}
\subfigure[]{\includegraphics[width=12pc]{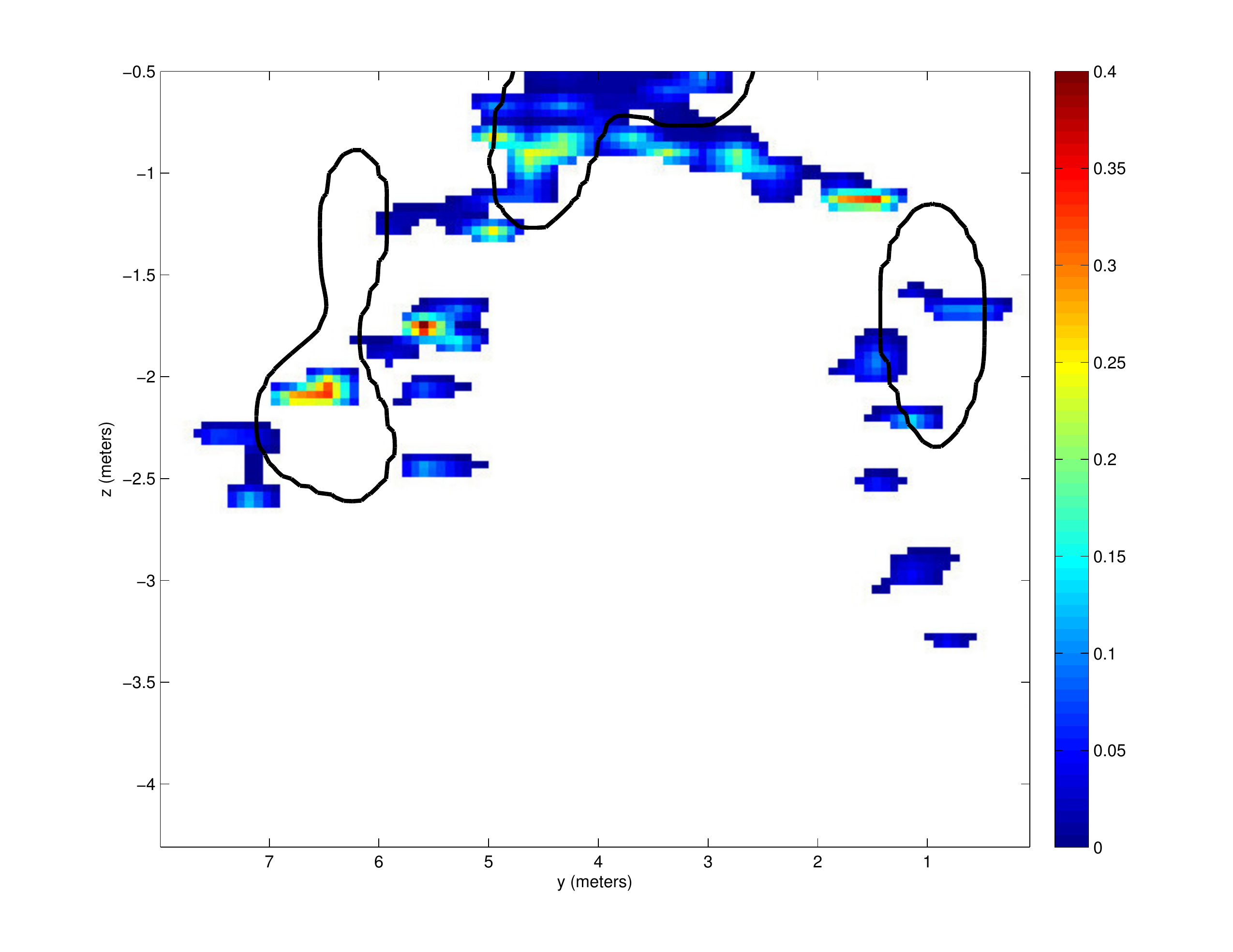}}
\\[-.2 cm]
\subfigure[]{\includegraphics[width=12pc]{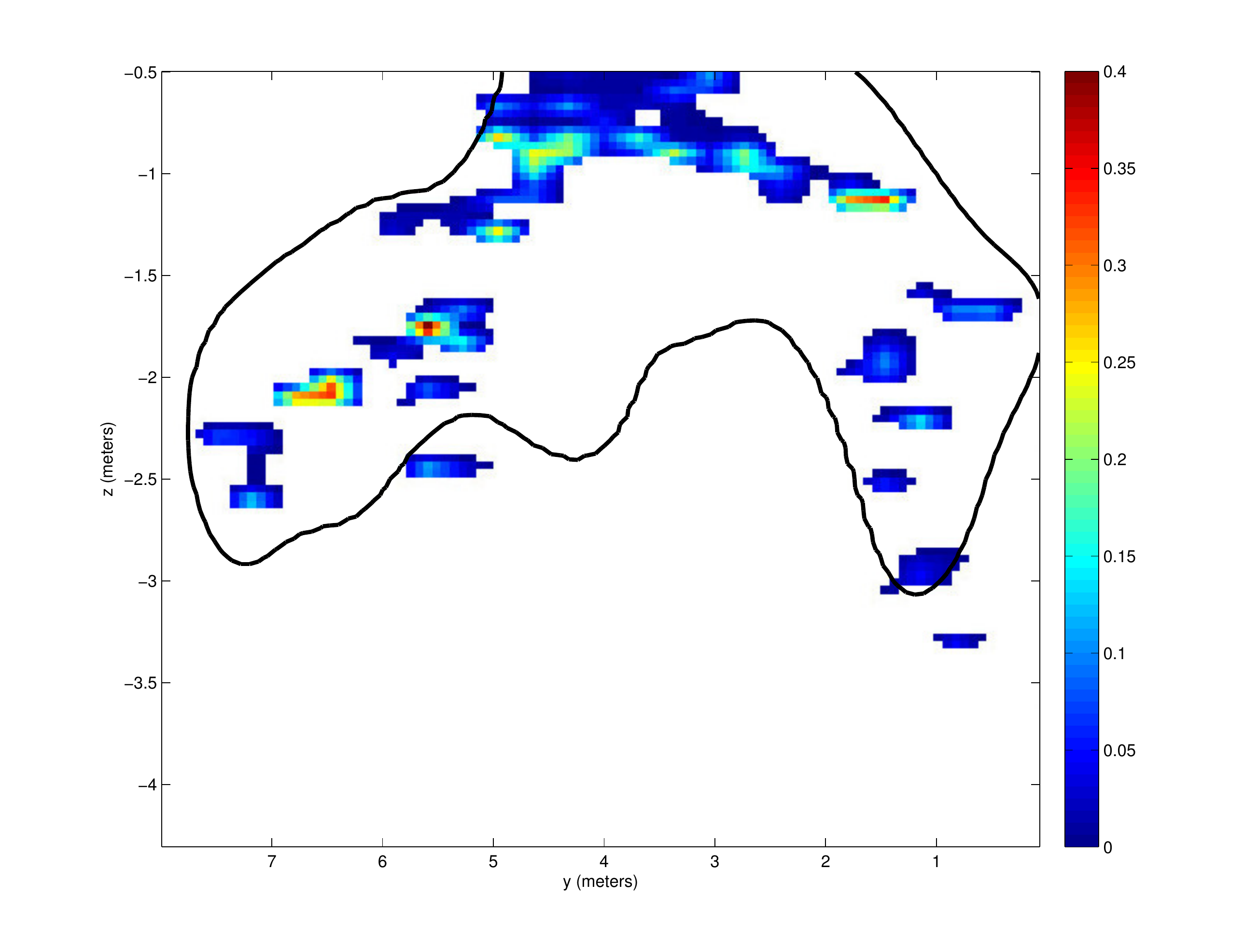}}
\subfigure[]{\includegraphics[width=12pc]{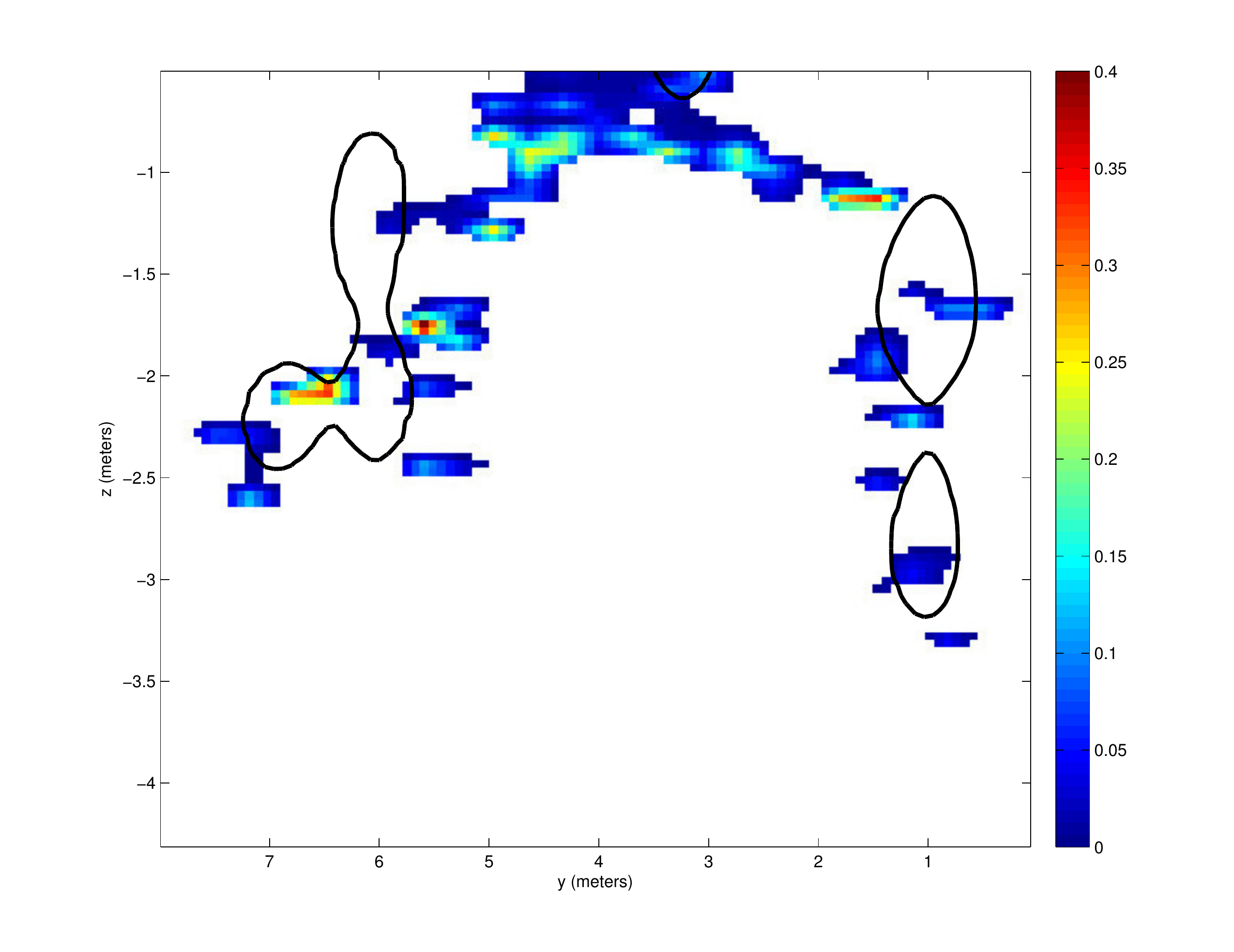}}
\\[-.2 cm]
\centering \subfigure[]{\includegraphics[width=16pc]{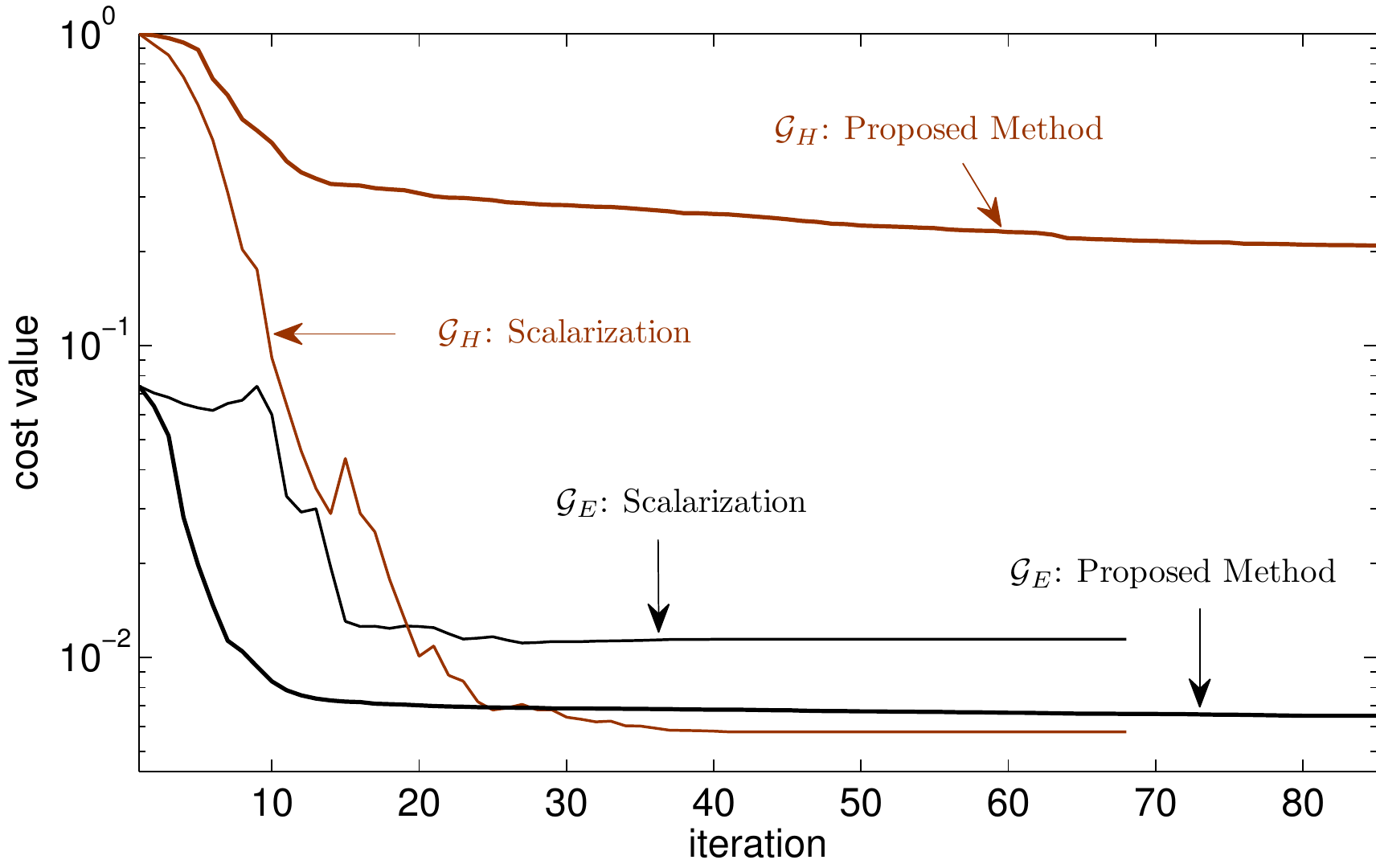}}

\caption{A slice of the saturation at $x=3.97$ m showing the DNAPL saturation in this plane and the corresponding reconstruction contour a) Using MONT; b) Using a scalarization approach; c) ERT-only data inversion; d) Hydrologic data inversion; e) Iterative performance of the scalarization approach compared to the proposed technique (MONT). Individual costs corresponding to every technique are marked }\label{fig6-2}
\end{figure}

The result of applying the multi-objective Newton-type (MONT) method is shown in Figure \ref{fig6}(c). The reconstruction visually compares well with the true source (Figure \ref{fig6}(a)) and contains many similar features (e.g., the multi-lobed shape is well represented and areas of DNAPL figuring are well captured). Besides the geometric comparison, in the simulations we calculate the relative error in the total reconstructed mass. This parameter is closely related to how successfully the reconstructed texture parameter captures the inhomogeneity of the shape. Based upon this metric, the joint inversion presented in Figure \ref{fig6}(c) is characterized by 5.5$\%$ error in the relative mass. To provide a more detailed picture of the reconstruction, in Figure \ref{fig6-2}(a) we have shown a transect of the saturation distribution and the corresponding reconstruction contour.

In contrast to MONT, reconstruction based upon the scalarization approach for joint inversion results in 8.1$\%$ relative mass error and a rather poor reconstruction as shown in Figure \ref{fig6}(d) and Figure \ref{fig6-2}(b). It is worth noting that considering different weights for the hydrologic and electrical misfit terms does not improve the results and this is mainly due to the thresholding phenomenon associated with the hydrologic model. In fact, the success of the scalarization approach in providing an acceptable mass error is thanks to the update of the texture parameter, $s_i$, based almost entirely on the ERT modality.

The reconstruction shown in Figure \ref{fig6}(c) is based upon the joint inversion of both electrical and hydrologic data. Thus, it is important to understand what each data set can provide when independently inverted. Reconstruction based on inverting only the ERT data is shown in Figures \ref{fig6}(e) and \ref{fig6-2}(c). The recovered mass error in this case is 1$\%$. Reconstruction based on inverting only the concentration data is shown in Figures \ref{fig6}(f) and \ref{fig6-2}(d) (76.4$\%$ relative mass error). The electrical-only reconstruction limits mass errors at the expense of shape matching (note the bulkier shape in Figure \ref{fig6}(e) and the loose contour in Figure \ref{fig6-2}(d)). In contrast, inversion based only on hydrologic data appears to be poor in both relative mass errors and shape matching. This is mainly due to the severe ill-posedness of the hydraulic model originating in the thresholding effect of aqueous solubility.

The iterative cost reductions for both the scalarized and MONT algorithms are shown in Figure \ref{fig6-2}(e). Both methods are initialized in identical states. In using a scalarization approach, the balance between the costs quickly changes and at every iteration the reduction of one cost term might be at the expense of increasing the other. The scalarization approach converges in about 45 iterations, but ends in a local minimum that is highly affected by the hydrologic model (compare Figures \ref{fig6}(d) and \ref{fig6}(f)). In using the MONT approach both individual cost terms monotonically decrease at every iteration. With the MONT approach, descent directions are determined based on the constrained problem (\ref{eq23}) and the number of iterations before reaching a steady state is about 75 iterations.

It is important to note that considering a constant hydraulic conductivity brings some level of uncertainty into the inversion model. Besides this, there is an added uncertainty associated with the thresholding effect of the hydrological model and the inaccuracy in the data. Based on these three sources of uncertainty, full matching of the hydrologic model with the data does not necessarily result in successful reconstructions (revisit Figure \ref{fig6}(f)). As observable in Figure \ref{fig6-2}(e), in using the scalarization approach, the ultimate cost value for the hydrologic model is less than that of MONT, originating from the dominance of the hydrologic cost in the inversion process. This dominance causes an early trapping in a local minimum. However, in case of MONT, roughly speaking, thanks to the mutual control that cost functions have over each other, only the ``useful'' portion of the hydrologic data is employed and by avoiding a full hydrologic data-model matching, an early local minimum is prevented. In this case the ERT cost value shows to be less than that of the scalarization approach and a more reliable iterative process is achieved.

\subsection{Using Two PaLS Functions for Characterization of Ganglia and Pools}
In the previous example a single level set function was used to reconstruct a shape representing the average DNAPL profile. When characterization efforts seek to distinguish areas of high saturation, a second PaLS function can be added to create a texture function (rather than a texture parameter as described above). A more general form of (\ref{eq24}) is
\begin{equation}\label{eq26}
s_n(\x)=s_i(\x)H_\epsilon \big( \phi_1 (\x,\bmu_{\p,1})\big)\ .
\end{equation}
The main difference between (\ref{eq24}) and (\ref{eq26}) is in the nature of $s_i$. In (\ref{eq24}), $s_i$ is only a scalar while in (\ref{eq26}), $s_i(\x)$ is a texture function defined within the source zone. This texture function is used to classify the source zone texture into low and high saturation values. More specifically we represent the texture as
\begin{equation}\label{eq27}
s_i(\x)=s_p H_\epsilon \big(\phi_2(\x,\bmu_{\p,2})\big) +s_g \Big(1-H_\epsilon \big(\phi_2(\x,\bmu_{\p,2})\big)\Big)\ ,
\end{equation}
where $s_p$ and $s_g$ are scalar values representing average saturation values for the pools and ganglia. Plugging (\ref{eq26}) into (\ref{eq27}) yields
\begin{align}
\hspace{.5cm}s_n(\x)=(s_p-s_g)H_\epsilon \big(\phi_2(\x,\bmu_{\p,2})\big) H_\epsilon \big(\phi_1(\x,\bmu_{\p,1})\big)+ s_g H_\epsilon \big(\phi_1(\x,\bmu_{\p,1})\big)\ .
\end{align}
In this representation the level set function $\phi_1$ characterizes the regions inside and outside the source zone. Within the source zone, the function $\phi_2$ characterizes the ganglia and the pools. Therefore this representation is capable of classifying regions as: zero saturation, low saturation or high saturation. Here rather than estimating $s_p$ and $s_g$, for simplicity we use fixed values with $s_g=$0.01 to define the extent of contamination and $s_p=$0.15 to then extract the regions containing pooled DNAPL. These values are arbitrary, but consistent with our prior work in defining ganglia and pools \cite{christ2006estimating}. Alternative values can be assigned to these two parameters, and the inversion will produce the corresponding iso-surfaces. PaLS parameters $[\bmu_{\p,1};\bmu_{\p,2}]$ are obtained through the joint inversion technique.

\begin{figure}[!htb]
\hspace{-.4cm}
\begin{tabular}{c}
\includegraphics[width=16pc]{fig6a}

\includegraphics[width=16pc]{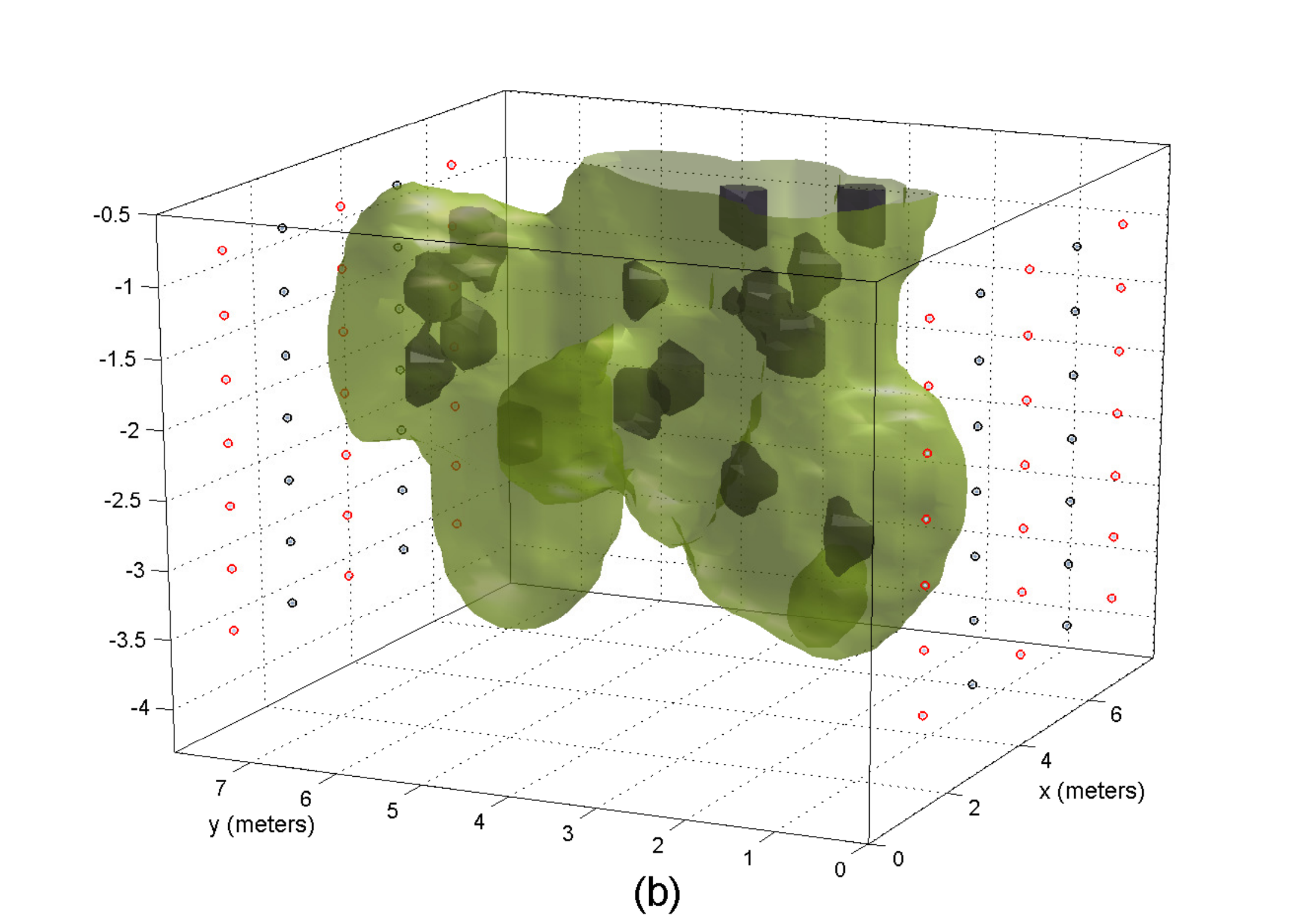}
\\
\includegraphics[width=19pc]{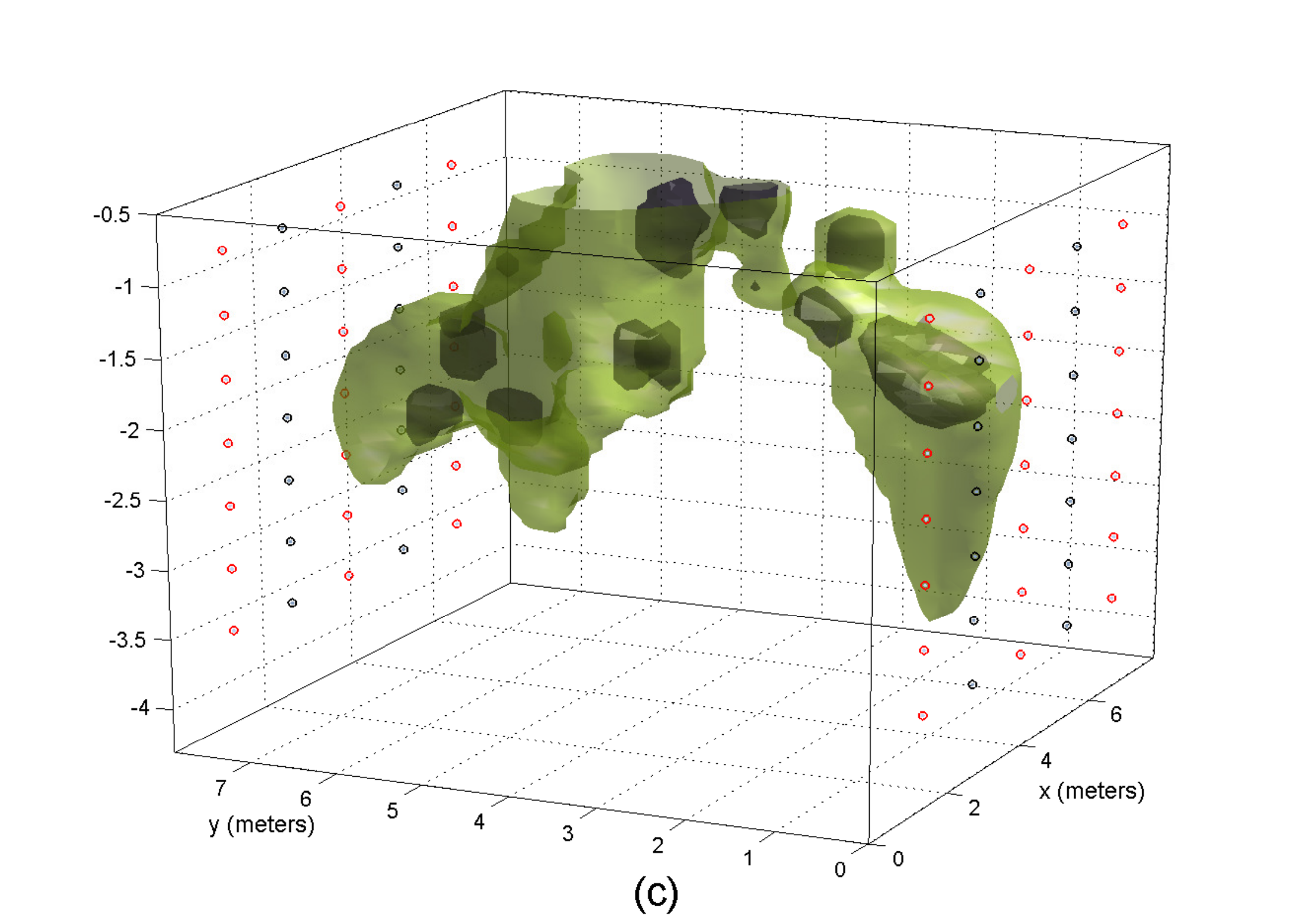}
\end{tabular}
\caption{Source zone reconstruction using two level set functions: a) original source zone structure with the iso-surfaces corresponding to 1$\%$ and 15$\%$ saturation values; b) initialization; c) reconstruction using the proposed joint inversion technique}\label{fig8}
\end{figure}

\begin{figure}[!htb]
\hspace{-.62cm}
\begin{tabular}{c}
\includegraphics[width=16pc]{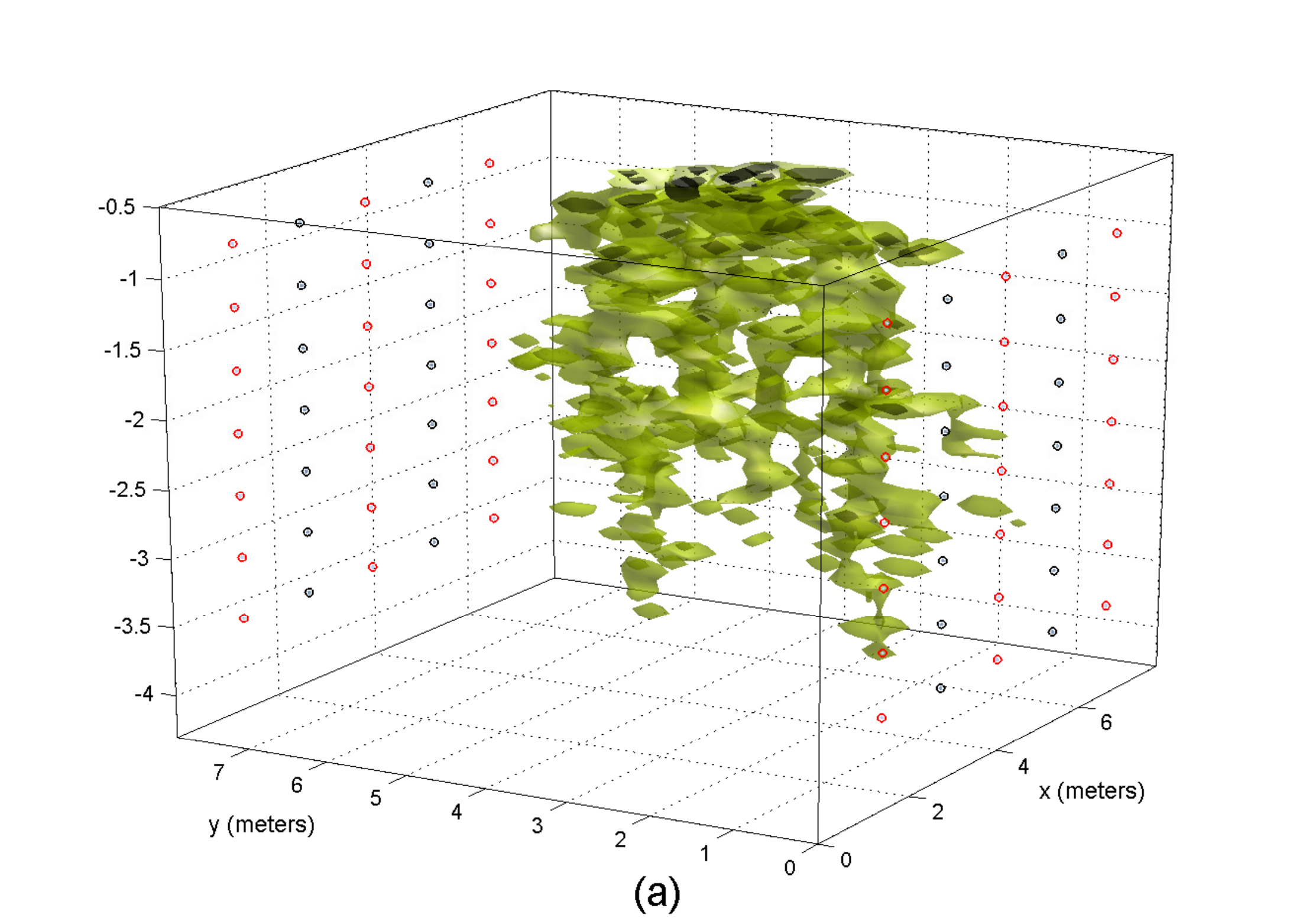}

\includegraphics[width=16pc]{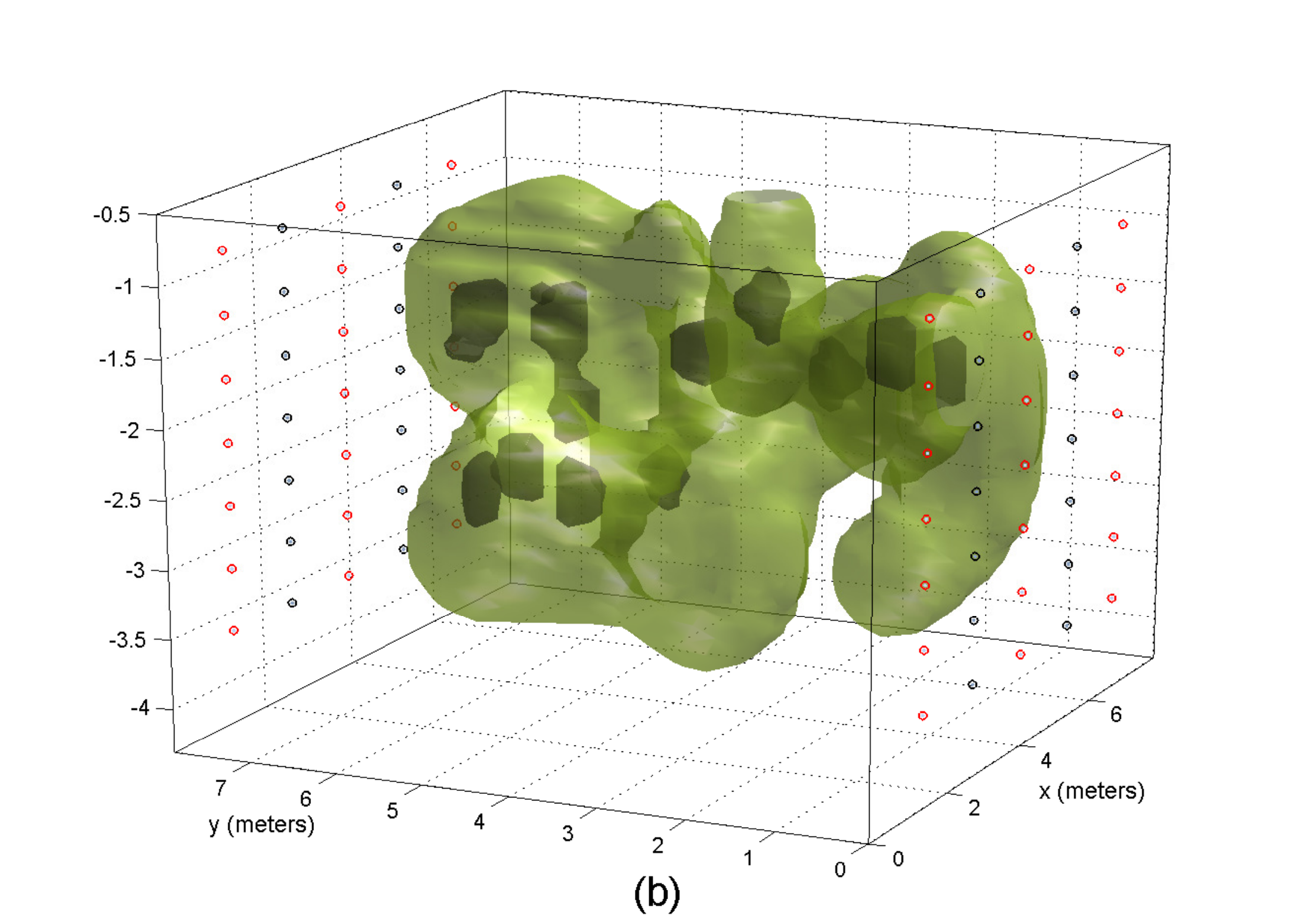}
\\
\includegraphics[width=19pc]{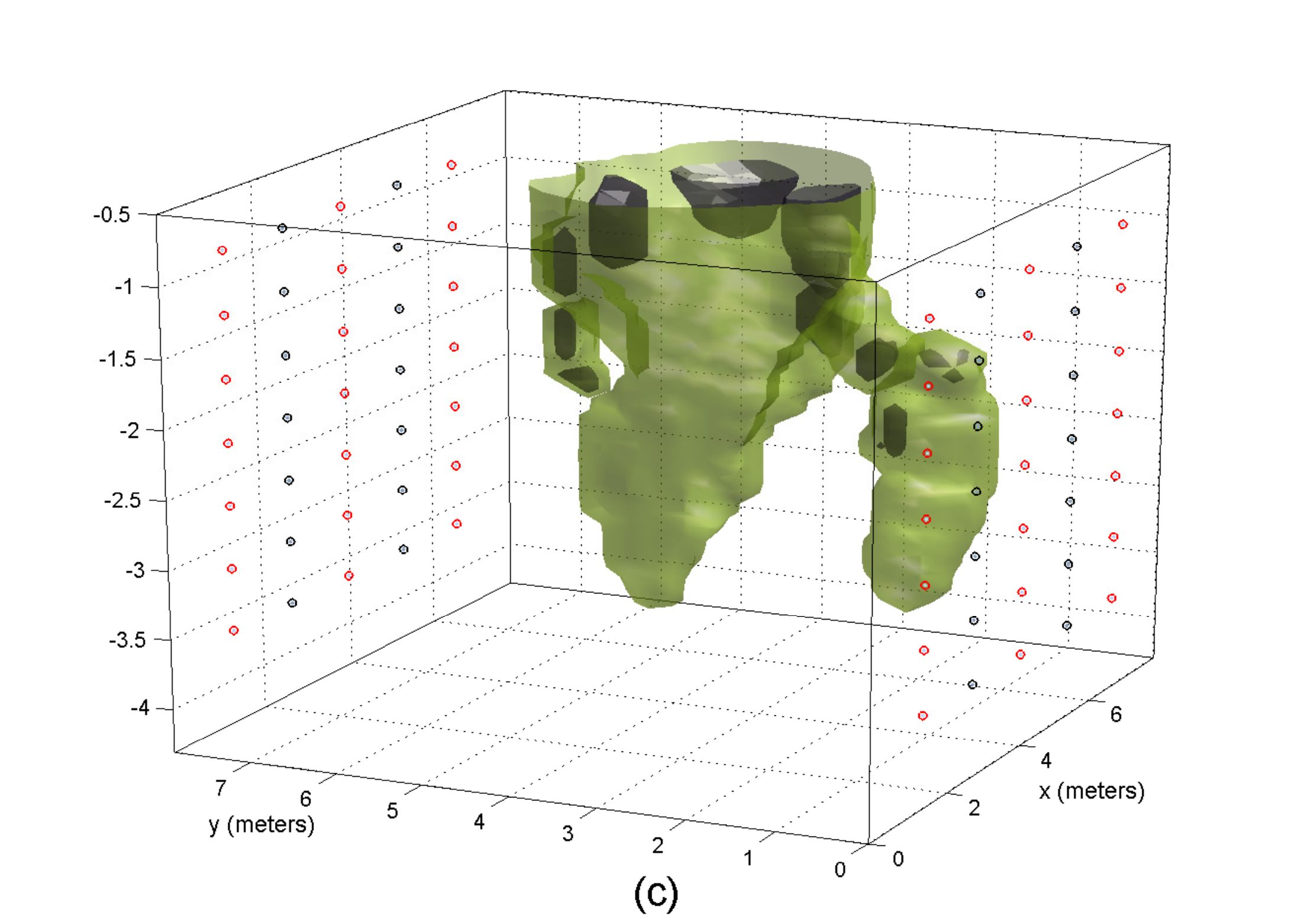}
\end{tabular}
\caption{Another source zone reconstruction using two level set functions: a) original source zone structure with the iso-surfaces corresponding to 1$\%$ and 15$\%$ saturation values; b) initialization; c) final reconstruction result }\label{fig9}
\end{figure}

\begin{figure}[!htb]\centering
\subfigure[]{\includegraphics[width=14pc]{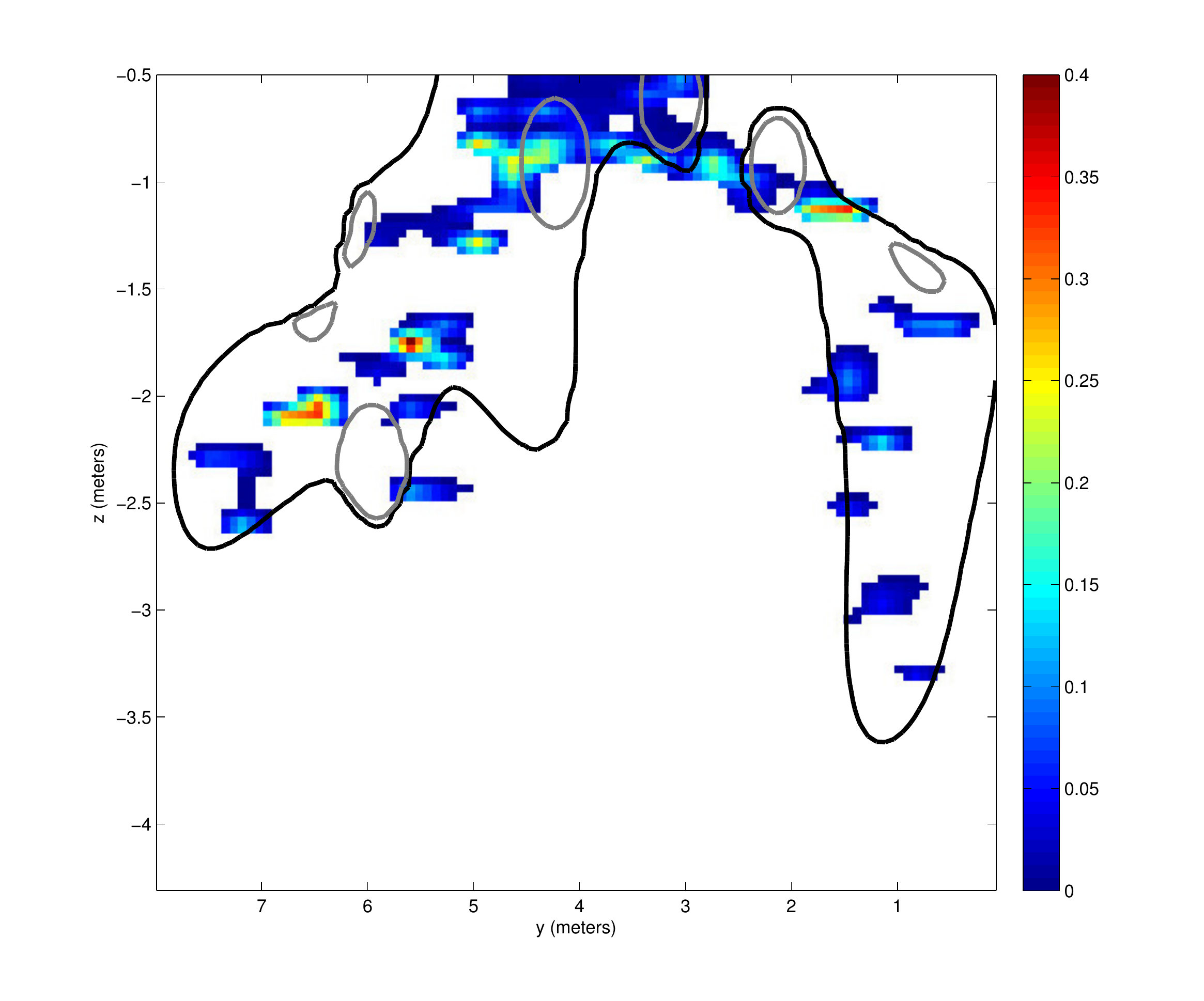}}
\subfigure[]{\includegraphics[width=14pc]{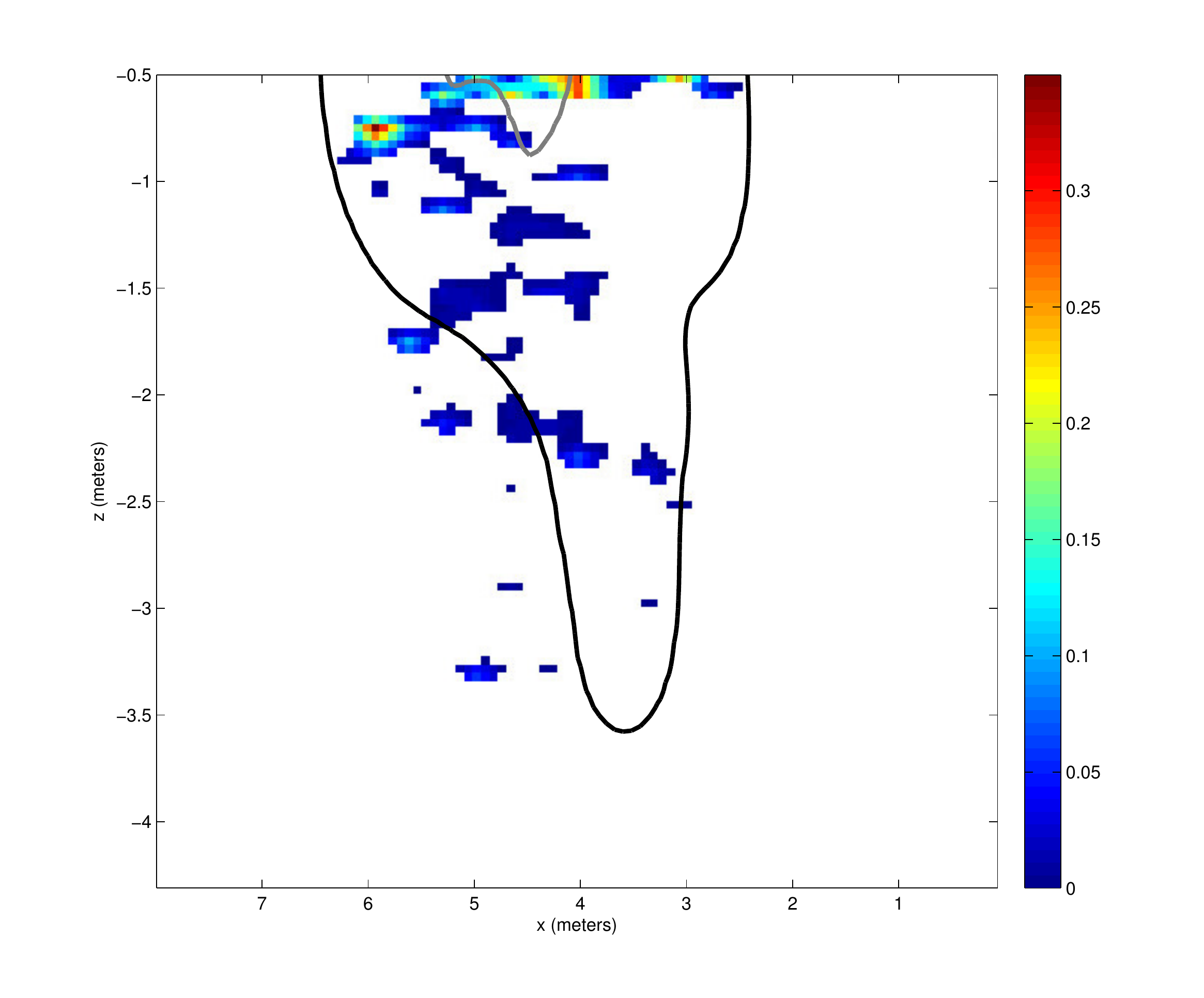}}
\caption{a) A slice of the saturation presented in Figure \ref{fig8}(a) at $x=3.97$ m; the ganglia contour is shown in black and the pools contour is in gray b) A slice of the saturation in Figure \ref{fig9}(a) at $y=3.97$ m, again showing the ganglia contour in black and the pools contour in gray}\label{fig10}
\end{figure}

As a first example in differentiating ganglia and pooled regions we use the same electrical and hydrological data sets as the single level set case considered in the previous section. In representing $\phi_1$ we again use 45 bumps with similar initialization parameters. To represent $\phi_2$ we use 35 bumps, as detailed in Table \ref{tab2}. Initial centers are chosen randomly amongst those associated with $\phi_1$, and the dilation factor is selected to be $\beta_i=2$ to result in smaller initial geometries for the pools. The overall initialization is shown in Figure \ref{fig8}(b) where the green iso-surface corresponds to the initial ganglia and the brown surface represents the initial pools.

Visual comparison of the joint inversion results (Figure \ref{fig8}(c)) with the original saturation profile (Figure \ref{fig8}(a)) suggests that the features corresponding to ganglia and pools are well reconstructed. The error in recovering the total mass for this reconstruction was 7.8$\%$. It is interesting to note that, in this case at least, use of the two level sets actually decreases overall performance of the joint inversion. Thus, the benefits of delineating the high saturation regions come at the expense of some accuracy. A transect of the source zone along with the underlying contours are shown in Figure \ref{fig10}(a). It can be observed that besides the good performance in characterizing the source zone, the pool contours are more or less accumulated around the denser pooling sites. The pools in this release are very sporadic and distributed all around the source zone. To more clearly illustrate the performance of the method in characterizing the pools, a second source zone architecture (Figure \ref{fig9}(a)) was created by simulating the release of a PCE-DNAPL within a different permeability field (not shown). For this example although still sporadic, most pooling sites are close to the ground surface. Initial PaLS parameters are the same as in the previous example (Table \ref{tab2}) though the initialization (Figure \ref{fig9}(b)) employed a new set of random centers.

The resulting source zone reconstruction is shown in Figure \ref{fig9}(c). Besides the successful reconstruction of the ganglia structure, this example more clearly shows the accumulation of reconstructed pools close to the surface where the actual pools are located. Despite the random initialization of the pool locations, in the course of inversion they are gradually pushed towards the actual locations. The error in recovering the total mass for this example was 2.5$\%$ and the reconstruction was performed in 32 iterations. As before, a 2D transect of the saturation and the reconstruction contours are presented in Figure \ref{fig10}(b), that help showing the appearance of the pools contour close to the actual pooling sites.

\section{Conclusion}
This paper basically provides a geometric approach to joint inversion. While a specific application was considered herein, the formulation and strategy in solving the inversion in a joint form is general. In fact, the joint inversion approach that is presented in this work can be adopted by other disciplines to combine various physical modalities. This is particularly important as many joint inversion techniques simply suggest using a scalarization approach for physically incompatible models.

Application of the proposed technique to the DNAPL characterization problem produces reasonable reconstructions of the source zone structure based on the fusion of electrical and hydraulic sensors. Using this technique we are able to reconstruct a detailed picture of the source zone along with useful information about its texture and the high saturation regions. In fact, our results suggest how hydrology information can be used in a controlled way to ``help'' improving the geophysical reconstructions.

Although application of geophysical techniques to 3D problems can be found in the literature, use of a fully 3D, multiphase flow and transport model within an inversion routine has not previously been demonstrated, due in large part to the computational burden and ill-posedness associated with the use of such a model. In the context of joint inversion, the problem becomes even more complex, since appropriately combining the modalities and making use of different data sets significantly exacerbates the problem. Our ability to employ 3D models is directly linked to the low-order nature of the PaLS and still its high flexibility in shape representation. Thanks to this low-order representation, applying the MONT technique becomes computationally tractable and a controlled inversion over incompatible data sets becomes possible.

The presented work is in fact a proof of concept to what can be done in the future. As stated before, a future work would be exploring other geophysical modalities, such as GPR, which are capable of providing a higher contrast at the expense of computational complexity. Moreover, to reduce the hydrologic model uncertainties, methods of estimating the hydraulic permeability prior to (or in line with) the inversion may be considered. Ultimately, the successful performance of the method in the challenging and rather realistic examples considered, makes it a reasonable technology to be tested with real field data.

\appendix
\section{Joint Minimization Algorithm}\label{app}
In case of having a single least squares cost as
\begin{equation}
\label{r1}
\mathcal{G}(\bmu)=\frac{1}{2}\|\boldsymbol{\mathcal{E}}(\bmu)\|_{\boldsymbol{R}}^2,
\end{equation}
the Levenberg-Marquardt (LM) algorithm suggests taking a damped Gauss-Newton step, where at every iteration $k$, a descent direction $\bdelta_k$ is acquired via
\begin{equation}\label{r2}
\bdelta_k=\argmin_{\bdelta} \frac{1}{2}\bdelta^T\!\Big(\boldsymbol{J}(\bmu_k)^T\boldsymbol{R}\boldsymbol{J}(\bmu_k)+\lambda_k\boldsymbol{I}\Big)\bdelta+ \bdelta^T\boldsymbol{J}(\bmu_k)^T\boldsymbol{R}\boldsymbol{\mathcal{E}} (\bmu_k).
\end{equation}
Here $\boldsymbol{J}=\partial\boldsymbol{\mathcal{E}}/\partial \bmu$ and the objective function is a second order Taylor approximation to $\mathcal{G}(\bmu_k+\bdelta)-\mathcal{G}(\bmu_k)$, where $\nabla\mathcal{G}=\boldsymbol{J}^T\boldsymbol{R}\boldsymbol{\mathcal{E}}$ and $\nabla^2  \mathcal{G}\simeq {\boldsymbol{J}}^T \boldsymbol{R}  \boldsymbol{J}$. For a positive damping factor $\lambda_k$, the minima to (\ref{r2}) is bounded and may be uniquely determined as
\begin{equation}\label{r3}
\bdelta_k=-\Big(\boldsymbol{J}(\bmu_k)^T\boldsymbol{R}\boldsymbol{J}(\bmu_k)+\lambda_k\boldsymbol{I}\Big)^{-1} \boldsymbol{J}(\bmu_k)^T\boldsymbol{R}\boldsymbol{\mathcal{E}}(\bmu_k).
\end{equation}
Determining a suitable value for $\lambda_{k+1}$ is based on a \emph{gain ratio} parameter \cite{more1978levenberg, madsen1999methods} defined as
\begin{equation}\label{r4}
\rho_k= \frac{\mathcal{G}(\bmu_k+\bdelta_k)-\mathcal{G}(\bmu_k)}{\frac{1}{2} \|\boldsymbol{J}(\bmu_k) \bdelta_k \|_{\boldsymbol{R}}^2+ \bdelta_k^T\nabla \mathcal{G}(\bmu_k)}.
\end{equation}
This parameter is in fact the ratio between the actual and predicted cost decrease (note the denominator in (\ref{r4}) is the same as the objective in (\ref{r2}) when the $\lambda_k$ term is neglected).

In the original version of the algorithm proposed by Marquardt \cite{marquardt1963algorithm}, small and negative values of $\rho$ indicate that $\lambda$ needs to increase in the next iteration and large values of $\rho$ indicate that $\lambda$ may be reduced. We specifically consider the more recent updating rule proposed in \cite{madsen1999methods} which is shown to outperform the original scheme. The updating strategy is\\[-1.4em]
\begin{flalign}\label{ra}
\nonumber\quad\;\;\;\;&\textbf{if}\;\;\rho_k>0\;\; \textbf{then}&\\[-.75em]\nonumber
&\quad\lambda_{k+1}\leftarrow\lambda_k\max  ( \frac{1}{3},1-(2\rho_{k}-1)^3 );&\\[-.55em]\nonumber &\quad v_k\leftarrow 2;&\\[-.1em]\nonumber &\textbf{else}&\\[-.35em]\nonumber & \quad\lambda_{k+1}\leftarrow v_k\lambda_k;&\\[-.05em]\nonumber & \quad v_{k+1}\leftarrow 2v_k;&\\[-.1em]\nonumber &\textbf{end}.&\\[-2em]
\end{flalign}
For more details, the interested reader is referred to \cite{madsen1999methods} and references therein.

We now focus on the generalization of the LM algorithm to the multi-objective case. Consider having multiple least squares costs as
\begin{equation}
\label{r5}
\mathcal{G}_j(\bmu)=\frac{1}{2}\|\boldsymbol{\mathcal{E}}_j(\bmu)\|_{\boldsymbol{R}_j}^2,\qquad j=1,2,\cdots ,m_c.
\end{equation}
Inspired by the LM algorithm for the single-objective case, in an iterative multi-objective scheme the Hessian for each cost may be approximated as
\begin{equation}\label{r8}
\tilde{\boldsymbol{H}}_j(\bmu_k)=\boldsymbol{J}_j(\bmu_k)^T\boldsymbol{R}_j\boldsymbol{J}_j(\bmu_k)+\lambda^{(j)}_k\boldsymbol{I},
\end{equation}
where the $\lambda^{(j)}_k$ quantities are the corresponding damping factors. According to (\ref{eq22}), at each iteration a potential direction $\bdelta_k$ is acquired via
\begin{equation}\label{r6}
\bdelta_k= \argmin_{\bdelta} F_k(\bdelta),
\end{equation}
where
\begin{equation}\label{r7}
F_k(\bdelta)=\maxi_{j=1,\cdots, m_c} \frac{1}{2}\bdelta^T\tilde{\boldsymbol{H}}_j(\bmu_k)\bdelta+ \bdelta^T\boldsymbol{J}_j(\bmu_k)^T\boldsymbol{R}_j\boldsymbol{\mathcal{E}}_j(\bmu_k).
\end{equation}
Verifying the descent property of $\bdelta_k$ for every individual cost is quite straightforward.\\[-.8em]

\noindent\textbf{Proposition A.1}. \emph{For $\lambda^{(j)}_k>0$, a nonzero direction $\bdelta_k$ acquired from (\ref{r6}) is a descent direction for every individual cost $\mathcal{G}_j$ where $j=1,2,\cdots,m_c$.}\\[-.5em]

\noindent\textbf{Proof}. From (\ref{r7}) we have $F_k(\boldsymbol{0})=0$ and since $\bdelta_k$ is a minima for the convex program (\ref{r6})
\begin{equation}\label{r8}
F_k(\bdelta_k)\leq F_k(\boldsymbol{0})=0,
\end{equation}
which results in
\begin{equation}\label{r9}
\frac{1}{2}\bdelta_k^T\tilde{\boldsymbol{H}}_j(\bmu_k)\bdelta_k+ \bdelta_k^T\boldsymbol{J}_j(\bmu_k)^T\boldsymbol{R}_j\boldsymbol{\mathcal{E}}_j(\bmu_k) \leq 0,\quad \forall j\in\{1,2,\cdots,m_c\}.
\end{equation}
Clearly
\begin{equation}\label{r10}
\lambda^{(j)}_k>0 \Rightarrow \tilde{\boldsymbol{H}}_j(\bmu_k)\succ 0,
\end{equation}
and therefore for $\bdelta_k\neq \boldsymbol{0}$ we have
\begin{align}\label{r11}
\nonumber\hspace{-3.5cm}\forall j\in\{1,2,\cdots,m_c\}: \qquad \bdelta_k^T\nabla \mathcal{G}_j(\bmu_k)&=\bdelta_k^T\boldsymbol{J}_j(\bmu_k)^T\boldsymbol{R}_j\boldsymbol{\mathcal{E}}_j(\bmu_k)\\\nonumber &\leq -\frac{1}{2}\bdelta_k^T\tilde{\boldsymbol{H}}_j(\bmu_k)\bdelta_k\\&< 0,
\end{align}
where the inequalities are thanks to (\ref{r9}) and (\ref{r10}) respectively. $\hspace{2.2cm}\square$\\[-.7em]

Now that $\bdelta_k$ is verified to have the descent property for every individual cost, our strategy to establish a multi-objective LM scheme is defining a separate gain ratio $\rho^{(j)}_k$ for each cost $\mathcal{G}_j$, which controls the corresponding damping factor. For each cost we consider a similar procedure as (\ref{ra}) to update the damping factor.

Algorithm \ref{alg} is a detailed generalization of the single-objective LM algorithm proposed in \cite{madsen1999methods} to a multi-objective LM scheme. For brevity we only considered two least squares costs $\mathcal{G}_\h$ and $\mathcal{G}_{\mathpzc{E}}$ as the case in this paper, however, a further generalization to multiple costs is straightforward.

\begin{algorithm}[!htb]\label{alg}
\caption{A Multi-Objective LM Algorithm}

\lnl{l1}$v_\h\leftarrow 2;$ $v_{\mathpzc{E}}\leftarrow 2;$ $\varepsilon\leftarrow\varepsilon_0{}^*$;  \quad\scriptsize{\%($\varepsilon_0$ may be as small as the machine precision)}\normalsize

\lnl{l1}$\bmu\leftarrow \bmu_0;$

\lnl{l2}$\lambda_\h\leftarrow\max( \mathrm{diag} (\boldsymbol{J}_{\!\h}(\bmu)^T\boldsymbol{R}_{\h}\boldsymbol{J}_{\!\h}(\bmu)));$

\lnl{l2}$\lambda_{\mathpzc{E}}\leftarrow\max( \mathrm{diag} (\boldsymbol{J}_{\!{\mathpzc{E}}}(\bmu)^T\boldsymbol{R}_{{\mathpzc{E}}}\boldsymbol{J}_{\!{\mathpzc{E}}}(\bmu)));$

\lnl{l2}$found :=\mbox{\textbf{false}}$;

\While{$\sim found$ }{

\lnl{l3}$\tilde{\boldsymbol{H}}_{\h}\leftarrow\boldsymbol{J}_{\!\h}(\bmu)^T\boldsymbol{R}_{\h}\boldsymbol{J}_{\!\h}(\bmu)+\lambda_{\h}\boldsymbol{I}$;

\lnl{l3}$\tilde{\boldsymbol{H}}_{{\mathpzc{E}}}\leftarrow\boldsymbol{J}_{\hspace{-.2mm}{\mathpzc{E}}}(\bmu)^T\boldsymbol{R}_{{\mathpzc{E}}}\boldsymbol{J}_{\hspace{-.2mm}{\mathpzc{E}}}(\bmu)+\lambda_{{\mathpzc{E}}}\boldsymbol{I}$;

\lnl{l3}\mbox{solve (\ref{r6}) to determine} $\bdelta$; \quad\scriptsize{\%(See \ref{app2} for a solution strategy)} \normalsize

\lnl{l3}$\rho_\h\leftarrow(\mathcal{G}_\h(\bmu+\bdelta)-\mathcal{G}_\h(\bmu) )  / ( \bdelta^T \nabla \mathcal{G}_\h(\bmu) + \frac{1}{2} \|\boldsymbol{J}_{\!\h}(\bmu) \bdelta \|_{\boldsymbol{R}_{\h}}^2)$;

\lnl{l3}$\rho_{\hspace{-.2mm}{\mathpzc{E}}}\leftarrow(\mathcal{G}_{\hspace{-.2mm}{\mathpzc{E}}}(\bmu+\bdelta)-\mathcal{G}_{\hspace{-.2mm}{\mathpzc{E}}}(\bmu) )  / ( \bdelta^T \nabla \mathcal{G}_{\hspace{-.2mm}{\mathpzc{E}}}(\bmu) + \frac{1}{2} \|\boldsymbol{J}_{\hspace{-.2mm}{\mathpzc{E}}}(\bmu) \bdelta \|_{\boldsymbol{R}_{\hspace{-.2mm}{\mathpzc{E}}}}^2)$;

\eIf{$\rho_\h>0$ \bf{and} $\rho_{\mathpzc{E}}>0$}{
\lnl{l18}$\bmu\leftarrow\bmu+\bdelta$\;
\lnl{l18}$\lambda_\h\leftarrow\lambda_\h\max  ( \frac{1}{3},1-(2\rho_\h-1)^3 )$\;
\lnl{l18}$\lambda_{\mathpzc{E}}\leftarrow\lambda_{\mathpzc{E}}\max  ( \frac{1}{3},1-(2\rho_{\mathpzc{E}}-1)^3 )$\;
\lnl{l18}$v_\h\leftarrow 2$\;
\lnl{l18}$v_{\mathpzc{E}}\leftarrow 2$\;
\lnl{l18}$found:=(\|\bdelta\|<\varepsilon)$\;
}{
\If{$\rho_\h\leq0$}{
\lnl{l19}$\lambda_\h \leftarrow v_\h \lambda_\h $\;
\lnl{l18}$v_\h \leftarrow 2v_\h$\;
}
\If{$\rho_\h\leq0$}{
\lnl{l19}$\lambda_{\mathpzc{E}} \leftarrow v_\h \lambda_{\mathpzc{E}}$\;
\lnl{l18}$v_{\mathpzc{E}} \leftarrow 2v_{\mathpzc{E}}$\;
}
}
}
\lnl{l19}\Return $\bmu$;

\end{algorithm}

\section{Determining the Descent Direction}\label{app2}
In this section we discuss a solution strategy to the convex program
\begin{equation}\label{r12}
       \left\{
     \begin{array}{c}
       \underset{{(\boldsymbol{y},z)}} {\min} \; z\\[.3cm]
       \mbox{s.t.:}\quad\y^T \boldsymbol{g}_j+\frac{1}{2}\y^T \boldsymbol{H}_j\y-z\leq 0\\[.2cm]
       j=1,2,\cdots,m_c
     \end{array}\ ,
   \right.
\end{equation}
where $\boldsymbol{g}_j\in\mathbb{R}^M$ and $\boldsymbol{H}_j\in\mathbb{R}^{M\times M}$ are known vectors and matrices. Moreover all $\boldsymbol{H}_j$ matrices are assumed to be symmetric positive definite, i.e.,
\begin{equation}\label{r13}
\qquad \boldsymbol{H}_j\succ \boldsymbol{0}, \qquad j=1,2,\cdots,m_c.
\end{equation}
The goal is to find a vector $\boldsymbol{x}=[\y;z]\in \mathbb{R}^{M+1}$ which globally minimizes (\ref{r12}).

Among the variety of techniques that can address (\ref{r12}) (e.g., see \cite{bertsekas1999nonlinear}), we consider the interior point method with a log-barrier function, which is quite straightforward to implement \cite{boyd2004convex}.

In the standard log-barrier method, using $\tau>0$, (\ref{r12}) is transformed into an unconstrained problem
\begin{equation}\label{r14}
\eta_{\tau}(\boldsymbol{x})=z+\frac{1}{\tau}\sum_{j=1}^{m_c}-\log(z-\y^T \boldsymbol{g}_j-\frac{1}{2}\y^T \boldsymbol{H}_j\y),
\end{equation}
where the constraints are incorporated into the cost via a log penalty function. As $\tau\to \infty$, the minima of $\eta_{\tau}(\boldsymbol{x})$ approaches the solution to (\ref{r12}) \cite{boyd2004convex}. Accordingly, the main strategy is to consider an increasing sequence of $\tau_k$ values. Employing a fast converging minimization scheme (such as a Newton-type method), for each $k$, $\eta_{\tau_k}(\boldsymbol{x})$ is minimized using the minima of $\eta_{\tau_{k-1}}(\boldsymbol{x})$ as a starting point. It is worth mentioning that if $\tau$ is picked to be large in the first trial, $\eta_\tau(\boldsymbol{x})$ becomes difficult to minimize by a Newton-type technique as the Hessian varies quickly near the boundaries of the feasible set \cite{boyd2004convex}. This is basically the intuition behind the gradual increment of the $\tau_k$ values.

Referring to (\ref{r14}), the domain of $\eta_\tau(.)$ is characterized as
\begin{equation}\label{r14.5}
\mbox{dom}(\eta_\tau)=\Big\{[\y;z]\in\mathbb{R}^{M+1}: \;z-\y^T \boldsymbol{g}_j-\frac{1}{2}\y^T \boldsymbol{H}_j\y>0,\; j=1,2,\cdots,m_c \Big\}.
\end{equation}
It is straightforward to show that for a given $\boldsymbol{x}=[\y;z]\in \mbox{dom}(\eta_\tau)$
\begin{equation}\label{r15}
\nabla \eta_\tau (\boldsymbol{x})= \left( \!\!\begin{array}{c}
\boldsymbol{0} \\
1 \end{array}  \!\!\right)+ \frac{1}{\tau}\sum_{j=1}^{m_c}a_j\left( \!\!\begin{array}{c}
\boldsymbol{k}_j \\
-1 \end{array}  \!\!\right)
\end{equation}
and
\begin{equation}\label{r16}
\nabla^2 \eta_\tau (\boldsymbol{x})= \frac{1}{\tau}\sum_{j=1}^{m_c}a_j^2\left( \!\!\begin{array}{c}
\boldsymbol{k}_j \\
-1 \end{array}  \!\!\right)\left( \!\!\begin{array}{c}
\boldsymbol{k}_j \\
-1 \end{array}  \!\!\right)^T+\frac{1}{\tau}\sum_{j=1}^{m_c}a_j\left( \!\!\begin{array}{cc}
\boldsymbol{H}_{\!j} & \boldsymbol{0}\\
\boldsymbol{0}^T & 0 \end{array}  \!\!\right),
\end{equation}
where $a_j=(z-\y^T \boldsymbol{g}_j-\frac{1}{2}\y^T \boldsymbol{H}_j\y)^{-1}$, $\boldsymbol{k}_j= \boldsymbol{H}_j\y+\boldsymbol{g}_j$, and $\boldsymbol{0}\in \mathbb{R}^M$ is a zero vector.

For every $\tau_k$, using (\ref{r15}) and (\ref{r16}), a Newton-type scheme may be used to minimize $\eta_\tau(\boldsymbol{x})$. In the sequel we show that $\nabla^2 \eta_\tau(\boldsymbol{x})$ is positive definite over $\mbox{dom}(\eta_\tau)$. As a result, standard linear solvers for symmetric positive definite matrices (e.g., the conjugate gradient method \cite{bertsekas1999nonlinear}) may be employed to efficiently obtain the Newton directions for each subproblem.\\[-.8em]

\noindent\textbf{Proposition B.1}. \emph{For $\tau>0$, and all $\boldsymbol{x}\in \mbox{dom}(\eta_\tau)$}
\[\nabla^2 \eta_\tau (\boldsymbol{x})\succ \boldsymbol{0}.
\]

\noindent\textbf{Proof}. Referring to (\ref{r14.5}), for any $\boldsymbol{x}\in \mbox{dom}(\eta_\tau)$, we clearly have $a_j>0$. Moreover, (\ref{r16}) states that we can write $\nabla^2 \eta_\tau$ in the following block form
\begin{equation}\label{r17}
\nabla^2 \eta_\tau =\frac{1}{\tau}\left( \!\!\begin{array}{cc}
\sum_{j=1}^{m_c}a_j^2\boldsymbol{k}_j \boldsymbol{k}_j^T + \sum_{j=1}^{m_c}a_j\boldsymbol{H}_j & -\sum_{j=1}^{m_c}a_j^2 \boldsymbol{k}_j\\
-\sum_{j=1}^{m_c}a_j^2 \boldsymbol{k}_j^T & \sum_{j=1}^{m_c}a_j^2 \end{array}  \!\!\right).
\end{equation}
The Schur complement of the lower-right matrix block in (\ref{r17}) is
\begin{equation}\label{r18}
\boldsymbol{S}_{\nabla^2\eta_\tau}=\frac{1}{\tau}\Big( \sum_{j=1}^{m_c}a_j^2\boldsymbol{k}_j \boldsymbol{k}_j^T + \sum_{j=1}^{m_c}a_j\boldsymbol{H}_j - \frac{\big(\sum_{j=1}^{m_c}a_j^2 \boldsymbol{k}_j\big)\big(\sum_{j=1}^{m_c}a_j^2 \boldsymbol{k}_j^T\big)}{\sum_{j=1}^{m_c}a_j^2}\Big).
\end{equation}
Since $\sum_{j=1}^{m_c}a_j^2>0$, to show $\nabla^2 \eta_\tau\succ \boldsymbol{0}$, we only need to show $\boldsymbol{S}_{\nabla^2\eta_\tau}\succ \boldsymbol{0}$ (see \S A.5.5 in \cite{boyd2004convex}). For this purpose we rewrite (\ref{r18}) as
\begin{equation}\label{r19}
\boldsymbol{S}_{\nabla^2\eta_\tau}= \frac{1}{\tau}\sum_{j=1}^{m_c}a_j\boldsymbol{H}_j + \frac{1}{\tau \sum_{j=1}^{m_c}a_j^2}\Big( \big( \sum_{j=1}^{m_c}a_j^2 \big)\big( \sum_{j=1}^{m_c}a_j^2\boldsymbol{k}_j \boldsymbol{k}_j^T\big)-\big(\sum_{j=1}^{m_c}a_j^2 \boldsymbol{k}_j\big)\big(\sum_{j=1}^{m_c}a_j^2 \boldsymbol{k}_j^T\big)\Big).
\end{equation}
Clearly,
\begin{equation}\label{r20}
\sum_{j=1}^{m_c}a_j\boldsymbol{H}_j \succ \boldsymbol{0},
\end{equation}
as $a_j>0$ and $\boldsymbol{H}_j\succ \boldsymbol{0}$ for $j=1,2,\cdots,m_c$. Furthermore
\begin{align}\label{r21}
\nonumber\big( \sum_{j=1}^{m_c}a_j^2 \big)\big( \sum_{j=1}^{m_c}a_j^2\boldsymbol{k}_j \boldsymbol{k}_j^T\big)&-\big(\sum_{j=1}^{m_c}a_j^2 \boldsymbol{k}_j\big)\big(\sum_{j=1}^{m_c}a_j^2 \boldsymbol{k}_j^T\big)=\sum_{i=1}^{m_c}\sum_{j=1}^{m_c}\big(a_i^2a_j^2\boldsymbol{k}_j \boldsymbol{k}_j^T - a_i^2a_j^2\boldsymbol{k}_i \boldsymbol{k}_j^T\big)\\ \nonumber &=\sum_{i,j: i\neq j} a_i^2a_j^2 (\boldsymbol{k}_j-\boldsymbol{k}_i)\boldsymbol{k}_j^T\\ \nonumber &= \hspace{-.45cm} \sum_{i,j: 1\leq i<j\leq m_c} \hspace{-.6cm}a_i^2a_j^2 (\boldsymbol{k}_j-\boldsymbol{k}_i)\boldsymbol{k}_j^T + \hspace{-.45cm}\sum_{i,j: 1\leq j<i\leq m_c} \hspace{-.6cm} a_i^2a_j^2 (\boldsymbol{k}_j-\boldsymbol{k}_i)\boldsymbol{k}_j^T\\ \nonumber &= \hspace{-.45cm} \sum_{i,j: 1\leq i<j\leq m_c} \hspace{-.6cm}a_i^2a_j^2 (\boldsymbol{k}_j-\boldsymbol{k}_i)\boldsymbol{k}_j^T + \hspace{-.45cm}\sum_{i,j: 1\leq i<j\leq m_c} \hspace{-.6cm} a_j^2a_i^2 (\boldsymbol{k}_i-\boldsymbol{k}_j)\boldsymbol{k}_i^T\\ \nonumber &=\hspace{-.45cm} \sum_{i,j: 1\leq i<j\leq m_c} \hspace{-.6cm} a_i^2a_j^2 (\boldsymbol{k}_j-\boldsymbol{k}_i)(\boldsymbol{k}_j-\boldsymbol{k}_i)^T\\ &\succeq \boldsymbol{0}.
\end{align}
A direct result of (\ref{r20}) and (\ref{r21}) is $\boldsymbol{S}_{\nabla^2\eta_\tau}\succ \boldsymbol{0}$, which completes the proof.$\hspace{.9cm}\square$\\[-.7em]

From an implementation perspective, for the simulations performed in this paper we used $\tau_0=1$ and $\tau_k=10\tau_{k-1}$ for $k=1,2,\cdots,10$, which led us to a reasonably accurate solution of (\ref{r12}). The Newton minimization for $\eta_{\tau_0}(\boldsymbol{x})$ was initialized by the feasible point $\boldsymbol{x}=[\boldsymbol{0};1]$. Thanks to the low-dimensionality of the PaLS framework, the linear systems that provide us with the Newton directions for $\eta_{\tau_k}(\boldsymbol{x})$ are also low-dimensional and computationally tractable.

\section*{Acknowledgement}
This material is based upon work supported by the National Science Foundation under Grant No. EAR 0838313. Any opinions, findings, and conclusions or recommendations expressed in this material are those of the authors and do not necessarily reflect the views of the National Science Foundation. The   authors acknowledge Drs. Lee Slater (Rutgers University) and Dimitrios Ntarlagiannis (Rutgers University) for their helpful insights related to ERT measurements, Dr. John Christ (USAFA) for his help in producing the DNAPL release simulations, and Mr. Zachary Donahue for his help in providing parameters for use in Archie’s Law.


\begin{thebibliography}{10}

\bibitem{abriola1989modeling}
{\sc L.M. Abriola}, {\em Modeling multiphase migration of organic chemicals in
  groundwater systems--a review and assessment.}, Environmental Health
  Perspectives, 83 (1989), p.~117.

\bibitem{abriola2005pilot}
{\sc L.M. Abriola, C.D. Drummond, E.J. Hahn, K.F. Hayes, T.C.G. Kibbey, L.D.
  Lemke, K.D. Pennell, E.A. Petrovskis, C.A. Ramsburg, and K.M. Rathfelder},
  {\em Pilot-scale demonstration of surfactant-enhanced {PCE} solubilization at
  the {Bachman} road site. 1. {S}ite characterization and test design},
  Environmental science \& technology, 39 (2005), pp.~1778--1790.

\bibitem{abubakar2012joint}
{\sc A~Abubakar, G~Gao, TM~Habashy, and J~Liu}, {\em Joint inversion approaches
  for geophysical electromagnetic and elastic full-waveform data}, Inverse
  Problems, 28 (2012), p.~055016.

\bibitem{acar1994analysis}
{\sc R.~Acar and C.R. Vogel}, {\em Analysis of bounded variation penalty
  methods for ill-posed problems}, Inverse Problems, 10 (1994), p.~1217.

\bibitem{aghasi2011parametric}
{\sc A.~Aghasi, M.~Kilmer, and E.L. Miller}, {\em Parametric level set methods
  for inverse problems}, SIAM Journal on Imaging Sciences, 4 (2011),
  pp.~618--650.

\bibitem{aghasi2011sensitivity}
{\sc A.~Aghasi and E.L. Miller}, {\em Sensitivity calculations for {P}oisson's
  equation via the adjoint field method}, Geoscience and Remote Sensing
  Letters, IEEE, 9 (2012), pp.~237 --241.

\bibitem{aghasi2013sparse}
{\sc Alireza Aghasi and Justin Romberg}, {\em Sparse shape reconstruction}, to
  appear in SIAM Journal on Imaging Sciences, preprint arXiv:1303.0018,
  (2013).

\bibitem{ajo2006survey}
{\sc Jonathan~B Ajo-Franklin, Jil~T Geller, and Jerry~M Harris}, {\em A survey
  of the geophysical properties of chlorinated dnapls}, Journal of applied
  geophysics, 59 (2006), pp.~177--189.

\bibitem{archie1942electrical}
{\sc GE~Archie}, {\em The electrical resistivity log as an aid in determining
  some reservoir characteristics}, Transactions of the American Institute of
  Mining, Metallurgical and Petroleum Engineers, 146 (1942), p.~54.

\bibitem{bear1988dynamics}
{\sc J.~Bear}, {\em Dynamics of fluids in porous media}, Dover publications,
  1988.

\bibitem{ben2007projection}
{\sc MK~Ben Hadj~Miled and EL~Miller}, {\em A projection-based level-set
  approach to enhance conductivity anomaly reconstruction in electrical
  resistance tomography}, Inverse Problems, 23 (2007), p.~2375.

\bibitem{bertsekas1999nonlinear}
{\sc D.P. Bertsekas}, {\em Nonlinear Programming}, Athena Scientific, Belmont,
  MA, 1999.

\bibitem{boyd2004convex}
{\sc Stephen~Poythress Boyd and Lieven Vandenberghe}, {\em Convex
  optimization}, Cambridge university press, 2004.

\bibitem{bradford1998flow}
{\sc Scott~A Bradford, Linda~M Abriola, and Klaus~M Rathfelder}, {\em Flow and
  entrapment of dense nonaqueous phase liquids in physically and chemically
  heterogeneous aquifer formations}, Advances in Water Resources, 22 (1998),
  pp.~117--132.

\bibitem{brewster1994ground}
{\sc Michael~L Brewster and A~Peter Annan}, {\em Ground-penetrating radar
  monitoring of a controlled dnapl release; 200 mhz radar}, Geophysics, 59
  (1994), pp.~1211--1221.

\bibitem{brooks1964hydraulic}
{\sc Royal~Harvard Brooks and Arthur~Thomas Corey}, {\em Hydraulic properties
  of porous media}, Hydrology Papers, Colorado State University, 1964.

\bibitem{burdine1953relative}
{\sc NT~Burdine}, {\em Relative permeability calculations from pore size
  distribution data}, Journal of Petroleum Technology, 5 (1953), pp.~71--78.

\bibitem{burger2005survey}
{\sc Martin Burger and Stanley~J Osher}, {\em A survey on level set methods for
  inverse problems and optimal design}, European Journal of Applied
  Mathematics, 16 (2005), pp.~263--301.

\bibitem{cardiff2009bayesian}
{\sc M.~Cardiff and PK~Kitanidis}, {\em Bayesian inversion for facies
  detection: {A}n extensible level set framework}, Water Resources Research, 45
  (2009), p.~W10416.

\bibitem{censor1977pareto}
{\sc Yair Censor}, {\em Pareto optimality in multiobjective problems}, Applied
  Mathematics and Optimization, 4 (1977), pp.~41--59.

\bibitem{chambers2004noninvasive}
{\sc JE~Chambers, MH~Loke, RD~Ogilvy, and PI~Meldrum}, {\em Noninvasive
  monitoring of {DNAPL} migration through a saturated porous medium using
  electrical impedance tomography}, Journal of Contaminant Hydrology, 68
  (2004), pp.~1--22.

\bibitem{chavent2009nonlinear}
{\sc G.~Chavent}, {\em Nonlinear least squares for inverse problems:
  theoretical foundations and step-by-step guide for applications}, Springer
  Verlag, 2009.

\bibitem{Christ2009}
{\sc J.A. Christ, L.D. Lemke, and L.M. Abriola}, {\em The influence of
  dimensionality on simulations of mass recovery from nonuniform dense
  non-aqueous phase liquid ({DNAPL}) source zones}, Advances in Water
  Resources, 32 (2009), pp.~401--412.

\bibitem{christ2006estimating}
{\sc J.A. Christ, C.A. Ramsburg, K.D. Pennel, and L.M. Abriola}, {\em
  Estimating mass discharge from dense nonaqueous phase liquid source zones
  using upscaled mass transfer coefficients: {A}n evaluation using multiphase
  numerical simulations}, Water Resources Research, 42 (2006).

\bibitem{christ2005comparison}
{\sc John~A Christ, Lawrence~D Lemke, and Linda~M Abriola}, {\em Comparison of
  two-dimensional and three-dimensional simulations of dense nonaqueous phase
  liquids (dnapls): Migration and entrapment in a nonuniform permeability
  field}, Water Resources Research, 41 (2005), p.~W01007.

\bibitem{daily1998electrical}
{\sc W.~Daily, A.~Ramirez, and R.~Johnson}, {\em Electrical impedance
  tomography of a perchloroethylene release}, Journal of Environmental \&
  Engineering Geophysics, 2 (1998), pp.~189--201.

\bibitem{dekker2000influence}
{\sc T.J. Dekker and L.M. Abriola}, {\em The influence of field-scale
  heterogeneity on the surfactant-enhanced remediation of entrapped nonaqueous
  phase liquids}, Journal of contaminant hydrology, 42 (2000), pp.~219--251.

\bibitem{Delshad2000}
{\sc M.~Delshad, GA~Pope, and K.~Sepehrnoori}, {\em {UTCHEM} version 9.0
  technical documentation}, Center for Petroleum and Geosystems Engineering,
  The University of Texas at Austin, Austin, Texas, 78751 (2000).

\bibitem{dennis1996numerical}
{\sc J.E. Dennis and R.B. Schnabel}, {\em Numerical methods for unconstrained
  optimization and nonlinear equations}, vol.~16, Society for Industrial
  Mathematics, 1996.

\bibitem{dey1979resistivity}
{\sc A.~Dey and HF~Morrison}, {\em Resistivity modeling for arbitrarily shaped
  three-dimensional structures}, Geophysics, 44 (1979), pp.~753--780.

\bibitem{dorn2006level}
{\sc O.~Dorn and D.~Lesselier}, {\em Level set methods for inverse scattering},
  Inverse Problems, 22 (2006), p.~R67.

\bibitem{dorn2008history}
{\sc Oliver Dorn and Rossmary Villegas}, {\em History matching of petroleum
  reservoirs using a level set technique}, Inverse Problems, 24 (2008),
  p.~035015.

\bibitem{enfield2005design}
{\sc C.G. Enfield, A.L. Wood, F.P. Espinoza, M.C. Brooks, M.~Annable, and PSC
  Rao}, {\em Design of aquifer remediation systems:(1) {D}escribing hydraulic
  structure and {NAPL} architecture using tracers}, Journal of Contaminant
  Hydrology, 81 (2005), pp.~125--147.

\bibitem{engl1996regularization}
{\sc H.W. Engl, M.~Hanke, and A.~Neubauer}, {\em Regularization of inverse
  problems}, vol.~375, Springer, 1996.

\bibitem{ewing2006dependence}
{\sc RP~Ewing and AG~Hunt}, {\em Dependence of the electrical conductivity on
  saturation in real porous media}, Vadose Zone Journal, 5 (2006),
  pp.~731--741.

\bibitem{feng2003curve}
{\sc H.~Feng, W.C. Karl, and D.A. Casta{\~n}on}, {\em A curve evolution
  approach to object-based tomographic reconstruction}, Image Processing, IEEE
  Transactions on, 12 (2003), pp.~44--57.

\bibitem{Finsterle2008Joint}
{\sc S.~Finsterle and M.B. Kowalsky}, {\em Joint hydrologicalgeophysical
  inversion for soil structure identification}, Vadose Zone, 7 (2008),
  pp.~287--293.

\bibitem{fliege2010newton}
{\sc J.~Fliege, LM~Gra{\~n}a~Drummond, and BF~Svaiter}, {\em Newton's method
  for multiobjective optimization}, SIAM Journal on Optimization, 20 (2010),
  p.~602.

\bibitem{gallardo2011structure}
{\sc Luis~A Gallardo and Max~A Meju}, {\em Structure-coupled multiphysics
  imaging in geophysical sciences}, Reviews of Geophysics, 49 (2011),
  p.~RG1003.

\bibitem{gill1981practical}
{\sc P.E. Gill, W.~Murray, and M.H. Wright}, {\em Practical optimization},
  vol.~1, Academic press, 1981.

\bibitem{goes2004effective}
{\sc BJM Goes and JAC Meekes}, {\em An effective electrode configuration for
  the detection of {DNAPL}s with electrical resistivity tomography}, Journal of
  Environmental and Engineering Geophysics, 9 (2004), p.~127.

\bibitem{Gorman1990}
{\sc T.~Gorman and WE~Kelly}, {\em Electrical-hydraulic properties of
  unsaturated {O}ttawa sands}, Journal of Hydrology, 118 (1990), pp.~1--18.

\bibitem{haber1999joint}
{\sc E~Haber and D~Oldenburg}, {\em Joint inversion: A structural approach},
  Inverse problems, 13 (1999), p.~63.

\bibitem{hamilton2013direct}
{\sc S.J. Hamilton and J.L. Mueller}, {\em Direct eit reconstructions of
  complex admittivities on a chest-shaped domain in 2-d}, Medical Imaging, IEEE
  Transactions on, 32 (2013), pp.~757--769.

\bibitem{hinnell2010improved}
{\sc AC~Hinnell, TPA Ferr{\'e}, JA~Vrugt, JA~Huisman, S~Moysey, J~Rings, and
  MB~Kowalsky}, {\em Improved extraction of hydrologic information from
  geophysical data through coupled hydrogeophysical inversion}, Water resources
  research, 46 (2010), p.~W00D40.

\bibitem{horvath1982halogenated}
{\sc AL~Horvath}, {\em Halogenated hydrocarbons: Solubility-miscibility with
  water}, M. Dekker (New York), 1982.

\bibitem{hoversten2006direct}
{\sc G~Michael Hoversten, Florence Cassassuce, Erika Gasperikova, Gregory~A
  Newman, Jinsong Chen, Yoram Rubin, Zhangshuan Hou, and Don Vasco}, {\em
  Direct reservoir parameter estimation using joint inversion of marine seismic
  ava and csem data}, Geophysics, 71 (2006), pp.~C1--C13.

\bibitem{hu2009joint}
{\sc Wenyi Hu, Aria Abubakar, and Tarek~M Habashy}, {\em Joint electromagnetic
  and seismic inversion using structural constraints}, Geophysics, 74 (2009),
  pp.~R99--R109.

\bibitem{hunt2004continuum}
{\sc AG~Hunt}, {\em Continuum percolation theory and {A}rchie's law},
  Geophysical research letters, 31 (2004), p.~L19503.

\bibitem{hyndman1994coupled}
{\sc David~W Hyndman, Jerry~M Harris, and Steven~M Gorelick}, {\em Coupled
  seismic and tracer test inversion for aquifer property characterization},
  Water Resources Research, 30 (1994), pp.~1965--1978.

\bibitem{jin1995partitioning}
{\sc M.~Jin, M.~Delshad, V.~Dwarakanath, D.C. McKinney, G.A. Pope,
  K.~Sepehrnoori, C.E. Tilburg, and R.E. Jackson}, {\em Partitioning tracer
  test for detection, estimation, and remediation performance assessment of
  subsurface nonaqueous phase liquids}, Water Resources Research, 31 (1995),
  pp.~1201--1211.

\bibitem{kilmer2003three}
{\sc M.E. Kilmer, E.~L. Miller, A.~Barbaro, and D.~Boas}, {\em
  Three-dimensional shape-based imaging of absorption perturbation for diffuse
  optical tomography}, Applied Optics, 42 (2003), pp.~3129--3144.

\bibitem{knight2001ground}
{\sc Rosemary Knight}, {\em Ground penetrating radar for environmental
  applications}, Annual Review of Earth and Planetary Sciences, 29 (2001),
  pp.~229--255.

\bibitem{koch2009joint}
{\sc K.~Koch, J.~Wenninger, S.~Uhlenbrook, and M.~Bonell}, {\em Joint
  interpretation of hydrological and geophysical data: electrical resistivity
  tomography results from a process hydrological research site in the {B}lack
  {F}orest {M}ountains, {G}ermany}, Hydrological Processes, 23 (2009),
  pp.~1501--1513.

\bibitem{lemke2004dense}
{\sc L.D. Lemke, L.M. Abriola, and P.~Goovaerts}, {\em Dense nonaqueous phase
  liquid ({DNAPL}) source zone characterization: Influence of hydraulic
  property correlation on predictions of {DNAPL} infiltration and entrapment},
  Water Resources Research, 40 (2004), p.~W01511.

\bibitem{linde2006improved}
{\sc N.~Linde, A.~Binley, A.~Tryggvason, L.B. Pedersen, and A.~Revil}, {\em
  Improved hydrogeophysical characterization using joint inversion of
  cross-hole electrical resistance and ground-penetrating radar traveltime
  data.}, Water Resources Research, 42 (2006), p.~W04410.

\bibitem{lu2006parameter}
{\sc Zhiming Lu and Bruce~A Robinson}, {\em Parameter identification using the
  level set method}, Geophysical research letters, 33 (2006), p.~L06404.

\bibitem{madsen1999methods}
{\sc K.~Madsen, HB~Nielsen, and O.~Tingleff}, {\em Methods for non-linear least
  squares problems}, 1999.

\bibitem{marquardt1963algorithm}
{\sc Donald~W Marquardt}, {\em An algorithm for least-squares estimation of
  nonlinear parameters}, Journal of the Society for Industrial \& Applied
  Mathematics, 11 (1963), pp.~431--441.

\bibitem{miller2012environmental}
{\sc Eric~L Miller, Linda~M Abriola, and Alireza Aghasi}, {\em Environmental
  remediation and restoration: Hydrological and geophysical processing
  methods}, Signal Processing Magazine, IEEE, 29 (2012), pp.~16--26.

\bibitem{moorkamp2010joint}
{\sc M~Moorkamp, AG~Jones, and S~Fishwick}, {\em Joint inversion of receiver
  functions, surface wave dispersion, and magnetotelluric data}, Journal of
  Geophysical Research, 115 (2010), p.~B04318.

\bibitem{more1978levenberg}
{\sc Jorge~J Mor{\'e}}, {\em The levenberg-marquardt algorithm: implementation
  and theory}, in Numerical analysis, Springer, 1978, pp.~105--116.

\bibitem{moreno2006influence}
{\sc E.~Moreno-Barbero and T.H. Illangasekare}, {\em Influence of dense
  nonaqueous phase liquid pool morphology on the performance of partitioning
  tracer tests: {E}valuation of the equilibrium assumption}, Water resources
  research, 42 (2006).

\bibitem{osher2003level}
{\sc S.~Osher and R.P. Fedkiw}, {\em Level set methods and dynamic implicit
  surfaces}, vol.~153, Springer Verlag, 2003.

\bibitem{osher1988fronts}
{\sc S.~Osher and J.A. Sethian}, {\em Fronts propagating with
  curvature-dependent speed: {A}lgorithms based on {H}amilton-{J}acobi
  formulations}, Journal of Computational Physics, 79 (1988), pp.~12--49.

\bibitem{parker1987model}
{\sc JC~Parker and RJ~Lenhard}, {\em A model for hysteretic constitutive
  relations governing multiphase flow: 1. saturation-pressure relations}, Water
  Resources Research, 23 (1987), pp.~2187--2196.

\bibitem{Parker2004}
{\sc JC~Parker and E.~Park}, {\em Modeling field-scale dense nonaqueous phase
  liquid dissolution kinetics in heterogeneous aquifers}, Water Resources
  Research, 40 (2004), p.~W05109.

\bibitem{pollock2008temporal}
{\sc D.~Pollock and O.A. Cirpka}, {\em Temporal moments in geoelectrical
  monitoring of salt tracer experiments}, Water Resources Research, 44 (2008),
  p.~W12416.

\bibitem{pollock2012fully}
\leavevmode\vrule height 2pt depth -1.6pt width 23pt, {\em Fully coupled
  hydrogeophysical inversion of a laboratory salt tracer experiment monitored
  by electrical resistivity tomography}, Water Resources Research, 48 (2012),
  p.~W01505.

\bibitem{Polydorides2012}
{\sc N.~Polydorides, A.~Aghasi, and E.~Miller}, {\em High-order regularized
  regression in electrical impedance tomography}, SIAM Journal on Imaging
  Sciences, 5 (2012), pp.~912--943.

\bibitem{polydorides2012high}
{\sc Nick Polydorides, Alireza Aghasi, and Eric~L Miller}, {\em High-order
  regularized regression in electrical impedance tomography}, SIAM Journal on
  Imaging Sciences, 5 (2012), pp.~912--943.

\bibitem{powers1992experimental}
{\sc Susan~E Powers, Linda~M Abriola, and Walter~J Weber}, {\em An experimental
  investigation of nonaqueous phase liquid dissolution in saturated subsurface
  systems: Steady state mass transfer rates}, Water Resources Research, 28
  (1992), pp.~2691--2705.

\bibitem{Ramsburg2001}
{\sc C.A. Ramsburg and K.D. Pennell}, {\em Experimental and economic assessment
  of two surfactant formulations for source zone remediation at a former dry
  cleaning facility}, Ground Water Monitoring \& Remediation, 21 (2001),
  pp.~68--82.

\bibitem{Ramsburg2004}
{\sc C.A. Ramsburg, K.D. Pennell, T.C.G. Kibbey, and K.F. Hayes}, {\em
  Refinement of the density-modified displacement method for efficient
  treatment of tetrachloroethene source zones}, Journal of contaminant
  hydrology, 74 (2004), pp.~105--131.

\bibitem{smith1995real}
{\sc Robert~WM Smith, Ian~Leslie Freeston, and BH~Brown}, {\em A real-time
  electrical impedance tomography system for clinical use-design and
  preliminary results}, Biomedical Engineering, IEEE Transactions on, 42
  (1995), pp.~133--140.

\bibitem{temples2001noninvasive}
{\sc Tom~J Temples, Michael~G Waddell, William~J Domoracki, and Jerome Eyer},
  {\em Noninvasive determination of the location and distribution of dnapl
  using advanced seismic relfection techniques}, Ground Water, 39 (2001),
  pp.~465--474.

\bibitem{Tikhonov1977}
{\sc A.N. Tikhonov and V.~Arsenin}, {\em Solutions of ill-posed problems},
  Winston Washington, DC, 1977.

\bibitem{EPA1986Background}
{\sc US-EPA}, {\em Background document for the ground-water screening procedure
  to support 40 cfr part 269: Land disposal, epa/530-sw-86-047}, 1986.

\bibitem{vozoff1975joint}
{\sc K~Vozoff and DLB Jupp}, {\em Joint inversion of geophysical data},
  Geophysical Journal of the Royal Astronomical Society, 42 (1975),
  pp.~977--991.

\bibitem{WeberJr1996}
{\sc WJ~Weber~Jr and F.A. DiGiano}, {\em Process dynamics in environmental
  systems}, John Wiley and Sons, New York,  (1996).

\bibitem{Yeh2007}
{\sc T.C.J. Yeh and J.~Zhu}, {\em Hydraulic/partitioning tracer tomography for
  characterization of dense nonaqueous phase liquid source zones}, Water
  Resources Research, 43 (2007), p.~W06435.

\bibitem{Zheng1999}
{\sc C.~Zheng and P.P. Wang}, {\em {MT3DMS}: a modular three-dimensional
  multispecies transport model for simulation of advection, dispersion, and
  chemical reactions of contaminants in groundwater systems; documentation and
  user's guide}, 1999.

\end{thebibliography}
\end{document}